\RequirePackage{lineno}
\documentclass[a4paper,11pt,epsfig]{article}
\pdfoutput=1 

\usepackage{jheppub} 
\usepackage[T1]{fontenc} 
\usepackage{graphicx,color,dcolumn,booktabs,bm}
\usepackage{longtable,lscape}
\usepackage{amsmath}
\usepackage{indentfirst}
\usepackage{epsfig}
\usepackage{epstopdf}   
\usepackage{slashed}  
\usepackage{color}
\usepackage{multirow}
\usepackage{soul}  
\usepackage{graphicx,color,dcolumn,booktabs,bm}
\usepackage{verbatim}
\usepackage{orcidlink}

\usepackage{listings}
\newcommand{\BR}{{\cal B}}

\title{\boldmath Improved measurement of Born cross sections for $\chi_{bJ}\,\omega$ and $\chi_{bJ}\,(\pi^+\pi^-\pi^0)_{\rm non-\omega}$ ($J$ = 0, 1, 2) at Belle and Belle II}

\preprint{\vbox{\hbox{}
                        \hbox{}
                        \hbox{}
                        \hbox{Belle II Preprint 2025-003}
                        \hbox{KEK Preprint 2024-52}}}
                        
\vspace{20cm}
  \author{I.~Adachi\,\orcidlink{0000-0003-2287-0173},} 
  \author{L.~Aggarwal\,\orcidlink{0000-0002-0909-7537},} 
  \author{H.~Ahmed\,\orcidlink{0000-0003-3976-7498},} 
  \author{H.~Aihara\,\orcidlink{0000-0002-1907-5964},} 
  \author{N.~Akopov\,\orcidlink{0000-0002-4425-2096},} 
  \author{M.~Alhakami\,\orcidlink{0000-0002-2234-8628},} 
  \author{A.~Aloisio\,\orcidlink{0000-0002-3883-6693},} 
  \author{N.~Althubiti\,\orcidlink{0000-0003-1513-0409},} 
  \author{M.~Angelsmark\,\orcidlink{0000-0003-4745-1020},} 
  \author{N.~Anh~Ky\,\orcidlink{0000-0003-0471-197X},} 
% ...
\collaboration{The Belle and Belle II Collaborations}

\abstract{We study the processes $\chi_{bJ}\,\omega$ and $\chi_{bJ}\,(\pi^+\pi^-\pi^0)_{\rm non-\omega}$ ($J$ = 0, 1, 2) at center-of-mass 
energies $\sqrt{s}$ from 10.73 to 11.02 GeV using a $142.5\,\mathrm{fb}^{-1}$ data sample collected with the Belle detector
at the KEKB asymmetric-energy $e^+ e^-$ collider; and at $\sqrt{s}\sim10.75$ GeV using a $19.8\,\mathrm{fb}^{-1}$ sample collected with Belle
II at SuperKEKB.
We find that the $\Upsilon(10753)$ state decays into $\chi_{bJ}\,\omega$ but not into $\chi_{bJ}\,(\pi^+\pi^-\pi^0)_{\rm non-\omega}$, while the $\Upsilon(10860)$ state, in contrast, decays into $\chi_{bJ}\,(\pi^+\pi^-\pi^0)_{\rm non-\omega}$ but not into $\chi_{bJ}\,\omega$. The mass and width of the $\Upsilon(10753)$ state are measured to be $(10756.1\pm3.4({\rm stat.})\pm2.7({\rm syst.}))$ MeV/$c^2$ and $(32.2\pm11.3({\rm stat.})\pm14.9({\rm syst.}))$ MeV.
The products of the partial width to $e^+e^-$ and branching fractions for $\Upsilon(10753)\to\chi_{b1}\,\omega$ and $\Upsilon(10753)\to\chi_{b2}\,\omega$ are ($1.57\pm0.27({\rm stat.})\pm 0.22({\rm syst.})$) eV and ($1.39\pm0.41({\rm stat.})\pm 0.33({\rm syst.})$) eV.}


\begin{document} 
\maketitle
\flushbottom

\section{Introduction}

In the energy dependence of the $e^+e^-\to\Upsilon(nS)\pi^+\pi^-$ $(n = 1, 2, 3)$ cross sections, the Belle
experiment observed a new structure, $\Upsilon(10753)$, with a global significance of 5.2$\sigma$~\cite{220}.
Its existence was further supported by fits to the `dressed' cross sections
$\sigma(e^+e^-\to b\bar b)$ at center-of-mass (c.m.) energies $\sqrt{s}$ from 10.6 to 11.2 GeV~\cite{083001}.
The $\Upsilon(10753)$ state has been interpreted as a conventional bottomonium~\cite{074007,034036,59,014020,014036,357,04049,135340,11915,103845,2501.15110,2508.18720}, hybrid~\cite{034019,1}, or tetraquark state~\cite{135217,074507,11475,123102,381,2503.00552}.
To confirm this state and to study its properties, the Belle II experiment performed an energy scan, collecting four samples at energies 10653, 10701, 10745, and 10805 MeV and integrated luminosities of 3.6, 1.6, 9.9, and 4.7 fb$^{-1}$, respectively.
The $\Upsilon(10753)$ state was confirmed in the $\Upsilon(nS)\pi^+\pi^-$ channel~\cite{2401.12021}; from a combined fit to the Belle and Belle II data, its mass and width are $(10756.6\pm2.7\pm0.9)$ MeV/$c^2$ and $(29.0\pm8.8\pm1.2)$ MeV, respectively, where the first and second uncertainties are statistical and systematic throughout this paper.
Resonant substructure in the $\Upsilon(10753)\to\Upsilon(nS)\pi^+\pi^-$ decays was studied; no signal of intermediate $Z_b(10610,10650)^+\to\Upsilon(nS)\pi^+$ states was found. In addition, contrary to expectations of the hybrid and tetraquark models~\cite{034019,135217}, no prominent $f_0(980)\to\pi^+\pi^-$ signal was observed.

Belle II searched for the $e^+e^-\to\eta_b(1S)\omega$ process at $\sqrt{s} = 10745\,\mathrm{MeV}$, reporting a null result~\cite{072013}.
Belle II measured the energy dependence of the $e^+e^-\to\chi_{bJ}(1P)\,\omega$ $(J=1,2)$ cross sections and observed an enhancement with a shape consistent with $\Upsilon(10753)$~\cite{091902}. The ratio of $\chi_{b2}$ and $\chi_{b1}$ was measured to be $1.3\pm0.6$ at $\sqrt{s}$ = 10.745 GeV, which is far below the expectation of 15 for a pure $D$-wave bottomonium state~\cite{172}. 
There is also a $1.8\sigma$ difference with the prediction of 0.2
for an $S$-$D$-mixed state~\cite{034036}. The results are consistent with the
$\chi_{bJ}(1P)\,\omega$ production at the $\Upsilon(10860)$ energy,
reported earlier by Belle~\cite{142001}, being due to the tail of the
$\Upsilon(10753)$ state and not due to the decay of the $\Upsilon(10860)$ itself.

In this paper, we perform a systematic study of the $e^+e^- \to \chi_{bJ}(1P)\,\omega$ and $e^+e^-\to\chi_{bJ}\,(\pi^+\pi^-\pi^0)_{\rm non-\omega}$ cross sections in the energy range from 10653 to 11020 MeV using both Belle and Belle II data. We update the Belle II measurement of ref.~\cite{091902} by exploiting the precise $E_{\rm cm}$ ($e^+e^-$ c.m.\ energy) calibration which recently became available~\cite{114}, and performing kinematic fits to better separate the $\chi_{bJ}(1P)$ and $\chi_{bJ}(2P)$ states. In the case of Belle, the energy dependence of $e^+e^- \to \chi_{bJ}\,\pi^+\pi^-\pi^0$ was reported earlier~\cite{091102} without considering the $\omega$ and $(\pi^+\pi^-\pi^0)_{\rm non-\omega}$ contributions separately. For brevity, we denote $\chi_{bJ}(1P)$ as $\chi_{bJ}$ throughout this paper.

The paper is organized as follows.
We describe the Belle and Belle II detectors in section~\ref{sec.2} and the data sets in section~\ref{sec.3}. The selection of events is presented in section~\ref{sec.4}. Sections~\ref{sec.5} and~\ref{sec.6} are devoted to the measurements of Born cross sections for $e^+e^-\to\chi_{bJ}\,\omega$ and $e^+e^-\to\chi_{bJ}\,(\pi^+\pi^-\pi^0)_{\rm non-\omega}$. 
The evaluation of the systematic uncertainties is described in section~\ref{sec.7}.
The fits to the energy dependences of Born cross sections for $e^+e^-\to\chi_{b1,b2}\,\omega$ and $e^+e^-\to\chi_{b1,b2}\,(\pi^+\pi^-\pi^0)_{\rm non-\omega}$ are described in section~\ref{sec.8}. The results are summarized in section~\ref{sec.9}.

\section{Belle and Belle II detectors}\label{sec.2}

We use data collected by the Belle detector~\cite{Belle1,Belle2} at the KEKB asymmetric energy $e^+e^-$ collider~\cite{KEKB1,KEKB2} and the Belle II detector~\cite{BelleII} at SuperKEKB~\cite{SuperKEKB}. 

The Belle detector was a large-solid-angle magnetic spectrometer that consisted of a silicon vertex detector, a 50-layer central drift chamber (CDC), an array of aerogel threshold Cherenkov counters (ACC), a barrel-like arrangement of time-of-flight scintillation counters (TOF), and an electromagnetic calorimeter comprised of CsI(Tl) crystals (ECL) located inside a superconducting solenoid coil that provides a 1.5 T magnetic field. An iron flux-return yoke instrumented with resistive plate chambers located outside the coil was used to detect $K^0_L$ mesons and identify muons (KLM). A detailed description of the Belle detector can be found in refs.~\cite{Belle1,Belle2}. 

The Belle II detector is an upgraded version of the Belle detector that contains several completely new subdetectors, as well as substantial upgrades to others. The innermost subdetector is the vertex detector (VXD), which uses position-sensitive silicon sensors to precisely sample the trajectories of charged particles (tracks) in the vicinity of the interaction point. The VXD includes two inner layers of pixel sensors and four outer layers of double-sided silicon microstrip sensors. 
Only one sixth of the second pixel layer had been installed for the data analyzed here.
Charged-particle momenta and charges are measured by a new large-radius, helium-ethane, small-cell CDC, which also gives charged-particle-identification information through a measurement of particles’ specific ionization. The Belle PID system has been replaced. A Cherenkov-light angle and time-of-propagation (TOP) detector surrounding the CDC provides charged-particle identification in the central detector volume, supplemented by a proximity-focusing aerogel ring-imaging Cherenkov (ARICH) detector in the forward region. 
The readout electronics of the ECL have been upgraded, and in the KLM, the resistive plate chambers in the endcaps and the inner two layers of the barrel have been replaced by scintillator strips.
A detailed description of the Belle II detector can be found in ref.~\cite{BelleII}. 

\section{Data sets}~\label{sec.3}

The results are obtained from Belle and Belle II data samples.
We use Belle energy scan data at 18 energy points with approximately 1 fb$^{-1}$ per point collected in the energy range from 10.73 GeV to 11.02 GeV. We also use a $\Upsilon(5S)$ energy region data sample collected at 10865.8 MeV with a total luminosity of 122 fb$^{-1}$.
Thus, there are 19 energy points in total at which we measure cross sections for $e^+e^-\to\chi_{bJ}\,\omega$ and $\chi_{bJ}\,(\pi^+\pi^-\pi^0)_{\rm non-\omega}$ at Belle. The energies and luminosities of these data samples are given in table~\ref{tabsumchib1}.
We also use Belle II data at $\sqrt{s}$ = 10.653, 10.701, 10.745, and 10.805 GeV, corresponding to integrated luminosities of 3.6, 1.6, 9.9, and 4.7 fb$^{-1}$, respectively.

We generate signal Monte Carlo (MC) events with c.m.\ energies from 10.73 to 11.02~GeV at Belle and 10.653 to 10.805 GeV at Belle II using the {\sc evtgen} generator~\cite{152} to determine the reconstruction efficiency and signal shape. 
Initial-state radiation (ISR) at next-to-leading order accuracy in quantum electrodynamics is simulated with {\sc phokhara}~\cite{71}.
Both $e^+e^- \to \chi_{bJ}\,\omega$ and $e^+e^- \to\chi_{bJ}\,(\pi^+\pi^-\pi^0)_{\rm non-\omega}$ samples are generated isotropically in the c.m.\ system; the $\chi_{bJ}\,(\pi^+\pi^-\pi^0)_{\rm non-\omega}$ sample is
generated according to four-body phase space, and then reweighted to
match the observed $M(\pi^+ \pi^- \pi^0)$ distribution for this sample in
data (see figure~\ref{figadd2} left plot below).
For $\omega\to\pi^+\pi^-\pi^0$, the {\sc omega\_dalitz} model is used~\cite{152}.
For $\chi_{bJ}\to\Upsilon(1S)\gamma$, the {\sc helamp} model is used~\cite{152}.
For $\Upsilon(1S)\to\ell^+\ell^-$, the {\sc vll} model is used~\cite{152}.
Generic MC samples of  $e^+e^-\to q\bar q$  ($q=u,d,s,c$) and $\Upsilon(5S)\to B_s^{(*)}\bar B_s^{(*)}$ produced with 4 times the luminosity of the data, are used to identify possible peaking backgrounds.
MC samples of
$e^+e^-\to\Upsilon(2S)\pi^+\pi^-$ with $\Upsilon(2S)\to\chi_{bJ}\gamma$, $e^+e^-\to\Upsilon_2(1D)\pi^+\pi^-$ with $\Upsilon_2(1D)\to\chi_{bJ}\gamma$, and $e^+e^-\to\Upsilon(1S)\pi^+\pi^-\pi^0\pi^0$ are produced to study possible backgrounds in $e^+e^-\to\chi_{bJ}\,(\pi^+\pi^-\pi^0)_{\rm non-\omega}$.
The simulated events are processed with a detector simulation based on {\sc geant3}~\cite{geant3} and {\sc geant4}~\cite{geant4} at Belle and Belle II, respectively.

\section{Selection criteria}~\label{sec.4}

We reconstruct the decay chain $e^+e^-\to\chi_{bJ}\,\pi^+\pi^-\pi^0$, $\chi_{bJ}\to\Upsilon(1S)\gamma$, and $\Upsilon(1S)\to\ell^+\ell^-~(\ell=e~{\rm or}~\mu)$, where the decay to $\pi^+\pi^-\pi^0$ may proceed via the $\omega$.

For Belle data, the following selection criteria are applied.
All charged tracks are required to originate from the vicinity of the interaction point; the impact parameters perpendicular to and along the beam direction with respect to the interaction point are required to be less than $1$~cm and $4$~cm, respectively. 
We require the number of charged tracks to be four.
Additionally for each charged track, information from different detector subsystems including specific ionization in the CDC, time measurements in the TOF, and the response of the ACC are combined to form a likelihood $\mathcal{L}_i$ for particle species $i$, where $i$ = $\pi$, $K$, or $p$~\cite{pid}. Charged tracks with $\mathcal{R}_\pi=\frac{\mathcal{L}_\pi}{\mathcal{L}_K+\mathcal{L}_\pi}>0.6$ are identified as pions.
With this selection, the pion identification efficiency is 94\% and the kaon misidentification rate is 6\%.

Similar likelihood ratios are defined for lepton identification: $\mathcal{R}_e=\frac{\mathcal{L}_e}{\mathcal{L}_e+\mathcal{L}_{{\rm non}-e}}$ for electrons~\cite{eid}, and $\mathcal{R}_\mu=\frac{\mathcal{L}_\mu}{\mathcal{L}_\pi+\mathcal{L}_K+\mathcal{L}_\mu}$ for muons~\cite{muid}.
For lepton candidates we apply the requirements $\mathcal{R}_e>0.1$ and $\mathcal{R}_\mu>0.8$, which have efficiencies of 97\% and 96\%, respectively.
For $\Upsilon(1S)\to e^+e^-$, in order to reduce the effect of bremsstrahlung and final state radiation, photons detected in the ECL within 50 mrad of the original $e^+$ or $e^-$ direction are included in the calculation of the $e^+$ or $e^-$ momentum. The $\Upsilon(1S)$ signal regions are 9.20 $<$ $M(e^+e^-)$ $<$ 9.61 GeV/$c^2$ and 9.31 $<$ $M(\mu^+\mu^-)$ $<$ 9.61 GeV/$c^2$ (approximately $\pm$2.0$\sigma$ with $\sigma$ being the mass resolution).

An ECL cluster is treated as a photon candidate if it does not match the extrapolation of any charged track and its energy is greater than
30 MeV. In reconstructing $\pi^0$ decays, we reject photon candidates if the ratio of energies deposited in the central 3 $\times$ 3 square of cells to that deposited in the enclosing 5 $\times$ 5 square of cells in its ECL cluster is less than 0.8. This suppresses photon candidates originating from neutral hadrons.
The signal region for $\pi^0$ candidates is $110<M(\gamma\gamma)<150$ MeV/$c^2$ (approximately $\pm$2.5$\sigma$ with $\sigma$ being the mass resolution). Hereinafter, the variable $M$ denotes invariant mass. To calibrate the photon energy resolution function, three control channels $D^{*0} \to D^0\gamma$, $\pi^0 \to \gamma\gamma$, and $\eta \to \gamma\gamma$ are used~\cite{232002}.

To improve track momenta and photon energy resolutions and to reduce background, a 6C kinematic fit is performed, where the four-momentum of the final state system is constrained to match the initial $e^+e^-$ c.m.\ system, the invariant mass of the lepton pair is constrained to the $\Upsilon(1S)$ nominal mass~\cite{PDG}, and the invariant mass of the two photons is constrained to the nominal $\pi^0$ mass~\cite{PDG}.
The $\chi^2_{\rm 6C}/{\rm n_{dof}}$ value of the 6C fit is required to be less than 10, which has an efficiency of 90\%.
This requirement removes events with one or more additional or missing particles in the final state.
In events with multiple candidates, only the candidate with the smallest value of $\chi^2_{\rm 6C}/{\rm n_{dof}}$ is retained.
The fraction of selected events that have multiple candidates is 7\% in signal MC samples. These values are consistent with the multiple candidate rates observed in the data. The fraction of correctly reconstructed candidates in the signal MC sample is 95\%.
The selection requirements are optimized using a figure of merit $S/\sqrt{S+B}$, where $S$ and $B$ are the numbers of expected signal and background events at $\sqrt{s}$ =10.866 GeV~\cite{142001}.

The event selection in Belle II is very similar to that in Belle. We use the same requirements as in the previous Belle II study~\cite{091902},
except for the kinematic fit. In ref.~\cite{091902}, a 4C fit was used to suppress the background and select a signal candidate, but the momenta
of all particles were not updated after the fit because a reliable $e^+e^-$ c.m.\ energy calibration was not yet available. In this paper, we
perform a 6C fit and update the momenta, which helps to improve the resolution of $M(\Upsilon(1S)\gamma)$ by a factor of 1.9.

At Belle II, all charged-particle tracks are required to originate from the vicinity of the interaction point.
We require four or five tracks to reduce backgrounds while allowing for increased efficiency for signal events with an additional misreconstructed track.
The identification of pions, electrons, and muons is based on likelihood information from subdetectors~\cite{2506.04355}.
The pion identification efficiency is 90\% and the kaon misidentification rate is 8\%.
When forming $\Upsilon(1S)\to\ell^+\ell^-$ candidates, we only impose a lepton identification requirement on one of the two tracks; the identification efficiency is 95\% for electrons and 90\% for muons.
To reduce the effects of bremsstrahlung and final-state radiation, photons within a 50 mrad cone of the initial electron or positron direction are included in the calculation of the particle four-momentum.
Energy deposits in the ECL are treated as photon candidates if they are not associated with charged particles.
We reject photon candidates if the ratio of energies deposited in the central $3\times3$ square of cells to that deposited 
in the enclosing $5\times5$ square of cells (with corner cells omitted) in its ECL cluster
is less than 0.8.
Photons used to reconstruct $\pi^0$ candidates are required to pass energy
requirements depending on the region of the ECL in which they are
reconstructed: greater than 25 MeV in the barrel ($32.2^\circ < \theta <
128.7^\circ$) and the forward endcap ($12.4^\circ < \theta < 31.4^\circ$), and
greater than 40 MeV in the backward endcap ($130.7^\circ < \theta <
155.1^\circ$). They are also required to satisfy $0.105<M(\gamma\gamma)<0.150$ GeV/$c^2$ (approximately~$\pm$2.5$\sigma$).

\section{$e^+e^-\to\chi_{bJ}\,\omega$ at Belle and Belle II}~\label{sec.5}

\begin{figure}[htbp]
\centering
\includegraphics[width=6cm]{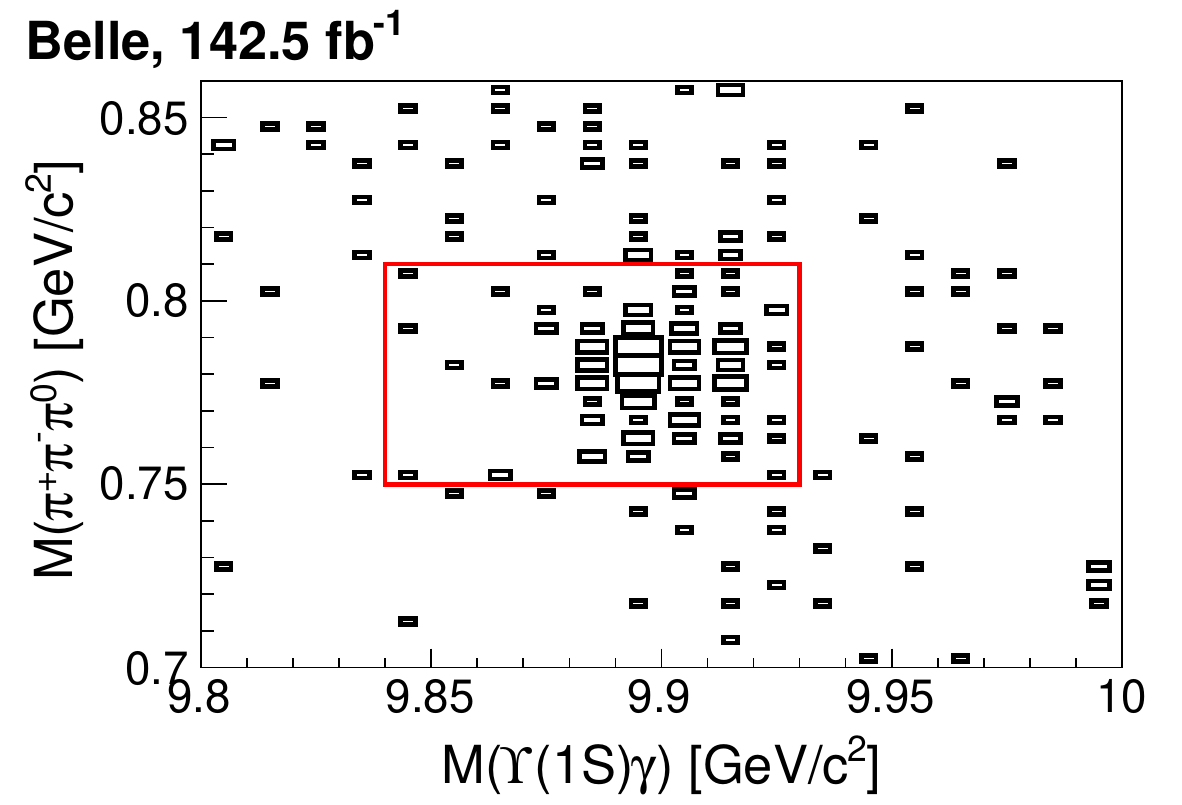}
\includegraphics[width=6cm]{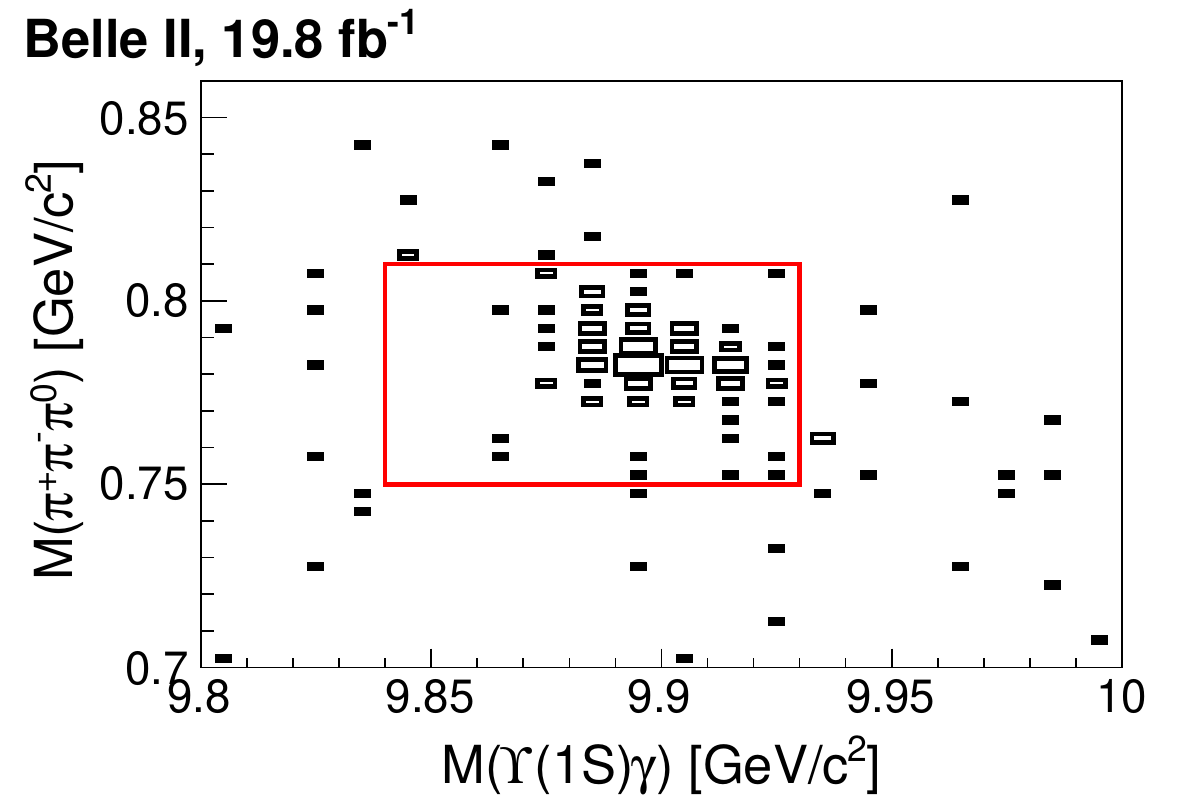}
\caption{Scatter plots of $M(\pi^+\pi^-\pi^0)$ versus $M(\Upsilon(1S)\gamma)$ for selected events in
Belle and Belle II data with all energies combined. The red boxes show the $\omega$ and $\chi_{bJ}$ signal regions.}\label{fig2}
\end{figure}

Figure~\ref{fig2} shows scatter plots of $M(\pi^+\pi^-\pi^0)$ versus $M(\Upsilon(1S)\gamma)$ for selected events in Belle and Belle II data, combining all energy points. There is a clear clustering of events in the
signal region, which is defined as 0.75 $<$ $M(\pi^+\pi^+\pi^0)$ $<$ 0.81 GeV/$c^2$ and 9.84 $<$ $M(\Upsilon(1S)\gamma)$ $<$ 9.93 GeV/$c^2$.

\begin{figure}[htbp]
\centering
\includegraphics[width=3.6cm]{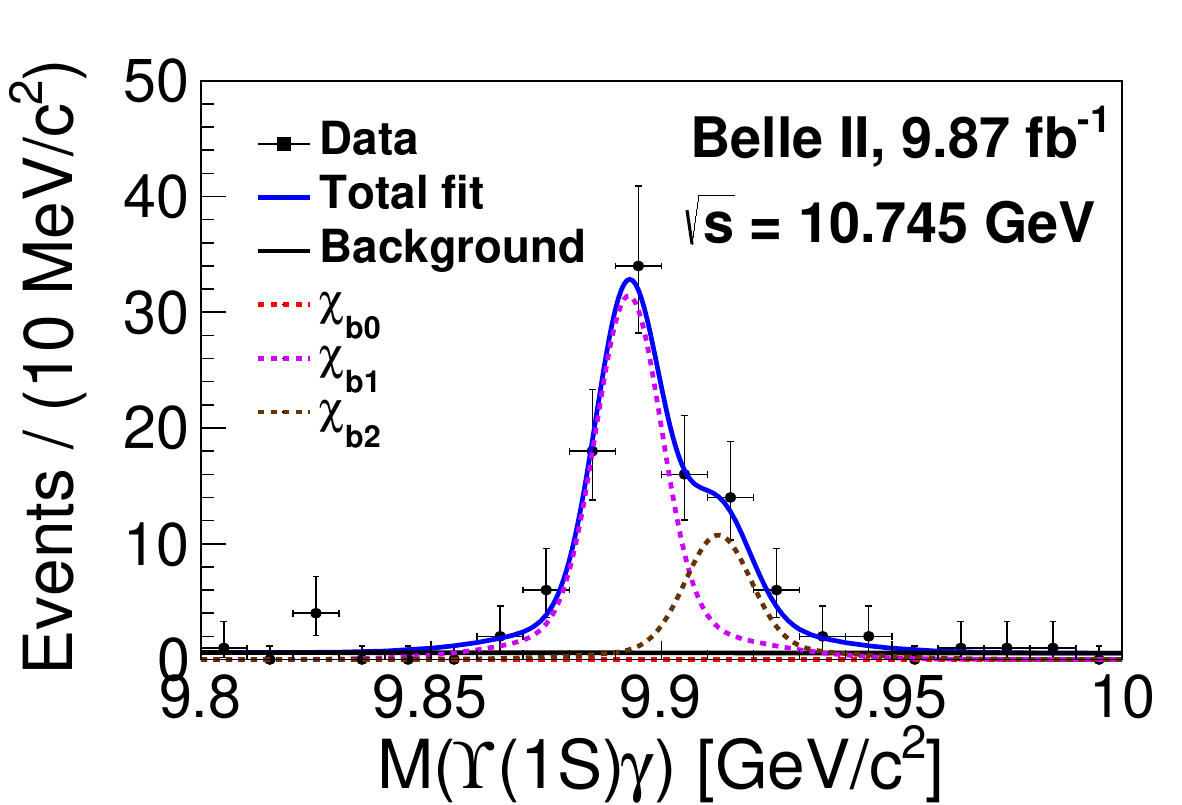}
\includegraphics[width=3.6cm]{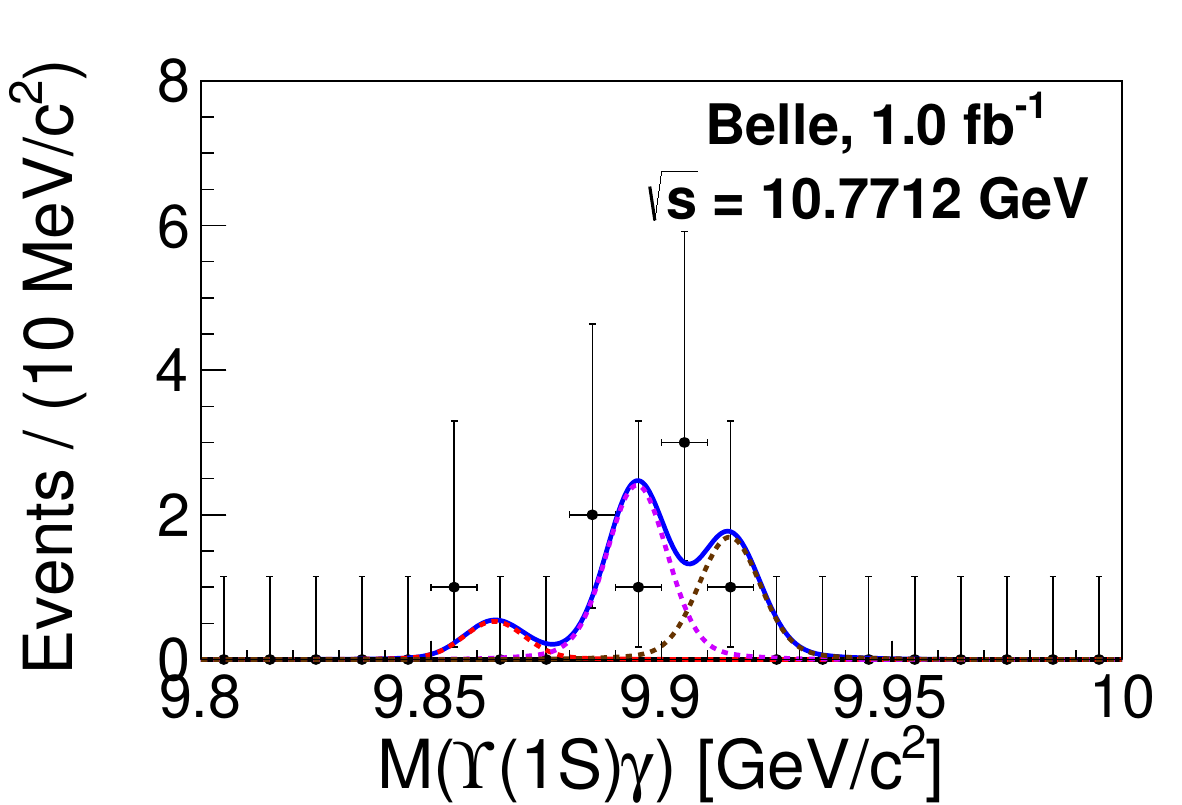}
\includegraphics[width=3.6cm]{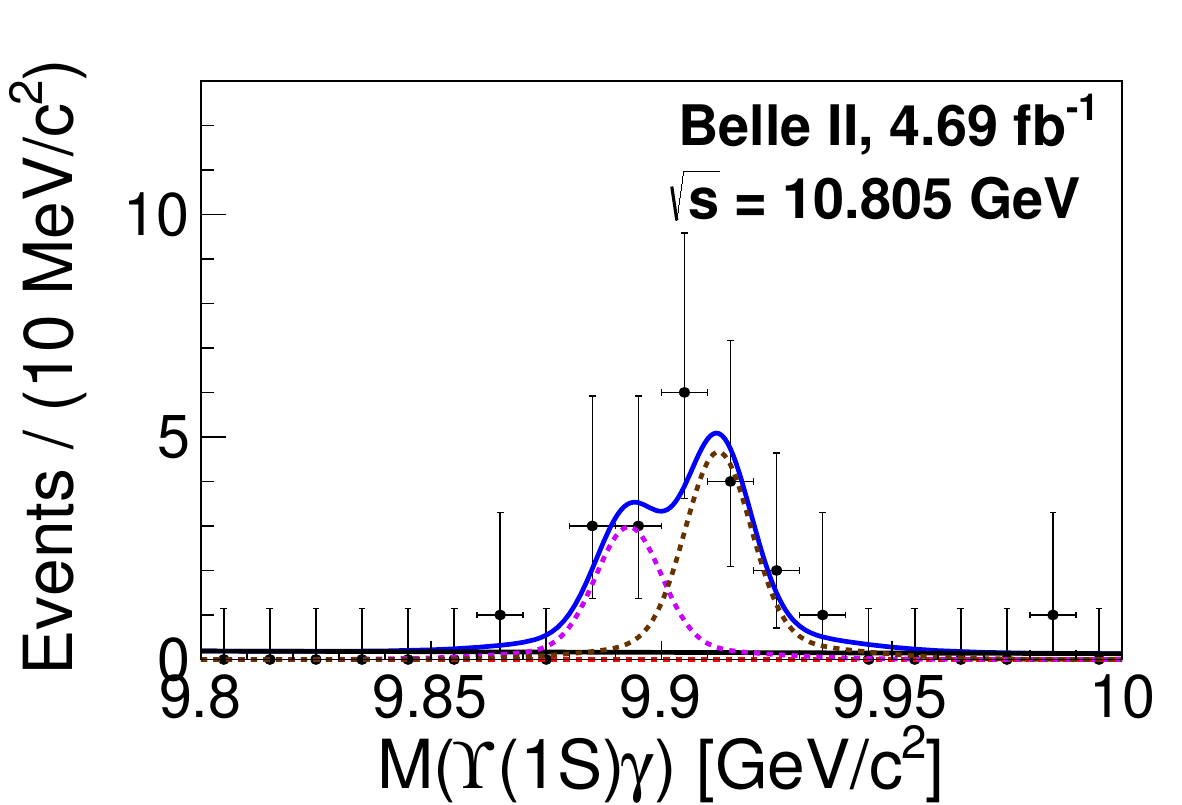}
\includegraphics[width=3.6cm]{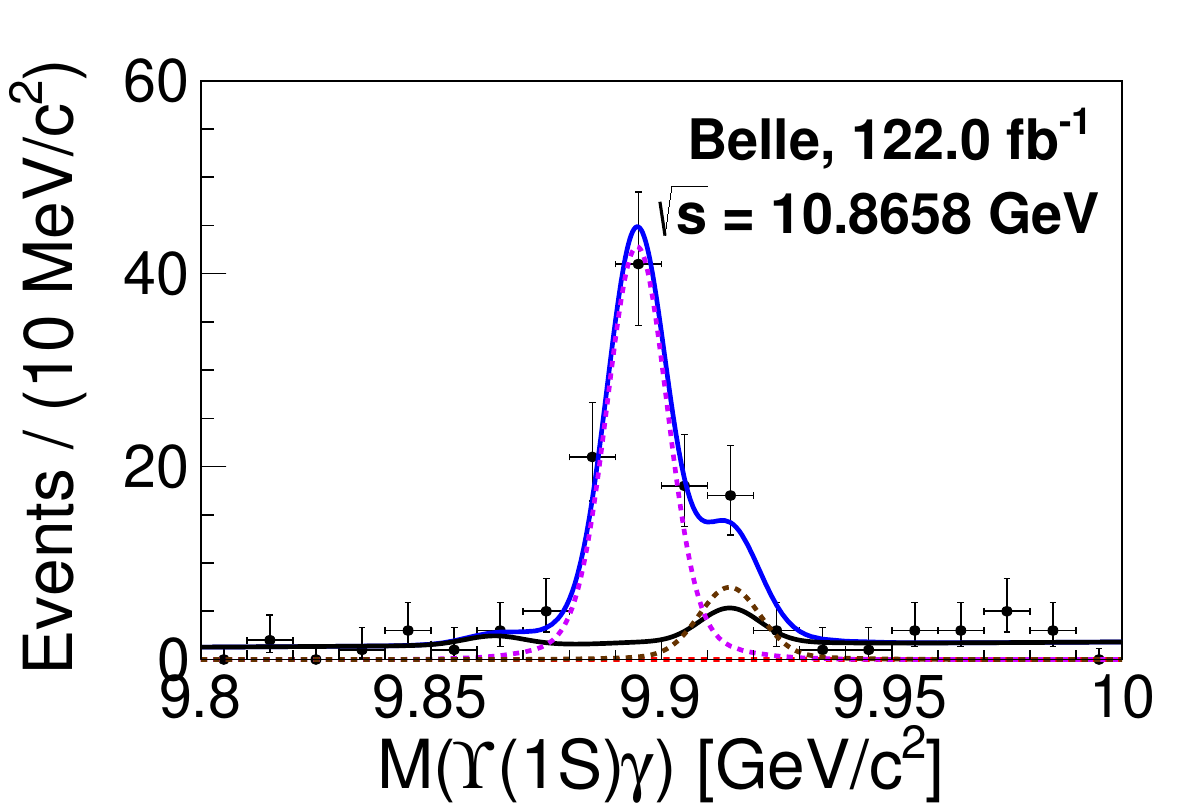}

\includegraphics[width=3.6cm]{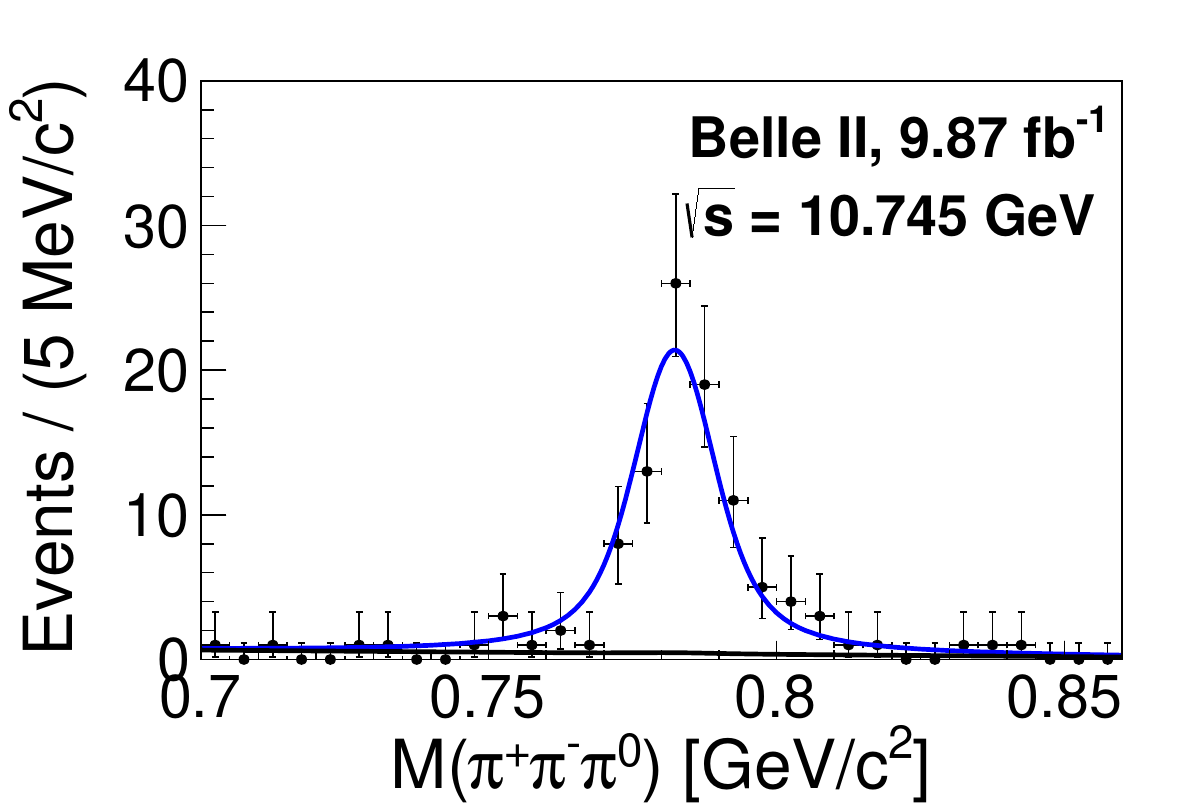}
\includegraphics[width=3.6cm]{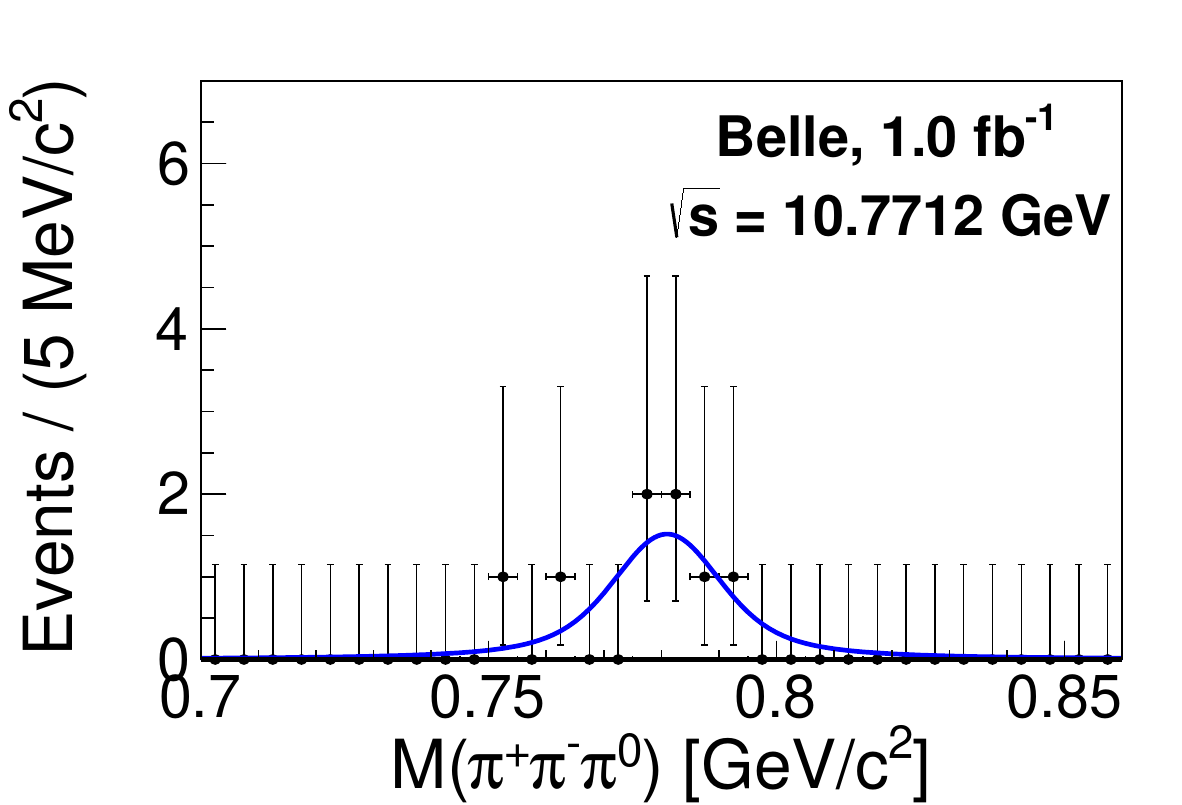}
\includegraphics[width=3.6cm]{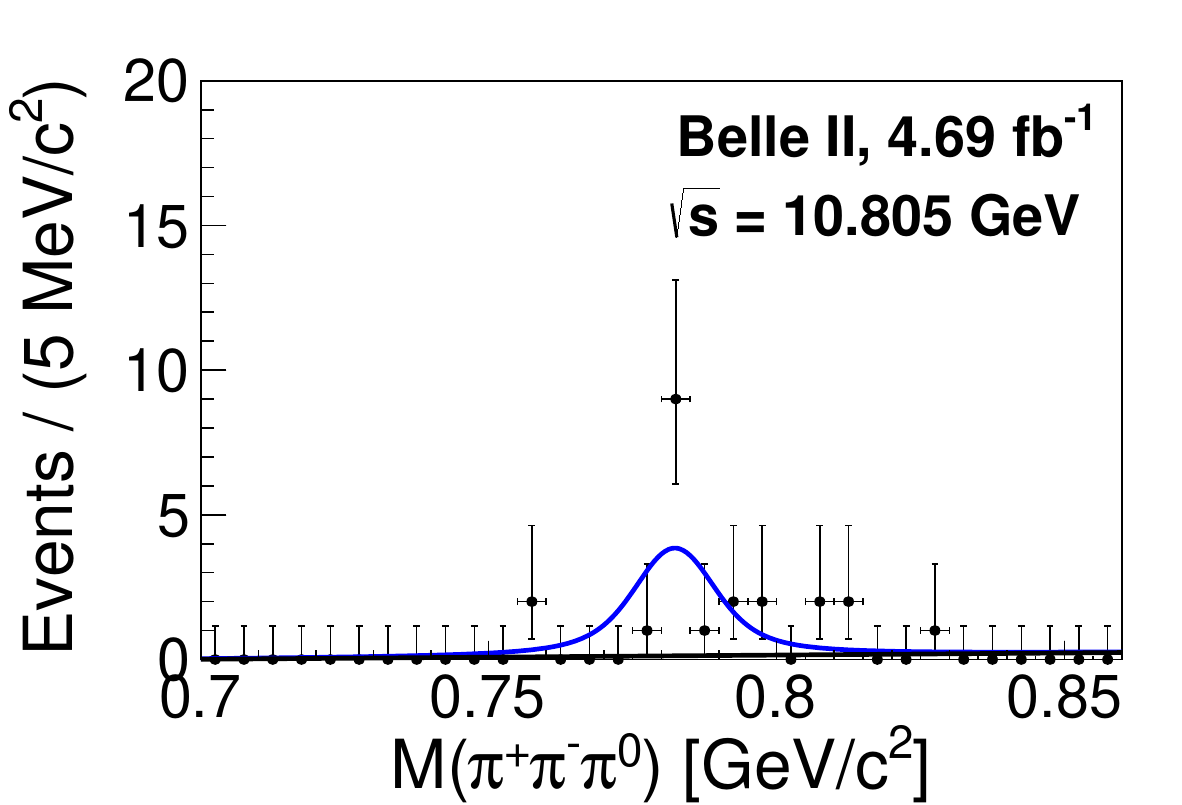}
\includegraphics[width=3.6cm]{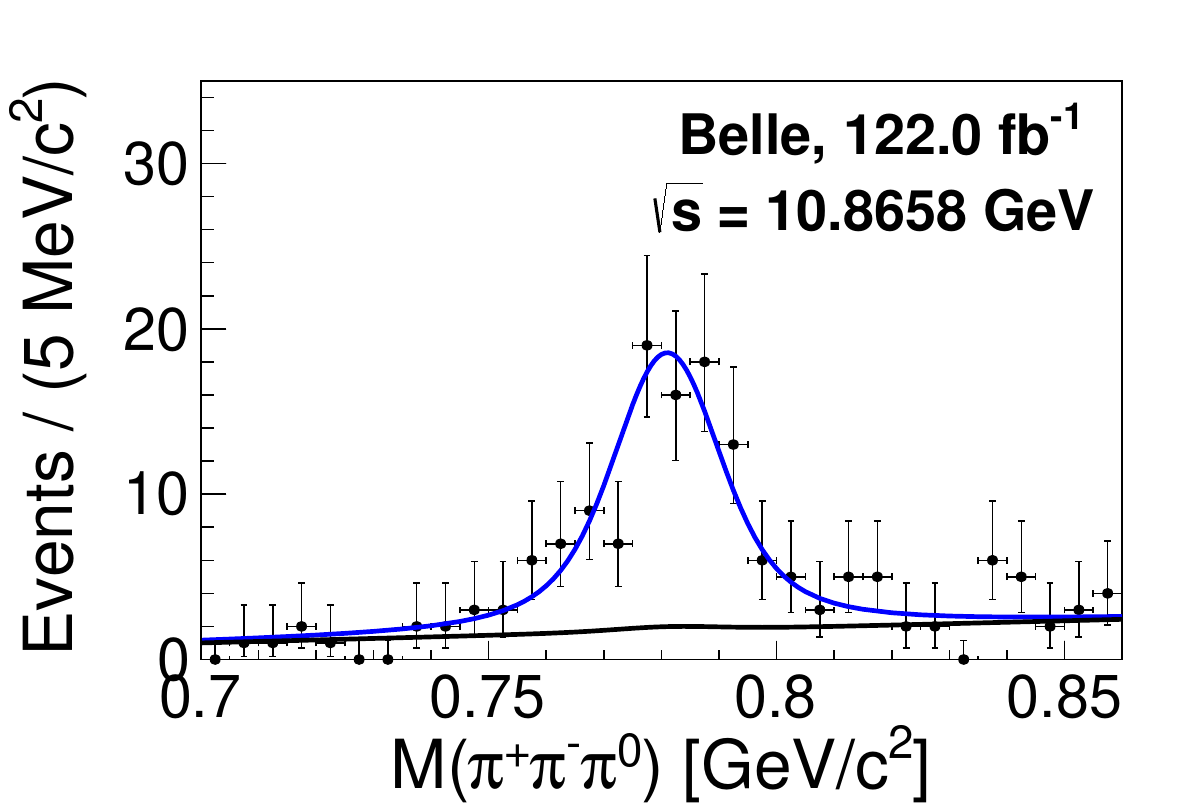}
\caption{Distributions of $M(\Upsilon(1S)\gamma)$ and $M(\pi^+\pi^-\pi^0)$ in Belle data at $\sqrt{s}$ = 10.7712 and 10.8658 GeV, and Belle II data at $\sqrt{s}$ = 10.745 and 10.805 GeV with 2D fit results overlaid. 
The solid blue and black curves show the total fit and total background; 
in the upper plots, the dashed red, violet, and brown curves show the $\chi_{b0}$, $\chi_{b1}$, and $\chi_{b2}$ signal components, respectively.
}\label{fig4}
\end{figure}

For well-populated samples ($\sqrt{s}$ = 10.7712 and 10.8658 GeV at Belle, and $\sqrt{s}$ = 10.745 and 10.805 GeV at Belle II), we perform a two dimensional (2D) unbinned extended maximum-likelihood fit to the $M(\Upsilon(1S)\gamma)$ and $M(\pi^+\pi^-\pi^0)$ distributions to extract signal yields.
The fitting function is a sum of four components: a 2D signal peak in $(M(\Upsilon(1S)\gamma),\,M(\pi^+\pi^-\pi^0))$, peaking background in the $M(\Upsilon(1S)\gamma)$ distribution from $e^+e^-\to\chi_{bJ}\,\pi^+\pi^-\pi^0$, peaking background in the $M(\pi^+\pi^-\pi^0)$ distribution 
from non-$\chi_{bJ}$ background with a $\omega$, and combinatorial background.
Each $\chi_{bJ}$ signal shape is described by a double Gaussian function, and the $\omega$ signal shape is described by a Breit-Wigner 
convolved with a Gaussian function. 
The mass resolutions for $\chi_{bJ}$ candidates at Belle and Belle II are 11 MeV/$c^2$ and 13 MeV/$c^2$, respectively.
Signal shape parameters are fixed according to signal MC simulations. 
A first-order polynomial function is used to describe the combinatorial background.
Fit projections for the signal region are shown in ﬁgure~\ref{fig4}.
Signal significances for each of
$\chi_{b0}\,\omega$, $\chi_{b1}\,\omega$, and $\chi_{b2}\,\omega$ are estimated
using $\sqrt{-2\ln(\mathcal{L}_0/\mathcal{L}_{\rm max})}$, where $\mathcal{L}_0$ and $\mathcal{L}_{\rm max}$ are the maximized likelihoods without and with the signal, respectively~\cite{Wilks}.

\begin{figure}[htbp]
\centering
\includegraphics[width=3.6cm]{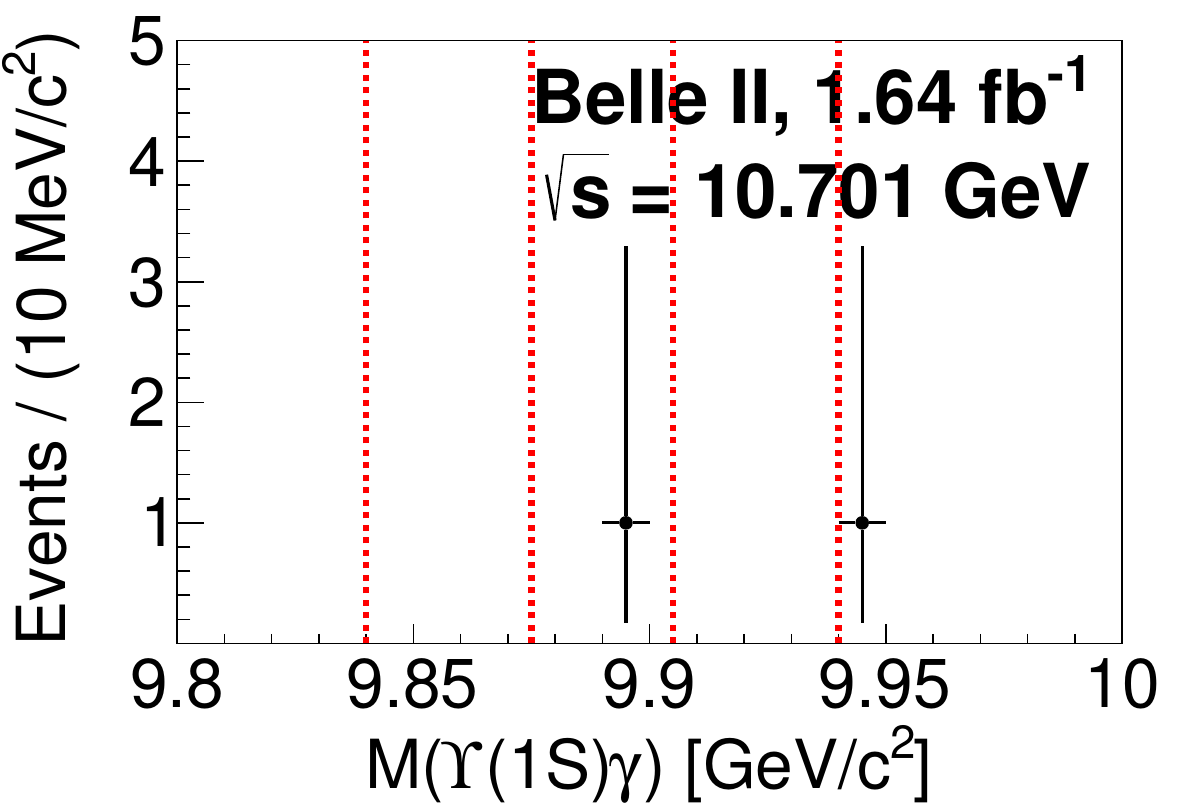}
\includegraphics[width=3.6cm]{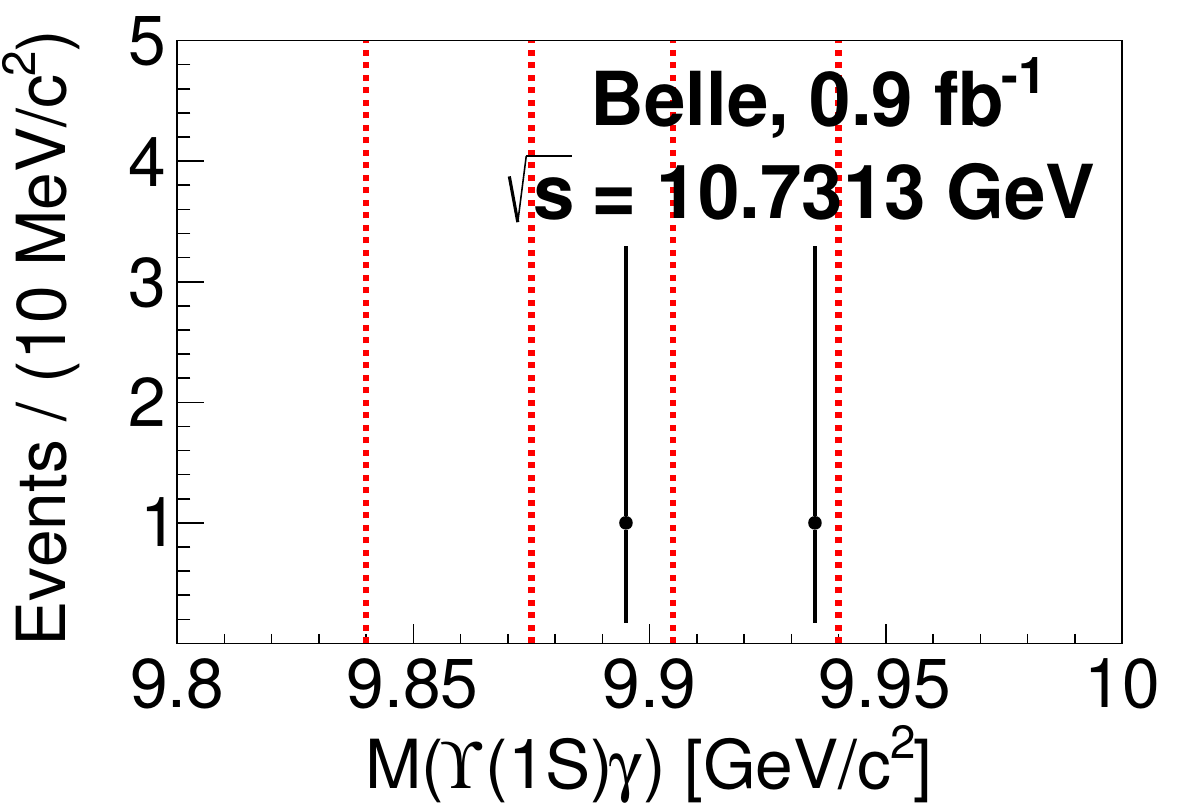}
\includegraphics[width=3.6cm]{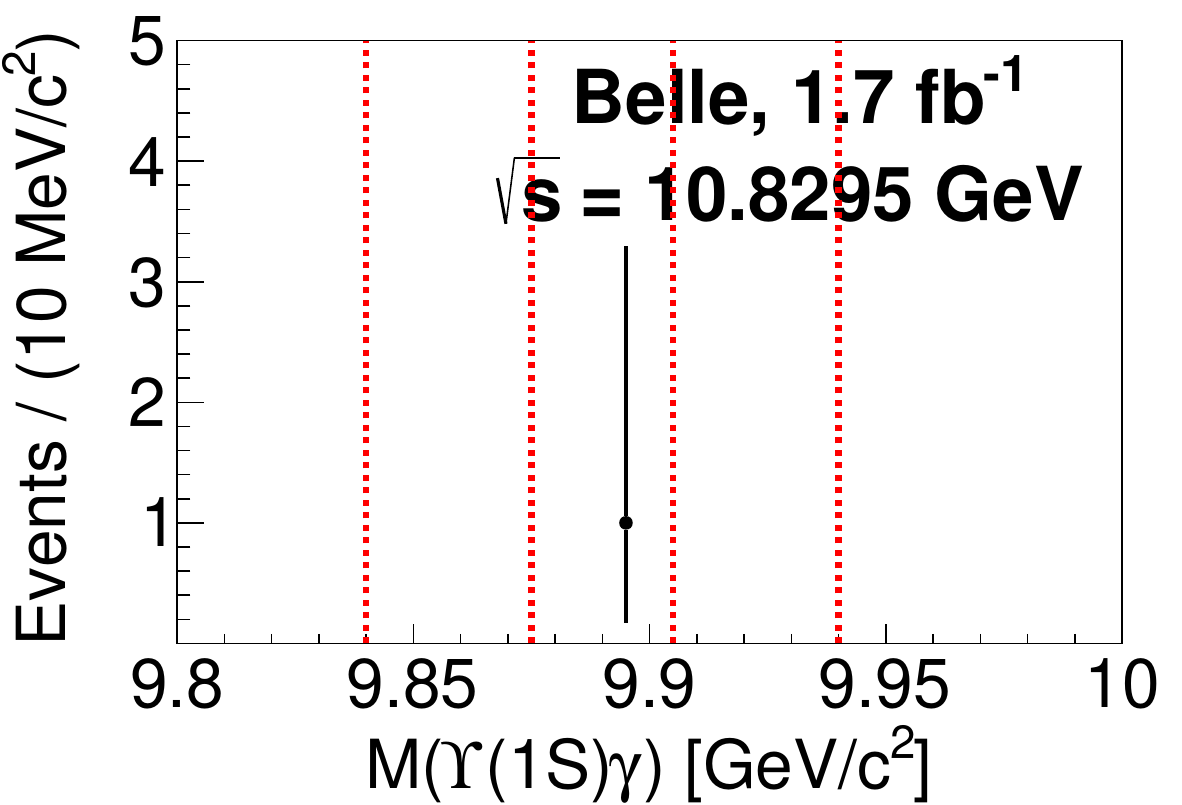}
\includegraphics[width=3.6cm]{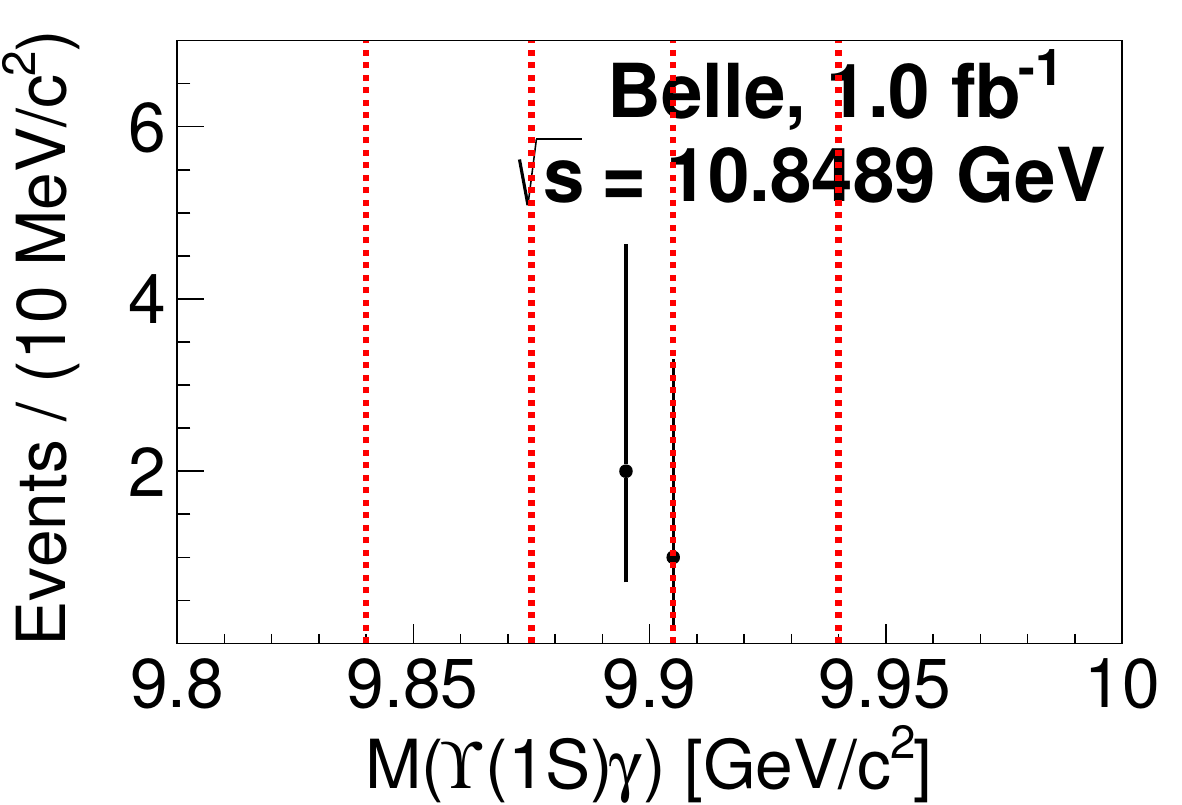}

\includegraphics[width=3.6cm]{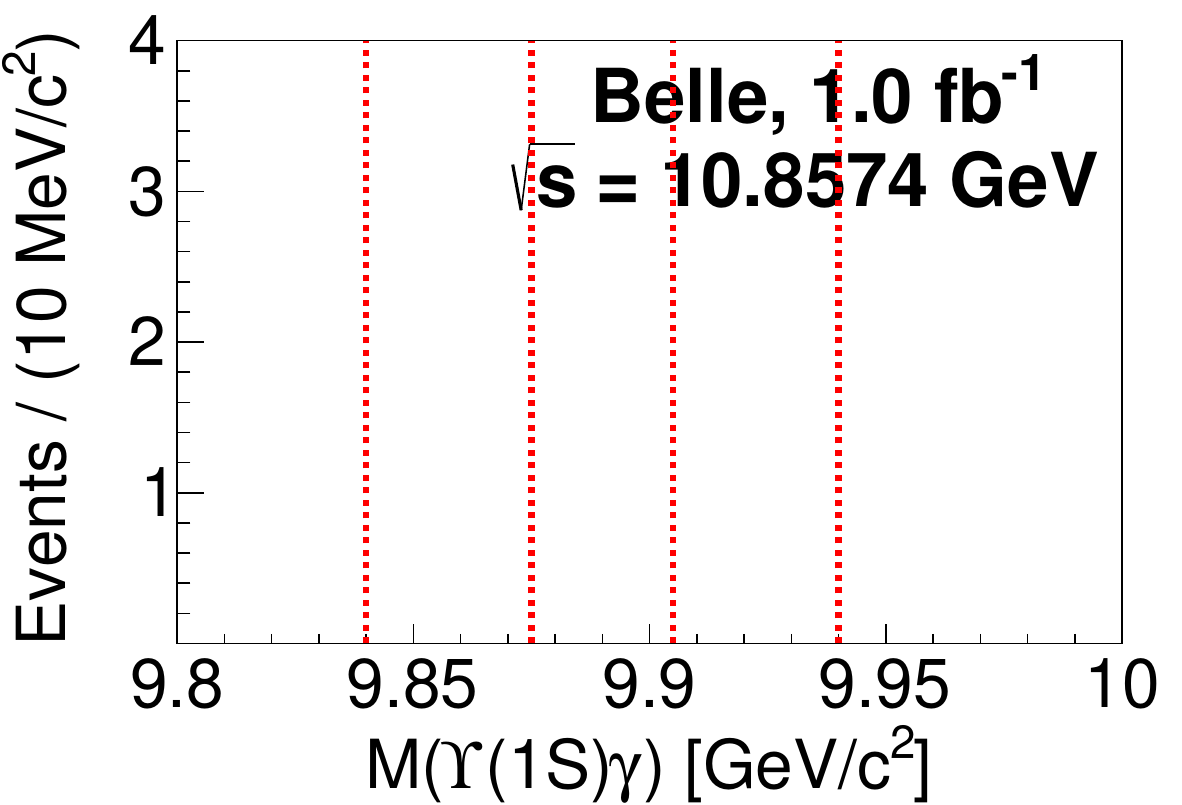}
\includegraphics[width=3.6cm]{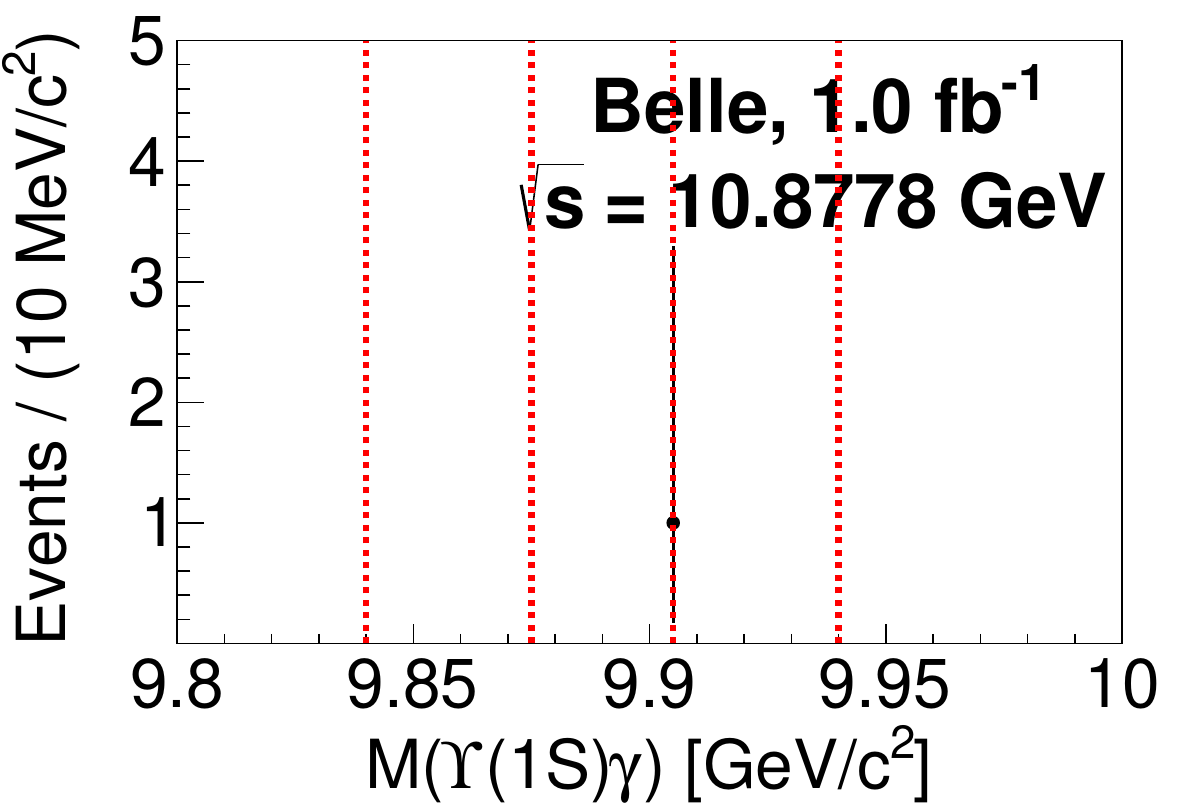}
\includegraphics[width=3.6cm]{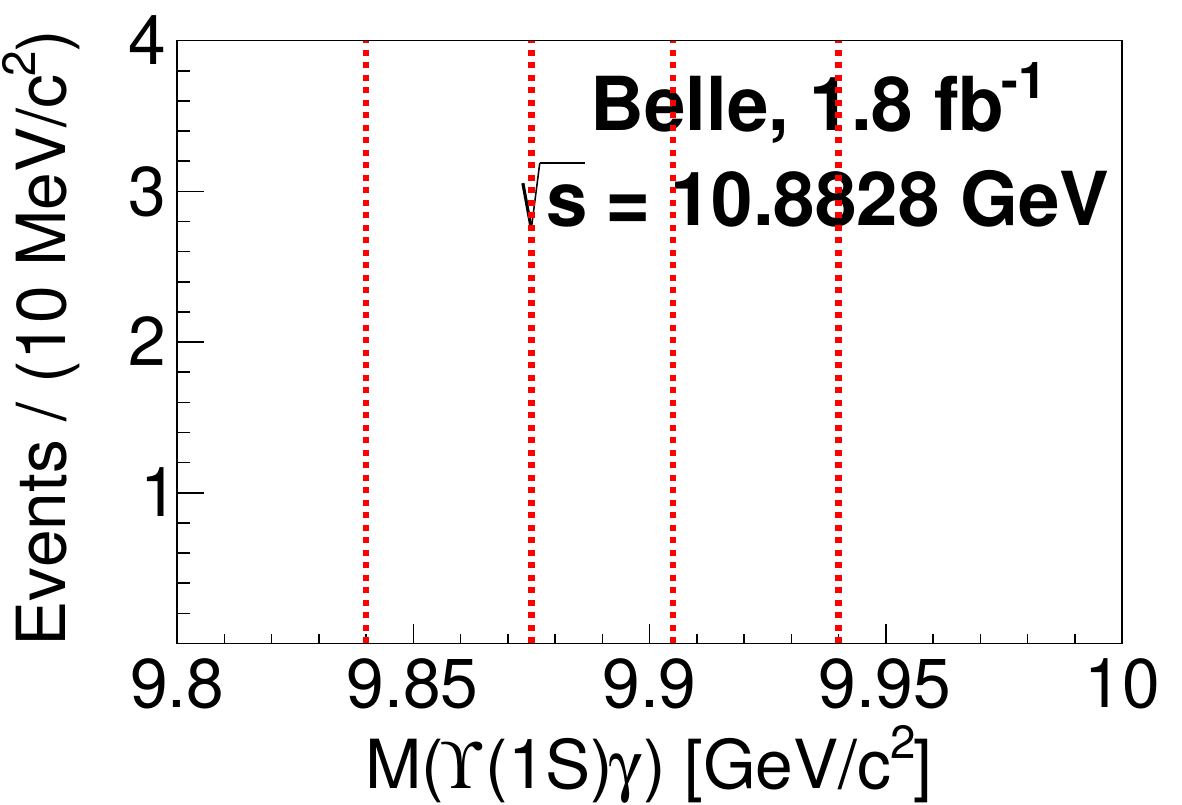}
\includegraphics[width=3.6cm]{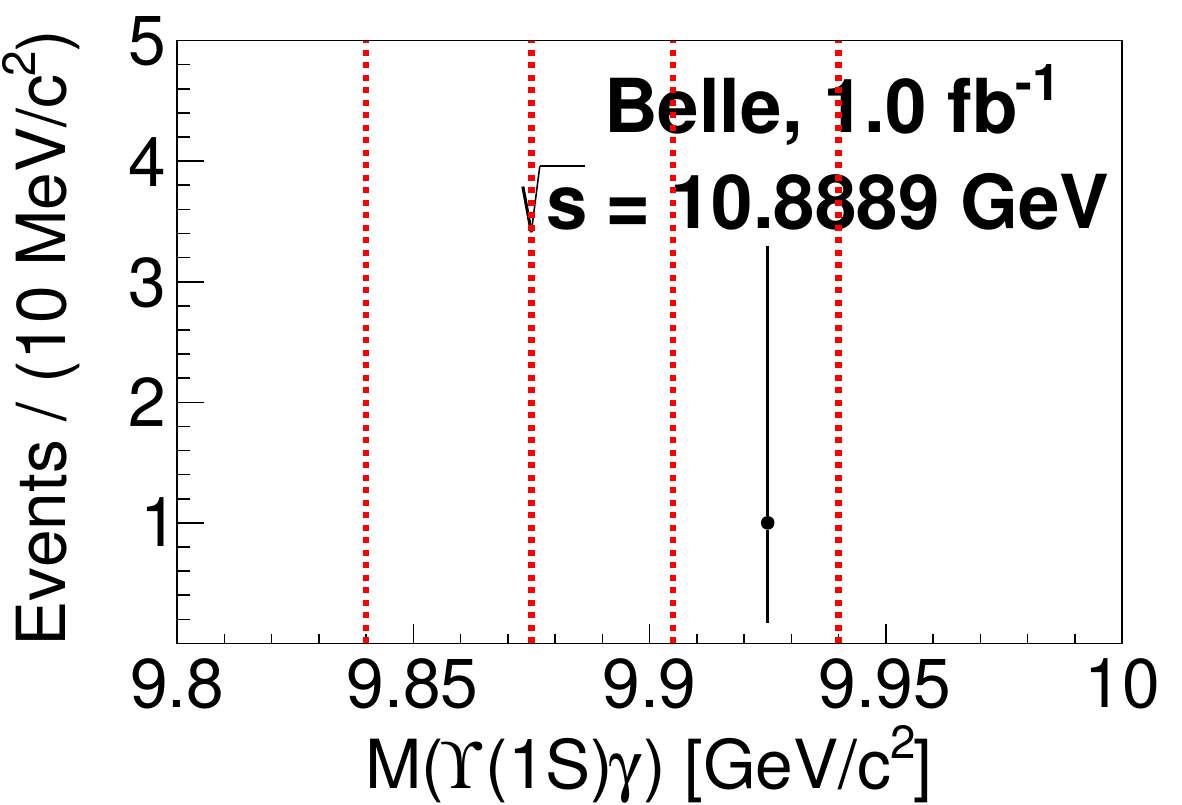}

\includegraphics[width=3.6cm]{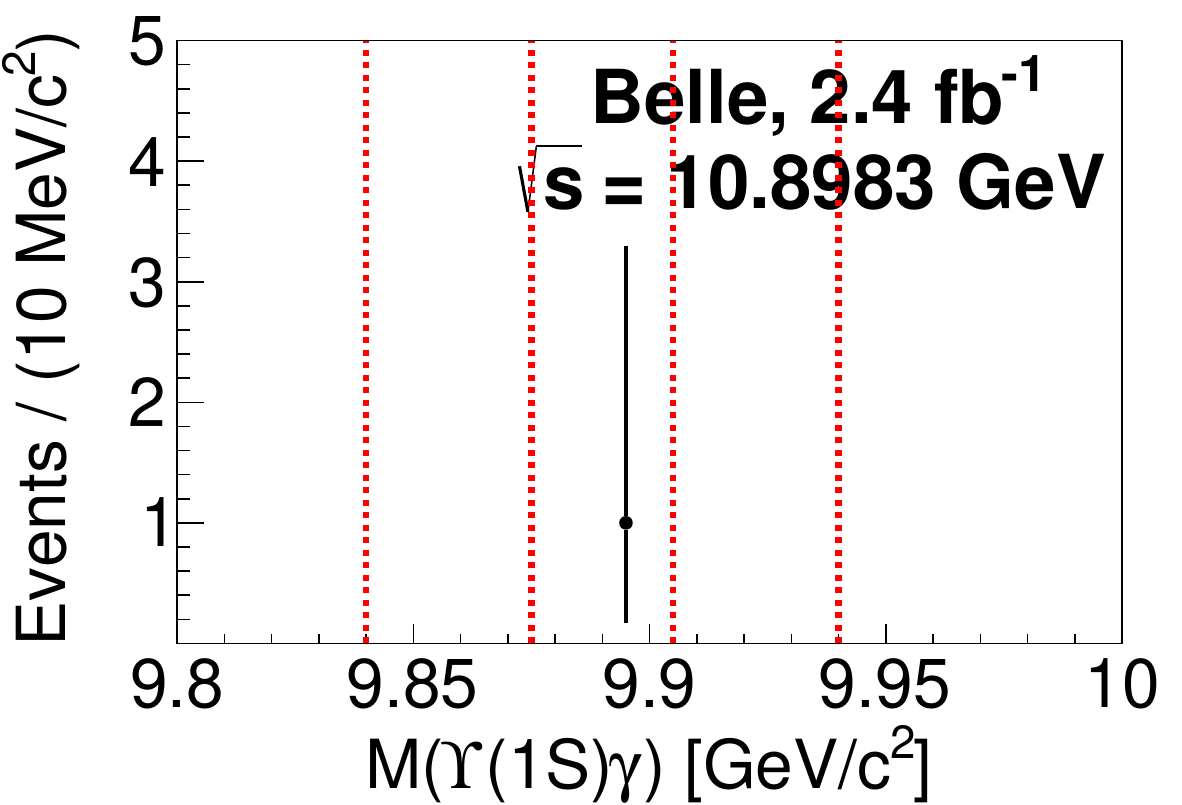}
\includegraphics[width=3.6cm]{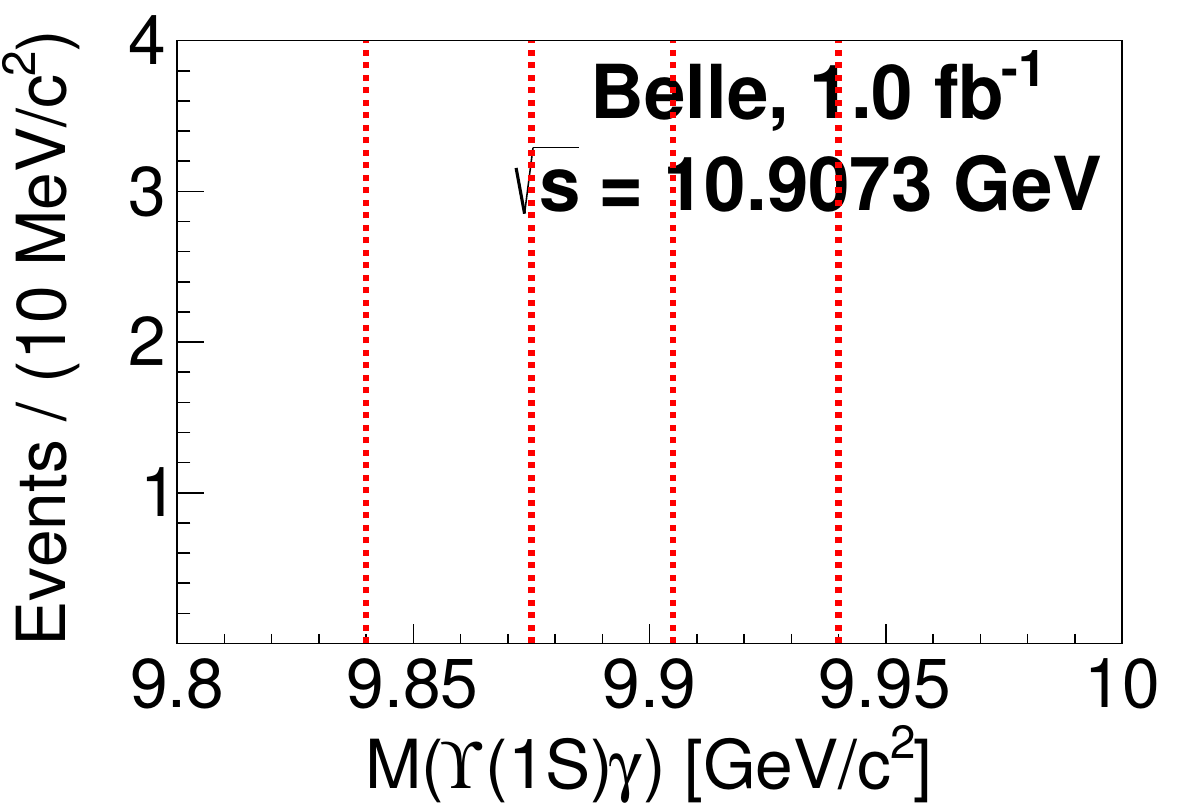}
\includegraphics[width=3.6cm]{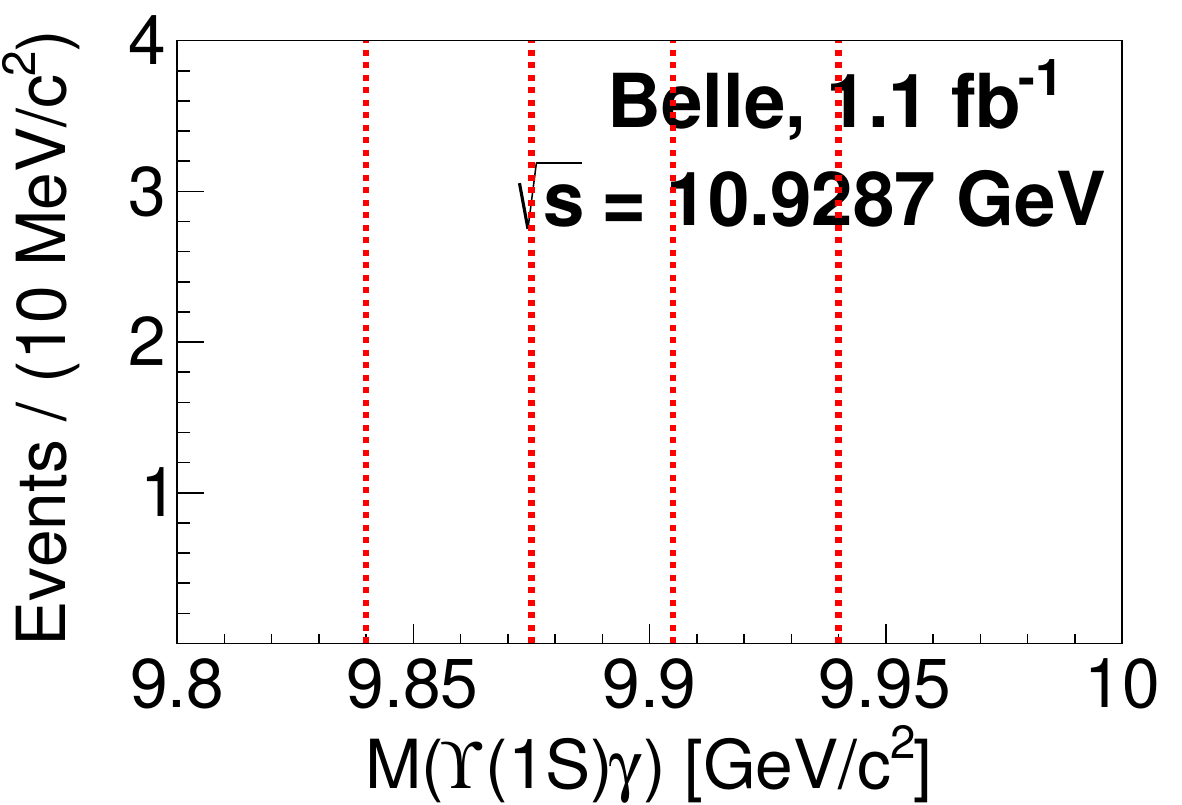}
\includegraphics[width=3.6cm]{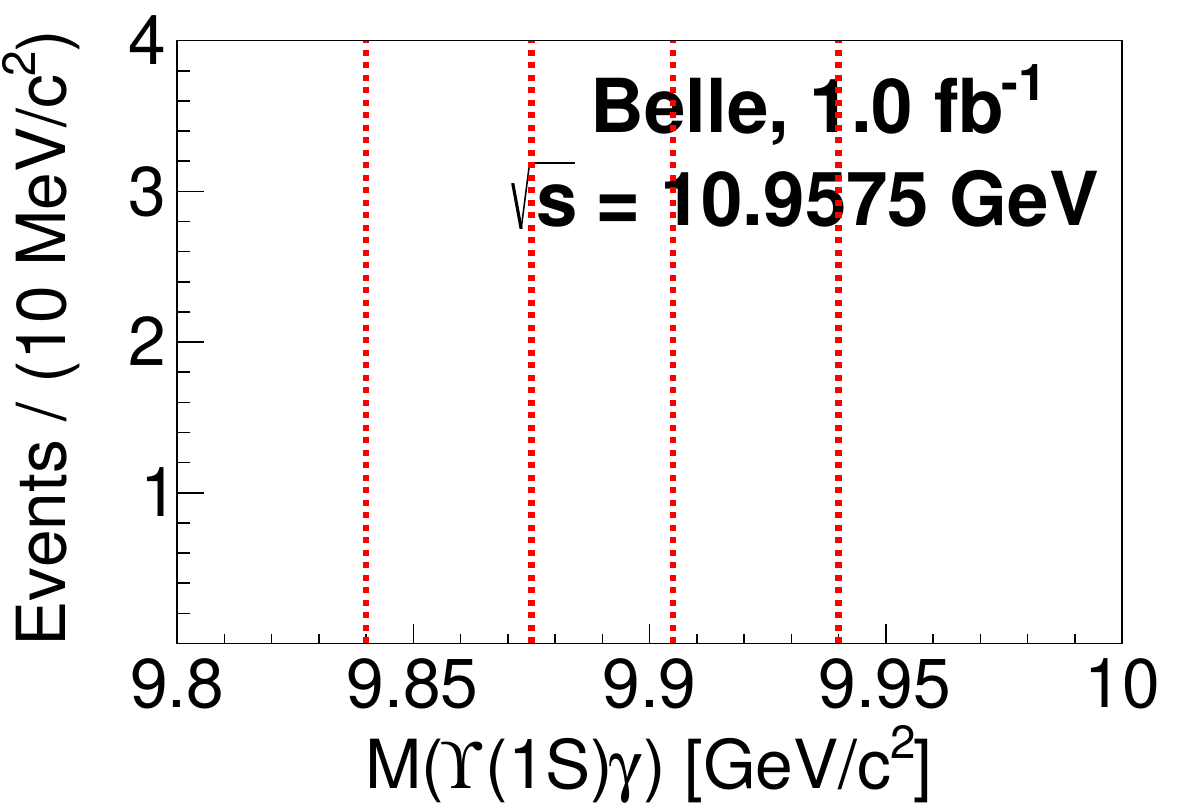}

\includegraphics[width=3.6cm]{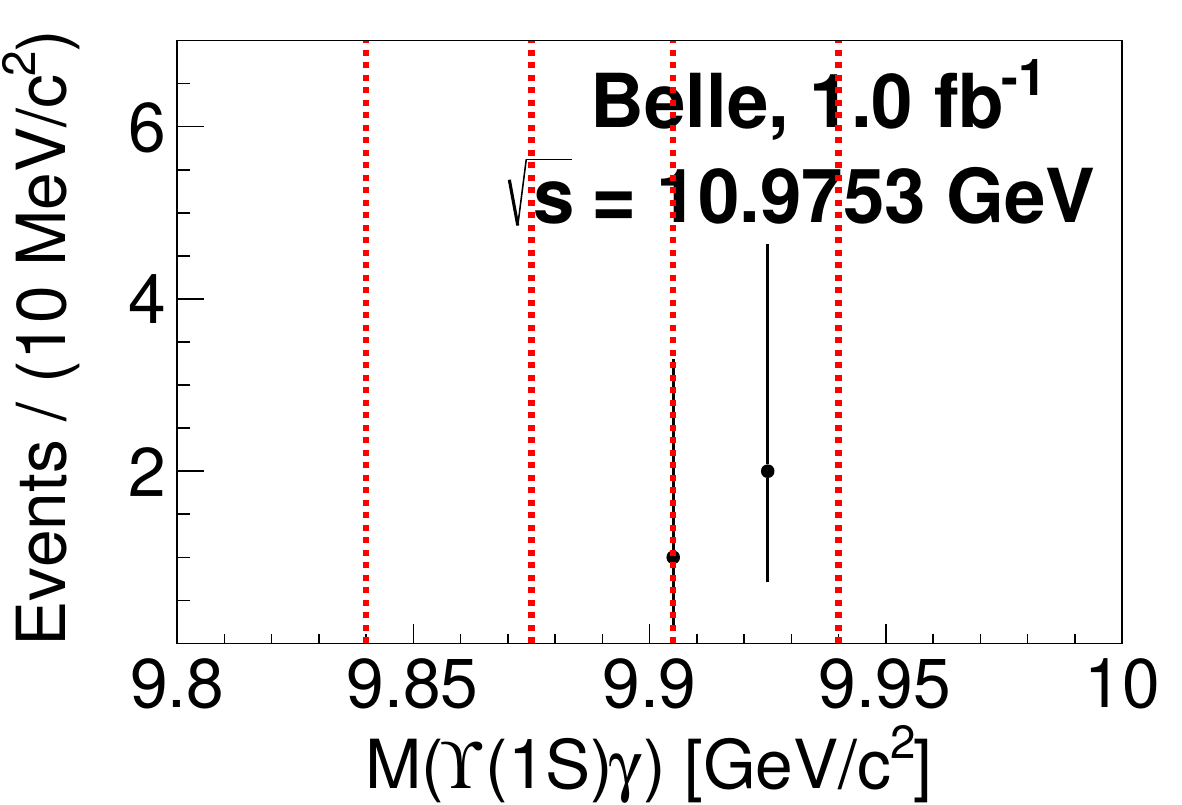}
\includegraphics[width=3.6cm]{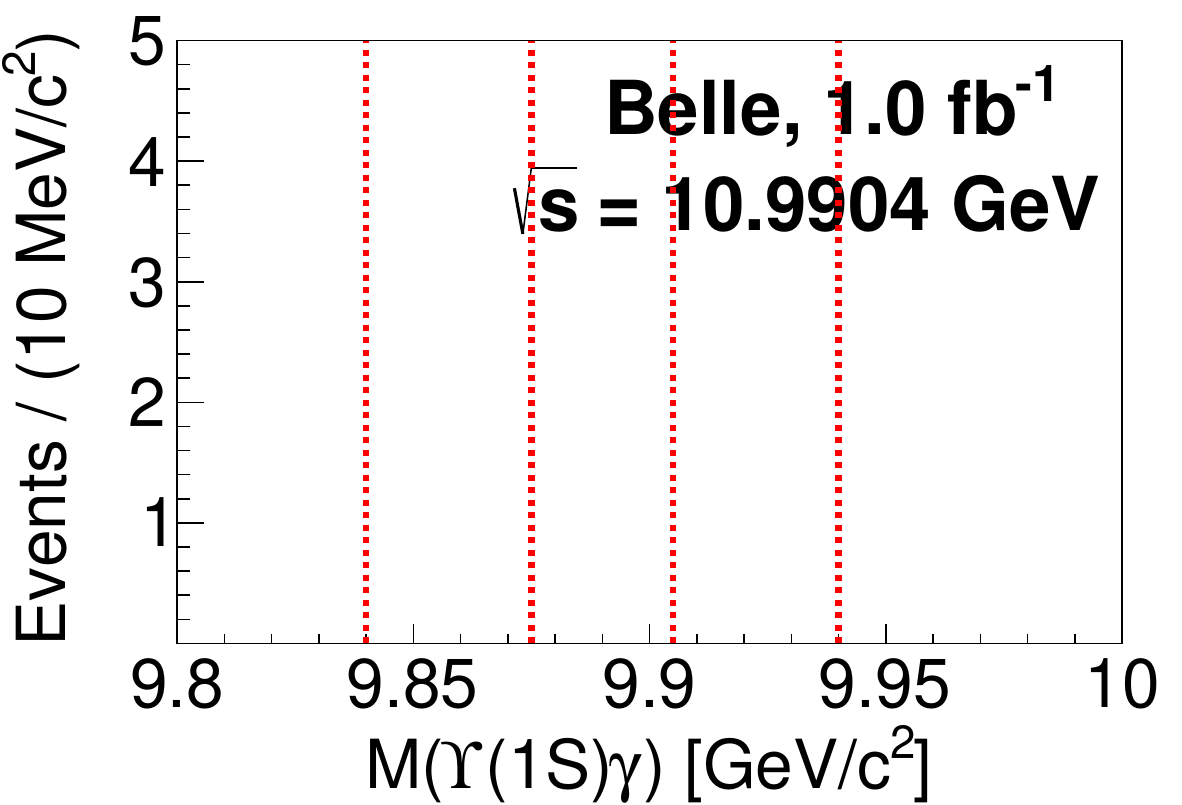}
\includegraphics[width=3.6cm]{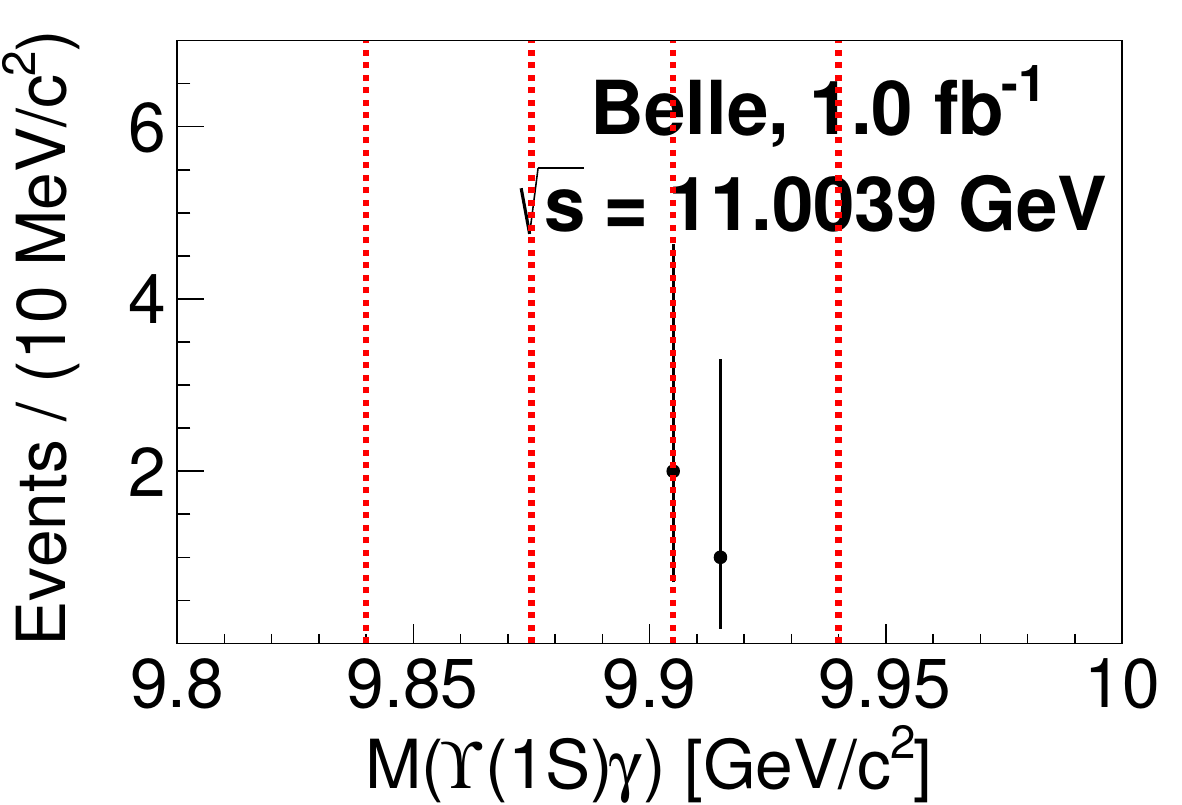}
\includegraphics[width=3.6cm]{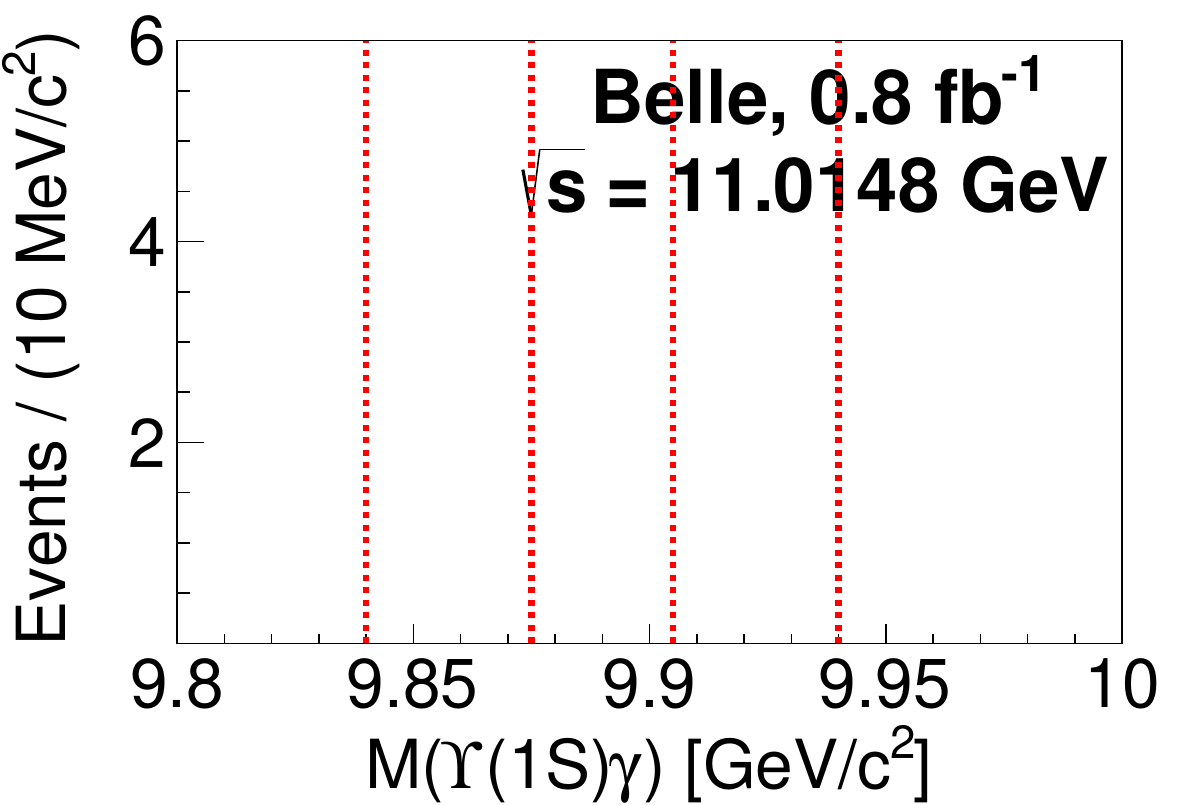}

\includegraphics[width=3.6cm]{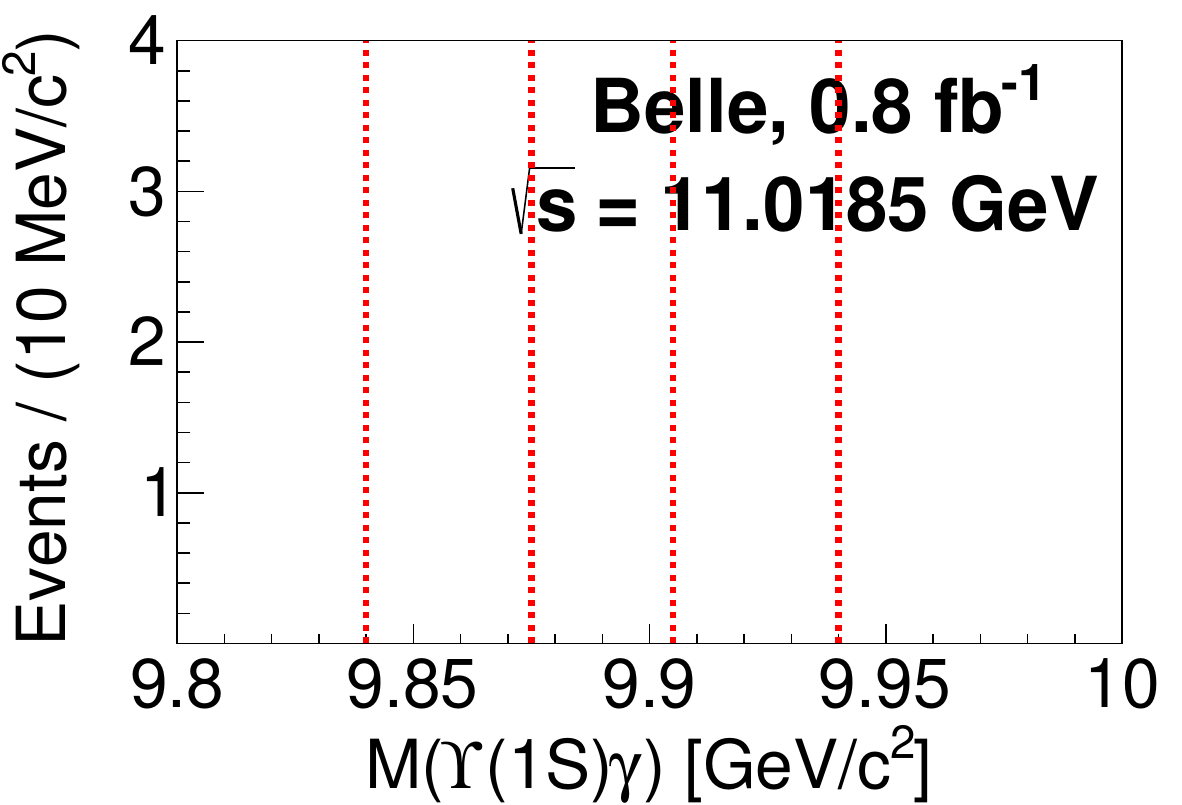}
\includegraphics[width=3.6cm]{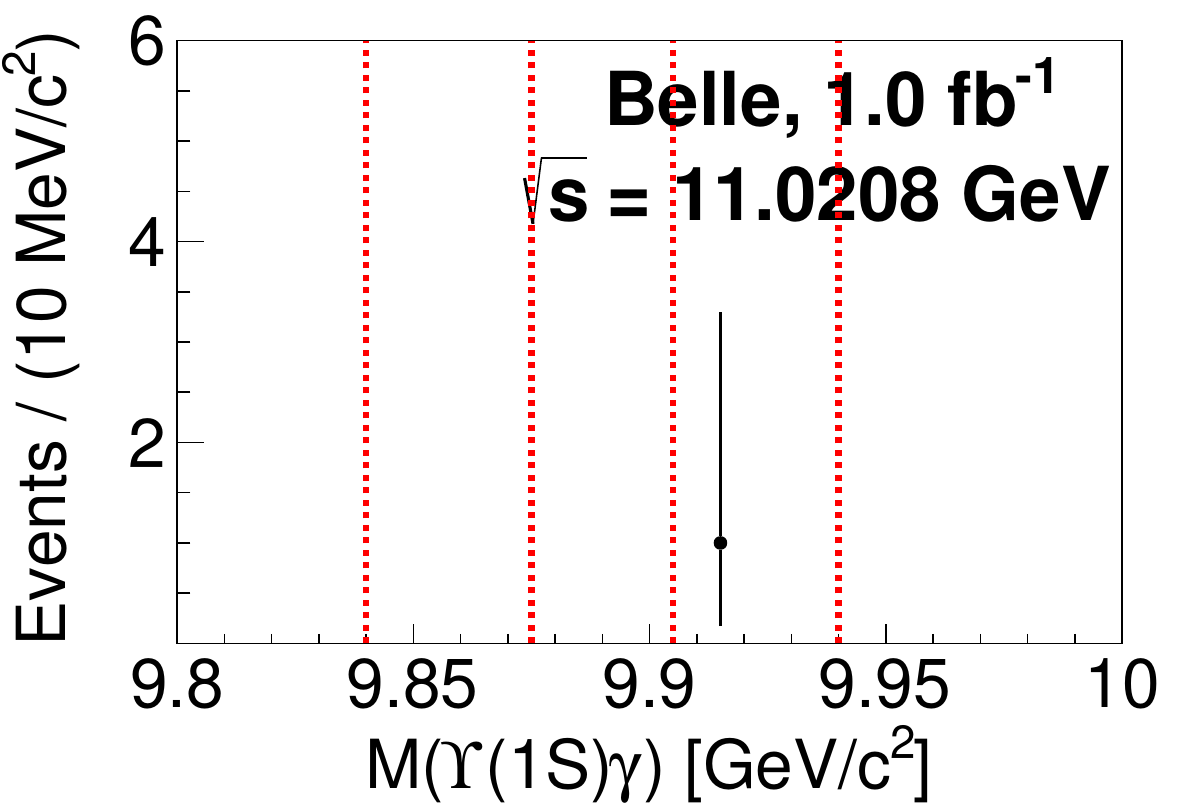}
\caption{The $M(\Upsilon(1S)\gamma)$ distributions after requiring events within the $\omega$ signal region in data at each energy point for the Belle and Belle II data samples. The vertical dashed lines (left to right) show the $\chi_{b0}$, $\chi_{b1}$, and $\chi_{b2}$ signal regions.
}\label{fig5}
\end{figure}

\begin{figure}[htbp]
\centering
\includegraphics[width=6cm]{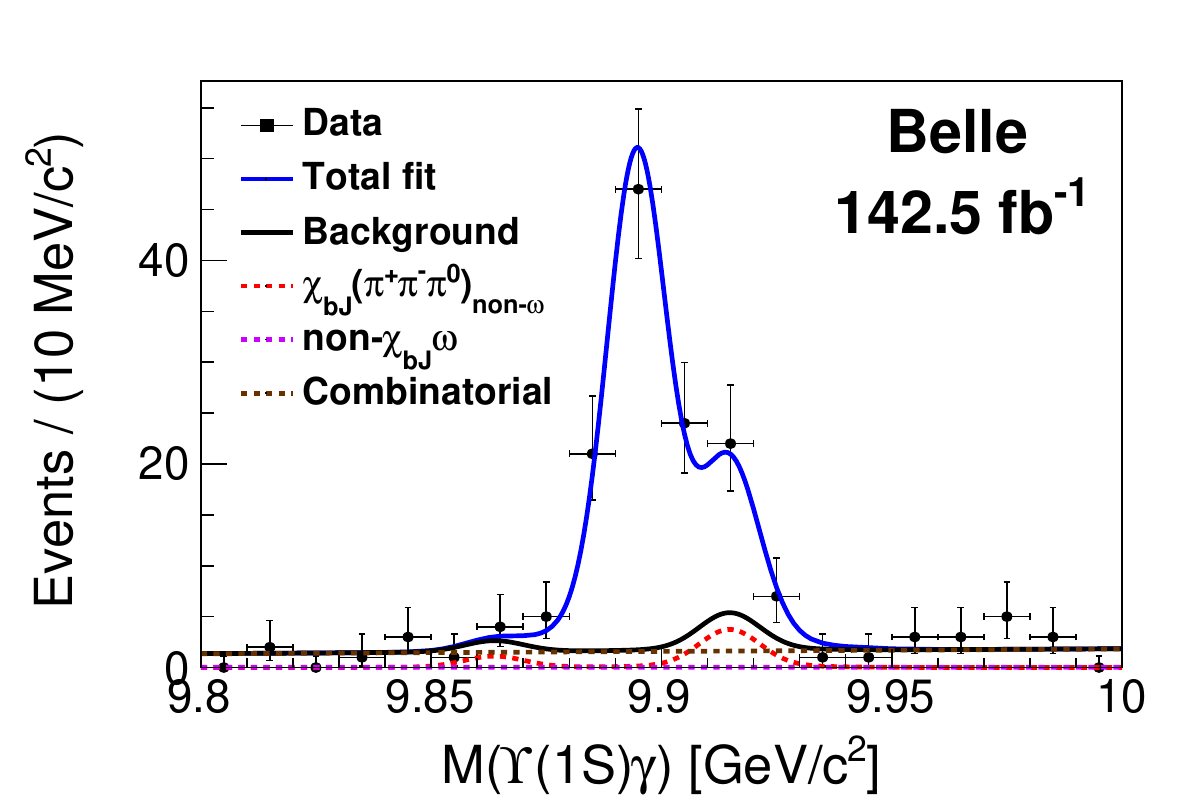}
\includegraphics[width=6cm]{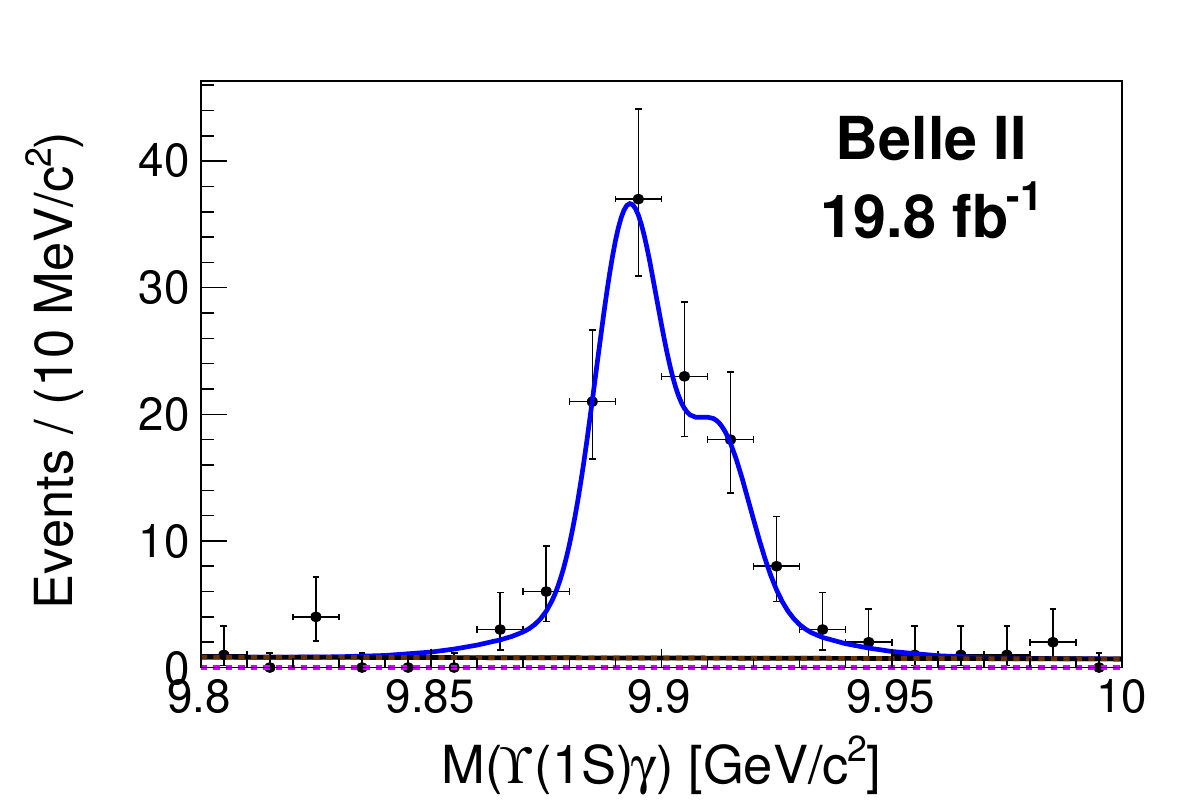}

\includegraphics[width=6cm]{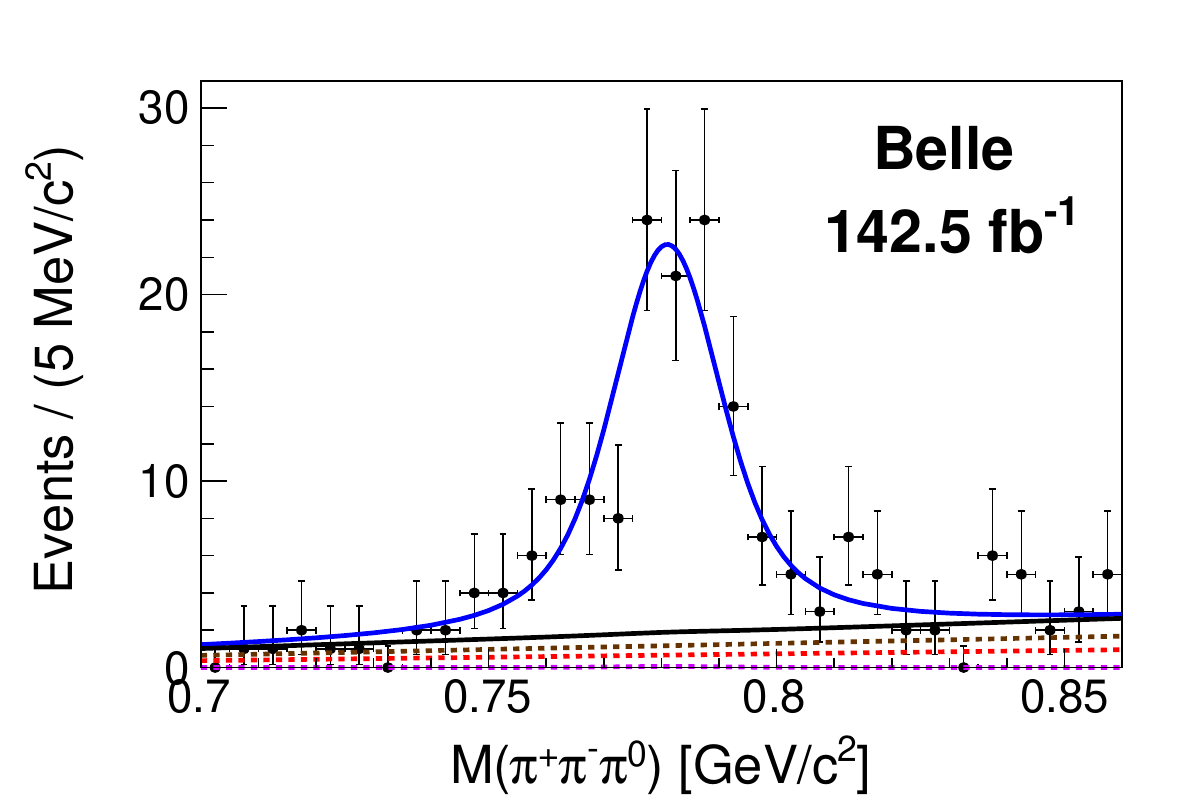}
\includegraphics[width=6cm]{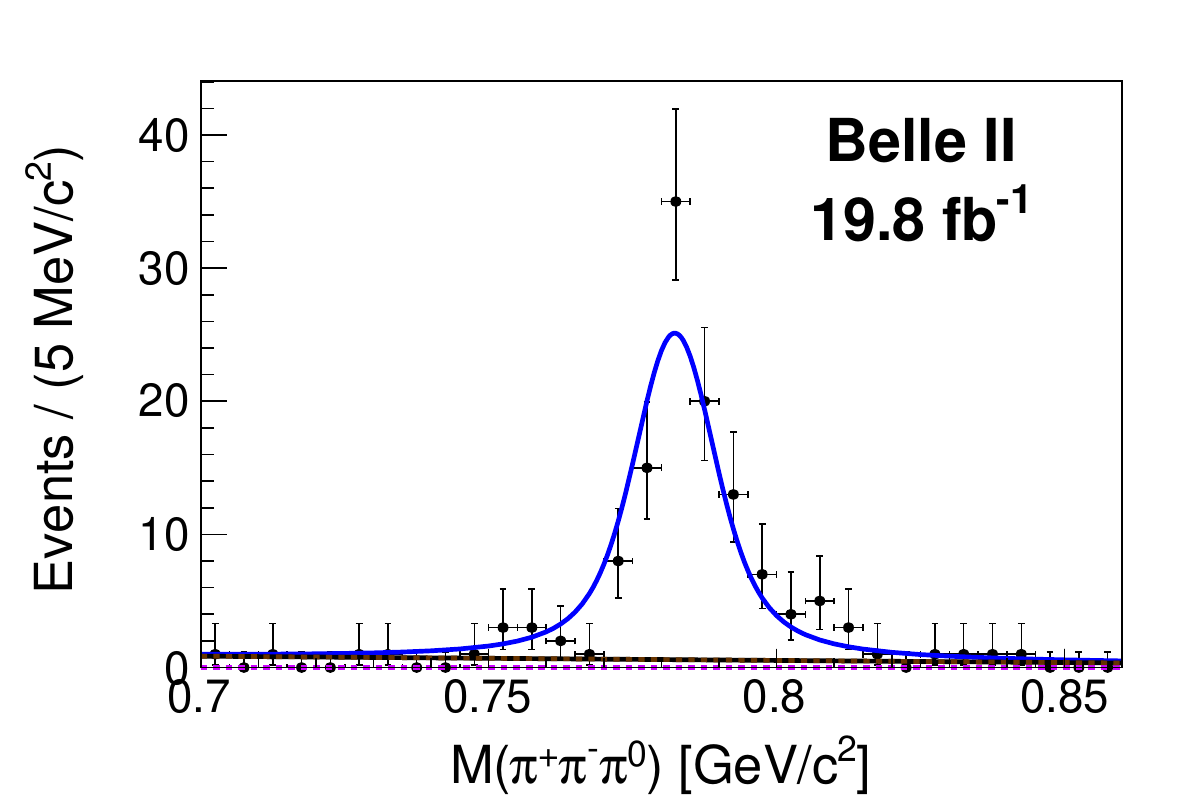}
\caption{
The 2D fits to $M(\Upsilon(1S)\gamma)$ and $M(\pi^+\pi^-\pi^0)$ distributions from the data sample that combines all energies at Belle (left) and
Belle II (right).
The solid blue and black curves show the total fit and total background; the dashed red, violet, and brown curves show the $\chi_{bJ}\,(\pi^+\pi^-\pi^0)_{\rm non-\omega}$ background, non-$\chi_{bJ}\,\omega$ background, and combinatorial background, respectively. At Belle II, only combinatorial background is seen.
}\label{figadd}
\end{figure}

For other energy points, the $M(\Upsilon(1S)\gamma)$ distributions after applying the $\omega$ mass-window requirement are shown in figure~\ref{fig5}. These distributions are very sparse, and instead of fitting, we use event counting. We count the numbers of events, $N^{\rm obs}$, in the $\chi_{b0}$, $\chi_{b1}$, and $\chi_{b2}$ signal regions of (9.84 -- 9.875) GeV/$c^2$, (9.875 -- 9.905) GeV/$c^2$, and (9.905 -- 9.94) GeV/$c^2$, respectively. The signal yield is defined as $N^{\rm sig}={\rm max}(0,N^{\rm obs}-N^{\rm bg})$. To determine the number of background events in each signal region, $N^{\rm bg}$, we perform a 2D fit to the data sample that combines all energies, as shown in figure~\ref{figadd}.
The combinatorial background is assumed to be non-resonant. For each energy point, the expected number of combinatorial background events is determined from the corresponding integrated luminosity and the background yield in the combined sample.
The number of $\chi_{bJ}\,(\pi^+\pi^-\pi^0)_{\rm non-\omega}$ background events is likewise assigned based on the luminosity, the yield in the combined sample, and the $e^+e^-\to\chi_{bJ}\,(\pi^+\pi^-\pi^0)_{\rm non-\omega}$ cross section obtained from the energy-dependence fit (section~\ref{sec.8}).
The statistical uncertainties on $N^{\rm sig}$ are
assigned based on 68.3\% confidence intervals provided by the Poissonian
limit estimator (POLE) program~\cite{012002}, with systematic uncertainties set
to zero; these are equivalent to the unified approach in ref.~\cite{3873}. 
These values are given in tables~\ref{tabsumchib1},~\ref{tabsumchib2}, and~\ref{tabsumchib0}.

The Born cross sections for $e^+e^-\to\chi_{bJ}\,\omega$ are calculated using

\vspace{-0.2cm}
\begin{equation} \label{eq:3}
\sigma_{\rm Born}(e^+e^-\to \chi_{bJ}\,\omega) = \frac{N^{\rm sig}}{{\cal L}\,\varepsilon\,\BR_{\rm int}\,(1+\delta_{\rm ISR})\,|1-\Pi|^2},
\end{equation}
where ${\cal L}$ is the integrated luminosity of the data sample, $\varepsilon$ is the reconstruction efficiency, $\BR_{\rm int}$ = $\BR(\omega\to\pi^+\pi^-\pi^0)\,\BR(\pi^0\to\gamma\gamma)\,\BR(\chi_{bJ}\to\Upsilon(1S)\gamma)\,\BR(\Upsilon(1S)\to\ell^+\ell^-)$ is the product of the branching fractions of the intermediate states to the reconstructed final states, $\frac{1}{|1-\Pi|^2}$ = 0.93 is the vacuum polarization factor~\cite{083001,585}, and $1+\delta_{\rm ISR}$ is the radiative correction factor~\cite{isr,113009,2605}.
In calculating the radiative correction factor, we use the energy dependence of the
Born cross sections for $e^+e^-\to\chi_{bJ}\,\omega$ measured in this work: see section~\ref{sec.8}.
(We initially assume the energy dependence measured in ref.~\cite{091902}; for the energy dependence of $\chi_{bJ}\,(\pi^+\pi^-\pi^0)_{{\rm non}-\omega}$ in section~\ref{sec.6}, we use the measurement of ref.~\cite{091102}. The procedure is then iterated. Two steps are sufficient for the result to converge.)
The Born cross sections, efficiencies, and $1+\delta_{\rm ISR}$ factors for $e^+e^-\to\chi_{bJ}\,\omega$ at each energy point are listed in tables~\ref{tabsumchib1},~\ref{tabsumchib2}, and~\ref{tabsumchib0}.
The Born cross sections for $e^+e^-\to\chi_{b1}\,\omega$ and $e^+e^-\to\chi_{b2}\,\omega$ at $\sqrt{s}$ = 10.745 GeV are $(3.8^{+0.6}_{-0.5}\pm0.4)$ pb and $(2.6^{+0.8}_{-0.7}\pm0.4)$ pb, which are consistent with the values $(3.6^{+0.7}_{-0.7}\pm0.5)$ pb and $(2.8^{+1.2}_{-1.0}\pm0.4)$ pb in ref.~\cite{091902}.
(Systematic uncertainties are discussed in section~\ref{sec.7}.)
The ratio $\sigma_{\rm Born}(e^+e^-\to\chi_{b1}\,\omega)/\sigma_{\rm Born}(e^+e^-\to\chi_{b2}\,\omega)$ is $1.5^{+0.5}_{-0.4}\pm0.2$ at $\sqrt{s}=10.745$ GeV, where we take into account that some sources of systematic uncertainty cancel in the ratio.
This ratio is consistent with the value $1.3^{+0.6}_{-0.5}\pm0.2$ in ref.~\cite{091902}.
The ratio $\sigma^{\rm UL}_{\rm Born}(e^+e^-\to\chi_{b0}\,\omega)/\sigma_{\rm Born}(e^+e^-\to\chi_{b1}\,\omega)$ is less than $1.3$ at 90\% C.L. at $\sqrt{s}=10.745$ GeV.

In cases where the signal counting method is used, the signal yields and Born cross sections ($\sigma^{\prime}_{\rm Born}$) are not corrected for the efficiency of the $\chi_{bJ}$ mass-window requirement and the cross-feed between channels. 
The effect of cross-feed (``migration'') is determined from the simulation to be
\begin{equation}\label{eq.6.1}
\begin{bmatrix}
0.86 & 0.05 \\
0.14 & 0.95
\end{bmatrix}
\begin{bmatrix}
\sigma_{1} \\
\sigma_{2}
\end{bmatrix}
=
\begin{bmatrix}
\sigma^{\prime}_{1} \\
\sigma^{\prime}_{2}
\end{bmatrix},
\end{equation}where subscripts 1 and 2 correspond to $\chi_{b1}$ and $\chi_{b2}$,
and $\sigma_i^\prime$ and $\sigma_i$ show the measured and underlying
yields, respectively.
The migration matrix is taken into account in the fit to the energy dependence of the cross sections (section~\ref{sec.8}).
Similarly, we estimate that 2.3\% of any $\chi_{b0}$ yield appears inside the $\chi_{b1}$ mass window, but due to the insignificant $\chi_{b0}$ signal, this contribution is negligible.

For insignificant signals, we set 90\% confidence level (C.L.) upper limits ($x^{\rm UL}$) on yields and cross sections. 
For energy points at which signal yields are determined from a fit, upper limits are found by solving the equation

\vspace{-0.3cm}
\begin{equation}\label{eq:2}
\int _0^{x^{\rm UL}}{\cal L}(x)dx/\int _0^{+\infty}{\cal L}(x)dx = 0.90,
\end{equation}
where $x$ is the assumed Born cross section, and ${\cal L}(x)$ is the corresponding maximized likelihood of the ﬁt.
To take into account the systematic uncertainties discussed in section~\ref{sec.7}, the above likelihood is convolved with a Gaussian function whose width equals the total multiplicative systematic uncertainty.
Additive systematic effects are included by choosing the variation that gives the most conservative limit.
For other energy points with small numbers of candidates, the upper
limits are calculated directly using POLE, with the uncertainty on the
background prediction set to the sum in quadrature of the statistical
and additive systematic uncertainty; the relative uncertainty on the
signal efficiency set to the total multiplicative uncertainty from table~\ref{sys1} (see also section~\ref{sec.7}); and the relative uncertainty on the
background efficiency set to the ``efficiency'' term from Table~\ref{sys1}. The
signal and background efficiency uncertainties are treated as
correlated.
The upper limits on the Born cross sections at each energy point are listed in tables~\ref{tabsumchib1},~\ref{tabsumchib2}, and~\ref{tabsumchib0}.

\begin{figure}[htbp]
\centering
\includegraphics[width=7cm]{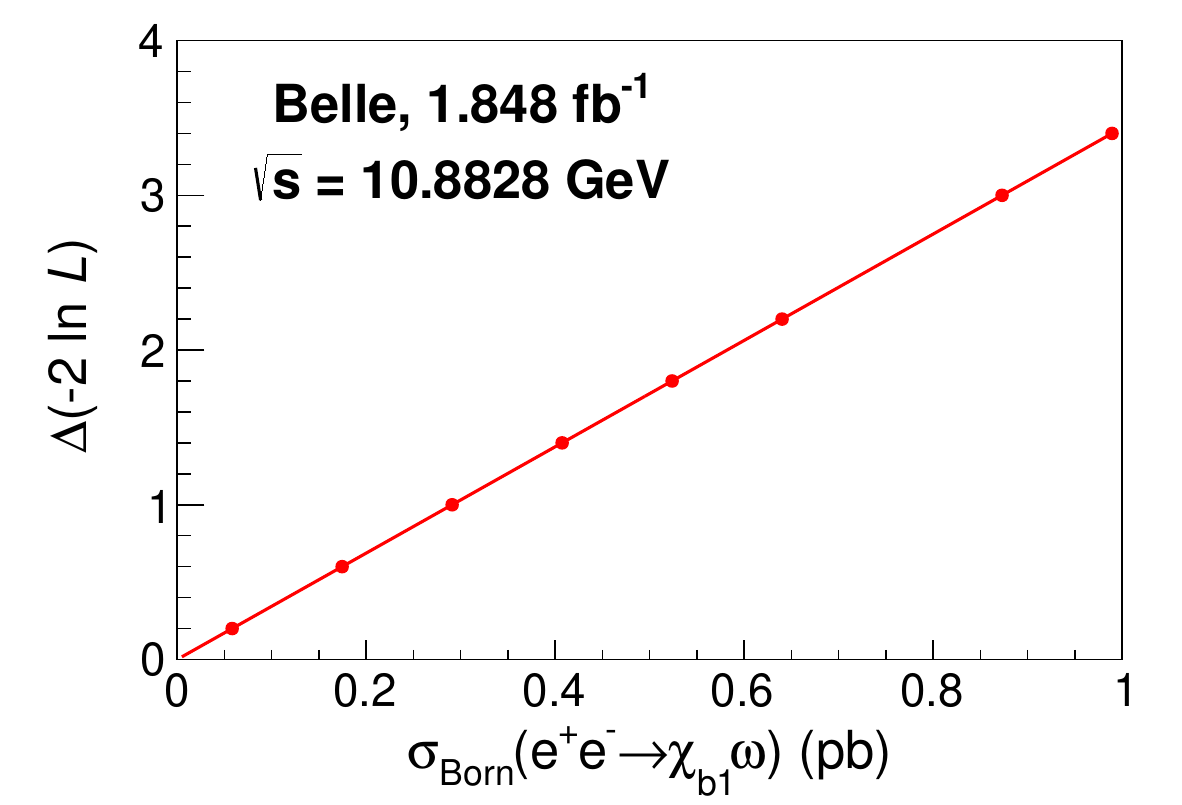}
\includegraphics[width=7cm]{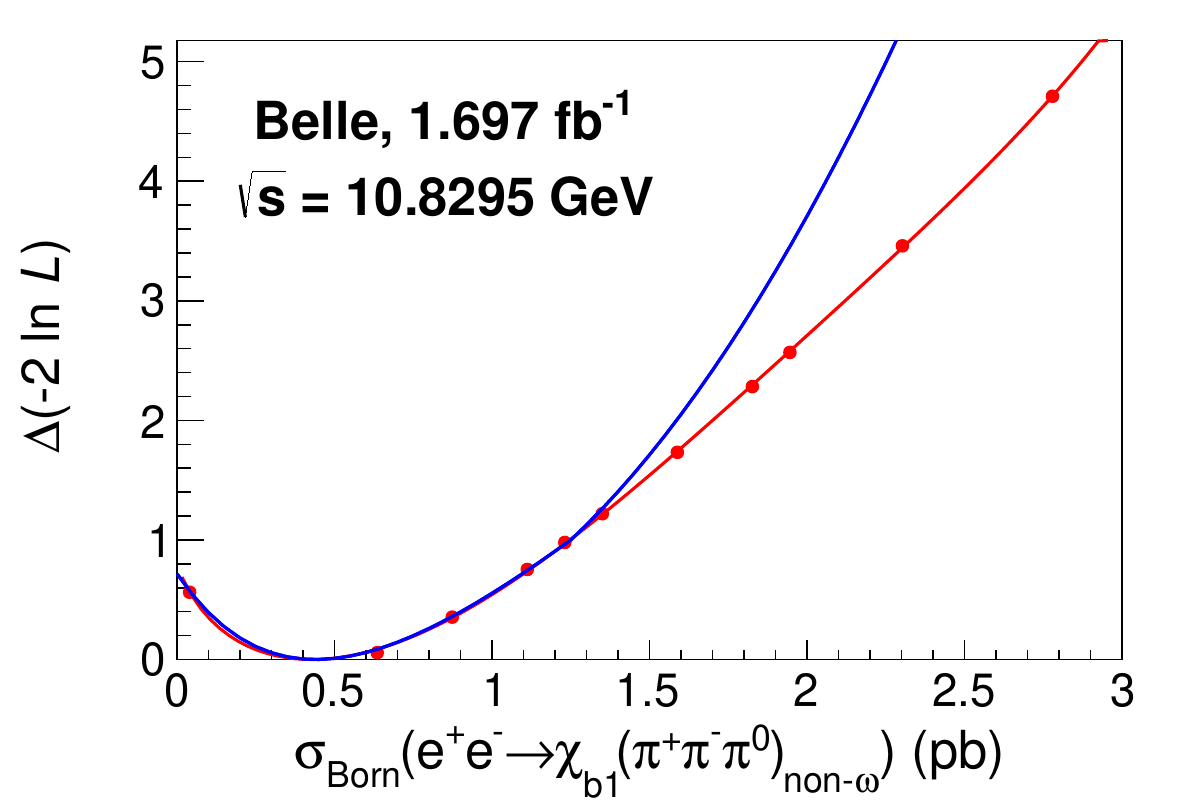}
\caption{The dependences of $\Delta(-2\ln\mathcal{L})$ on $\sigma_{\rm Born}(e^+e^- \to \chi_{b1}\,\omega)$
(red dots) at $\sqrt{s}$ = 10882.8 MeV (left) with $N^{\rm obs}=0$ and $N^{\rm bg}=0.3$ and on $\sigma_{\rm Born}(e^+e^- \to \chi_{b1}\,(\pi^+\pi^-\pi^0)_{\rm non-\omega})$ at $\sqrt{s}$ = 10829.5 MeV (right) with $N^{\rm obs}=1$ and $N^{\rm bg}=0.3$. The red curves show the results of the fit discussed in the text. The blue curve in (b) shows the approximation using asymmetric Gaussian uncertainties.}\label{likelihood}
\end{figure}

The results at each energy are used in section~\ref{sec.8} to fit the energy
dependence of the cross section. At c.m.\ energies with small numbers of
candidates, Gaussian uncertainties give a poor description of the
contribution to the likelihood. Figure~\ref{likelihood} shows examples of the
dependences of $\Delta(-2\ln\mathcal{L})$ on $\sigma_{\text{Born}}(e^+e^- \to \chi_{b1}\,\omega)$ and $\sigma_{\rm Born}(e^+e^- \to \chi_{b1}\,(\pi^+\pi^-\pi^0)_{\rm non-\omega})$ for samples with
$N^{\rm obs}=0$ and $N^{\rm obs}=1$. The cross section(s) corresponding to
different $\Delta(-2\ln\mathcal{L})$ values are obtained by requesting
intervals at various confidence levels from POLE, in the approximation
where the confidence level $(1-\alpha) = F_{\chi^2}( \Delta(-2 \ln
\mathcal{L}) ; 1)$. In POLE, the uncertainty on the background
prediction is set to the sum in quadrature of the statistical and
additive systematic uncertainties; 
the relative uncertainty on the signal efficiency is set to the uncertainty on the radiative correction factor (which is treated as uncorrelated --- see section~\ref{sec.7});
the relative uncertainty on the background efficiency
is set to zero.

For $N^{\rm obs}=0$, the dependence is parameterized using the proportionality function
\begin{equation}\label{eq:likelihood1}
f(x)=ax.
\end{equation}
When $N^{\rm obs}=1$ or higher, the dependence is parameterized by the function~\cite{220}
\begin{equation}\label{eq:likelihood2}
f(x)=2(p_2x + p_3 - p_1 + p_1 {\rm ln} \frac{p_1}{p_2x + p_3}) P_4(x),
\end{equation}
where $P_4=1+q_1 x+q_2 x^2+q_3 x^3+q_4 x^4$. 
The parameters $a$, $p_i$, and $q_i$ of eqs.~(\ref{eq:likelihood1}) and (\ref{eq:likelihood2}) are determined from fits. 
When fitting the energy dependence of the cross sections (section~\ref{sec.8}), instead of Gaussian uncertainties, we use eqs.~(\ref{eq:likelihood1}) and (\ref{eq:likelihood2}) to account for the contributions of low-population scan samples to the function being minimized. A similar procedure was used in ref.~\cite{220}. The fitted values of $a$, $p_i$, and $q_i$ are provided in the supplemental material (see appendix B).

\begin{table*}[htbp]
\renewcommand\arraystretch{1.2}
\setlength{\tabcolsep}{2pt}
\small
\centering
\caption{Results for $e^+e^-\to \chi_{b1}\,\omega$ as a function of c.m.\ energy at Belle II (points marked with an asterisk) and Belle (all other points). 
Uncertainties in the $N^{\rm sig}$ and $\sigma_{\rm Born}$ columns are statistical only; $\sigma^{\rm add}_{\rm syst}$ is the
additive systematic uncertainty; and $\sigma^{\rm UL}_{\rm Born}$ is the upper
limit on the Born cross section.
The underlined $N^{\rm sig}$ values are obtained by
fitting; corresponding statistical significances $\Sigma$ are also shown.
The remaining $N^{\rm sig}$ values are obtained using event counting: these
values, and the corresponding Born cross sections, are not corrected for
the efficiency
of the $\chi_{bJ}$ mass window requirement and cross-feed among the $\chi_{bJ}$ channels. These corrections are included in the energy-dependence fit (section~\ref{sec.8}).
}\label{tabsumchib1}
\vspace{0.2cm}
\begin{tabular}{ccccccccccc}
\hline\hline
&$\sqrt{s}$ (MeV) & $\cal L$ (fb$^{-1}$) & $\varepsilon$ & $1+\delta_{\rm ISR}$ & $N^{\rm sig}$ & $\Sigma(\sigma)$ & $\sigma_{\rm Born}$ (pb) & $\sigma^{\rm add}_{\rm syst}$ (pb) & $\sigma^{\rm UL}_{\rm Born}$ (pb) \\\hline
$\ast$&10701.0	&	1.640	&	0.146	&	0.640	&	$	0.4	^{+	1.4	}_{-	0.4	}$ &-	&	$	0.2	^{+	0.5	}_{-	0.2	}			$	&0.1&	1.4 \\
&10731.3	&	0.946	&	0.100	&	0.634	&	$	0.8	^{+	1.5	}_{-	0.7	}$	    &-	&	$	0.8	^{+	1.5	}_{-	0.7	}			$	&0.1&	4.1	\\
$\ast$&10745.0    &	9.870	&	0.178	&	0.630	&	\underline{$	71.8	^{+	10.3	}_{-	9.5	}$}	&10	&	$	3.8	^{+	0.6	}_{-	0.5	}	$	&0.3&	--	\\
&10771.2	&	0.955	&	0.106	&	0.784	&	\underline{$	4.7	^{+	2.9	}_{-	2.2	}$}	&2.8	&	$	3.6	^{+	2.3	}_{-	1.7	}	$	&0.7&	7.3	\\
$\ast$&10805.0    &	4.690  	&	0.177	&	0.940  	&	\underline{$	6.8	^{+	4.0	}_{-	3.0	}$}	&2.6	&	$	0.5	^{+	0.3	}_{-	0.2	}	$	&0.1&	1.0	\\
&10829.5	&	1.697	&	0.100	&	0.941	&	$	0.7	^{+	1.4	}_{-	0.7	}$	&-	&	$	0.3	^{+	0.5	}_{-	0.3	}			$	&0.1&	1.5	\\
&10848.9	&	0.989	&	0.101	&	0.924	&	$	2.8	^{+	1.9	}_{-	1.4	}$	&-	&	$	1.8	^{+	1.2	}_{-	0.9	}			$	&0.1&	4.6	\\
&10857.4	&	0.988	&	0.101	&	0.916	&	$	0.0	^{+	0.5	}_{-	0.2	}$	&-	&	$	0.0	^{+	0.3	}_{-	0.1	}			$	&0.1&	1.4	\\
&10865.8	&	122.0	&	0.109	&	0.909	&	\underline{$	82.8	^{+	10.9	}_{-	10.8	}$}	&12	&	$	0.4	^{+	0.1	}_{-	0.1	}	$	&0.1&	--	\\
&10877.8	&	0.978	&	0.104	&	0.899	&	$	0.8	^{+	1.5	}_{-	0.7	}$	&-	&	$	0.5	^{+	1.0	}_{-	0.5	}			$	&0.1&	2.7	\\
&10882.8	&	1.848	&	0.104	&	0.895	&	$	0.0	^{+	0.5	}_{-	0.2	}$	&-	&	$	0.0	^{+	0.2	}_{-	0.1	}			$	&0.1&	0.7	\\
&10888.9	&	0.990	&	0.104	&	0.891	&	$	0.0	^{+	0.5	}_{-	0.2	}$	&-	&	$	0.0	^{+	0.3	}_{-	0.1	}			$	&0.1&	1.4	\\
&10898.3	&	2.408	&	0.104	&	0.885	&	$	0.6	^{+	1.4	}_{-	0.6	}$	&-	&	$	0.2	^{+	0.4	}_{-	0.2	}			$	&0.1&	1.1	\\
&10907.3	&	0.980	&	0.106	&	0.879	&	$	0.0	^{+	0.5	}_{-	0.2	}$	&-	&	$	0.0	^{+	0.3	}_{-	0.1	}			$	&0.1&	1.4	\\
&10928.7	&	1.149	&	0.106	&	0.869	&	$	0.0	^{+	0.5	}_{-	0.2	}$	&-	&	$	0.0	^{+	0.3	}_{-	0.1	}			$	&0.1&	1.2	\\
&10957.5	&	0.969	&	0.106	&	0.859	&	$	0.0	^{+	0.5	}_{-	0.2	}$	&-	&	$	0.0	^{+	0.3	}_{-	0.1	}			$	&0.1&	1.6	\\
&10975.3	&	0.999	&	0.106	&	0.854	&	$	0.0	^{+	0.5	}_{-	0.2	}$	&-	&	$	0.0	^{+	0.3	}_{-	0.1	}			$	&0.1&	1.4	\\
&10990.4	&	0.985	&	0.107	&	0.851	&	$	0.0	^{+	0.5	}_{-	0.2	}$	&-	&	$	0.0	^{+	0.3	}_{-	0.1	}			$	&0.1&	1.4	\\
&11003.9	&	0.976	&	0.107	&	0.849	&	$	0.8	^{+	1.5	}_{-	0.7	}$	&-	&	$	0.5	^{+	1.0	}_{-	0.5	}			$	&0.1&	2.6	\\
&11014.8	&	0.771	&	0.107	&	0.848	&	$	0.0	^{+	0.5	}_{-	0.2	}$	&-	&	$	0.0	^{+	0.4	}_{-	0.2	}			$	&0.1&	2.0	\\
&11018.5	&	0.859	&	0.107	&	0.847	&	$	0.0	^{+	0.5	}_{-	0.2	}$	&-	&	$	0.0	^{+	0.4	}_{-	0.2	}			$	&0.1&	1.8	\\
&11020.8	&	0.982	&	0.108	&	0.847	&	$	0.0	^{+	0.5	}_{-	0.2	}$	&-	&	$	0.0	^{+	0.3	}_{-	0.1	}			$	&0.1&	1.4	\\
\hline\hline
\end{tabular}
\end{table*} 

\begin{table*}[htbp]
\renewcommand\arraystretch{1.2}
\setlength{\tabcolsep}{2pt}
\small
\centering
\caption{Results for $e^+e^-\to \chi_{b2}\,\omega$ at each energy point at Belle and Belle II. Column descriptions are the same as in table~\ref{tabsumchib1}.}\label{tabsumchib2}
\vspace{0.2cm}
\begin{tabular}{ccccccccccc}
\hline\hline
&$\sqrt{s}$ (MeV) & $\cal L$ (fb$^{-1}$) & $\varepsilon$ & $1+\delta_{\rm ISR}$ & $N^{\rm sig}$ & $\Sigma(\sigma)$ & $\sigma_{\rm Born}$ (pb) & $\sigma^{\rm add}_{\rm syst}$ (pb) & $\sigma^{\rm UL}_{\rm Born}$ (pb) \\\hline
$\ast$&10701.0	&	1.640	&	0.146	&	0.620	&	$	0.0	^{+	0.5	}_{-	0.2	}$	& - &	$	0.0	^{+	0.4	}_{-	0.2	}			$	&0.1&	1.3 \\
&10731.3	&	0.946	&	0.095	&	0.629	&	$	0.8	^{+	1.5	}_{-	0.7	}$	&	-& $	1.7	^{+	3.2	}_{-	1.5	}			$	&0.1&	8.6	\\
$\ast$&10745.0    &	9.870	&	0.179	&	0.630	&	\underline{$	24.6	^{+	7.5\phantom{2}	}_{-	6.6	}$}	& 4.6 &	$	2.6	^{+	0.8	}_{-	0.7	}	$	&0.4&	--	\\
&10771.2	&	0.955	&	0.106	&	0.782	&	\underline{$	3.3	^{+	2.6	}_{-	1.8	}$}	& 2.4 &	$	5.0	^{+	3.9	}_{-	2.7	}	$	&1.7&	11.7\phantom{1}	\\
$\ast$&10805.0    &	4.690  	&	0.178	&	0.940  	&	\underline{$	10.7	^{+	4.4\phantom{2}	}_{-	3.6	}$}	&3.5 &	$	1.6	^{+	0.7	}_{-	0.5	}	$	&0.3&	2.6	\\
&10829.5	&	1.697	&	0.096	&	0.936	&	$	0.0	^{+	0.5	}_{-	0.2	}$	& -&	$	0.0	^{+	0.4	}_{-	0.2	}			$	&0.1&	1.5	\\
&10848.9	&	0.989	&	0.096	&	0.920	&	$	0.0	^{+	0.5	}_{-	0.2	}$	& -&	$	0.0	^{+	0.7	}_{-	0.3	}			$	&0.1&	2.8	\\
&10857.4	&	0.988	&	0.097	&	0.912	&	$	0.0	^{+	0.5	}_{-	0.2	}$	& -&	$	0.0	^{+	0.7	}_{-	0.3	}			$	&0.1&	2.8	\\
&10865.8	&	122.0	&	0.109	&	0.905	&	\underline{$	14.7	^{+	7.3\phantom{2}	}_{-	6.4	}$}	&2.5&	$	0.1	^{+	0.1	}_{-	0.1	}	$	&0.1&	0.2	\\
&10877.8	&	0.978	&	0.099	&	0.895	&	$	0.0	^{+	0.5	}_{-	0.2	}$	& -&	$	0.0	^{+	0.7	}_{-	0.3	}	                  $	&0.1&	2.9	\\
&10882.8	&	1.848	&	0.099	&	0.892	&	$	0.0	^{+	0.5	}_{-	0.2	}$	& -&	$	0.0	^{+	0.4	}_{-	0.1	}			$	&0.1&	1.4	\\
&10888.9	&	0.990	&	0.100	&	0.887	&	$	0.8	^{+	1.5	}_{-	0.7	}$	& -&	$	1.1	^{+	2.0	}_{-	0.9	}			$	&0.1&	5.5	\\
&10898.3	&	2.408	&	0.100	&	0.882	&	$	0.0	^{+	0.5	}_{-	0.2	}$	& -&	$	0.0	^{+	0.3	}_{-	0.1	}			$	&0.1&	0.9	\\
&10907.3	&	0.980	&	0.101	&	0.876	&	$	0.0	^{+	0.5	}_{-	0.2	}$	& -&	$	0.0	^{+	0.7	}_{-	0.3	}			$	&0.1&	2.9	\\
&10928.7	&	1.149	&	0.101	&	0.866	&	$	0.0	^{+	0.5	}_{-	0.2	}$	& -&	$	0.0	^{+	0.6	}_{-	0.2	}			$	&0.1&	2.5	\\
&10957.5	&	0.969	&	0.101	&	0.857	&	$	0.0	^{+	0.5	}_{-	0.2	}$	& -&	$	0.0	^{+	0.7	}_{-	0.3	}			$	&0.1&	3.0	\\
&10975.3	&	0.999	&	0.102	&	0.852	&	$	2.8	^{+	1.9	}_{-	1.4	}$	& -&	$	3.8	^{+	2.6	}_{-	1.9	}			$	&0.1&	9.7	\\
&10990.4	&	0.985	&	0.102	&	0.849	&	$	0.0	^{+	0.5	}_{-	0.2	}$	& -&	$	0.0	^{+	0.7	}_{-	0.3	}			$	&0.1&	2.9	\\
&11003.9	&	0.976	&	0.102	&	0.847	&	$	1.7	^{+	1.8	}_{-	1.1	}$	& -&	$	2.4	^{+	2.5	}_{-	1.5	}			$	&0.1&	7.7	\\
&11014.8	&	0.771	&	0.102	&	0.846	&	$	0.0	^{+	0.5	}_{-	0.2	}$	& -&	$	0.0	^{+	0.9	}_{-	0.4	}			$	&0.1&	3.7	\\
&11018.5	&	0.859	&	0.103	&	0.845	&	$	0.0	^{+	0.5	}_{-	0.2	}$	& -&	$	0.0	^{+	0.8	}_{-	0.3	}			$	&0.1&	3.3	\\
&11020.8	&	0.982	&	0.104	&	0.845	&	$	0.8	^{+	1.5	}_{-	0.7	}$	& -&	$	1.1	^{+	2.1	}_{-	1.0	}			$	&0.1&	5.6	\\
\hline\hline
\end{tabular}
\end{table*}

\section{$e^+e^-\to\chi_{bJ}\,(\pi^+\pi^-\pi^0)_{\rm non-\omega}$ at Belle and Belle II}~\label{sec.6}

To measure the $e^+e^-\to\chi_{bJ}\,(\pi^+\pi^-\pi^0)_{\rm non-\omega}$ cross sections, we require the mass of $\pi^+\pi^-\pi^0$ combinations to be outside the $\omega$ signal region, $M(\pi^+\pi^-\pi^0)$ $<$ 0.75 GeV/$c^2$ or $M(\pi^+\pi^-\pi^0)$ $>$ 0.81 GeV/$c^2$.
In principle, the $\chi_{bJ}\,(\pi^+\pi^-\pi^0)_{\rm non-\omega}$ final state may be contaminated by background from
higher-$\Upsilon$ processes such as $e^+e^-\to\Upsilon(2S)\pi^+\pi^-$ and $\Upsilon_2(1D)\pi^+\pi^-$.~We check this by examining the spectrum of
the $\pi^+\pi^-$ recoil mass in combined data samples: no $\Upsilon(2S)$ or $\Upsilon_2(1D)$ signals are seen.
To further study and check possible backgrounds, we produce MC samples for each of the following processes: $e^+e^-\to\Upsilon(2S)\pi^+\pi^-$ with $\Upsilon(2S)\to\chi_{bJ}\gamma$, $e^+e^-\to\Upsilon_2(1D)\pi^+\pi^-$ with $\Upsilon_2(1D)\to\chi_{bJ}\gamma$, and $e^+e^-\to\Upsilon(1S)\pi^+\pi^-\pi^0\pi^0$. After applying all the signal selection criteria of $e^+e^-\to\chi_{bJ}\,(\pi^+\pi^-\pi^0)_{\rm non-\omega}$, the efficiencies are very low ($<10^{-3}$).
The events from the process $e^+e^-\to\Upsilon(1S)\pi^+\pi^-\pi^0\pi^0$ do not peak in the $M(\Upsilon(1S)\gamma)$ distribution. Thus, we 
neglect all the above backgrounds.

\begin{figure}[htbp]
\centering
\includegraphics[width=7cm]{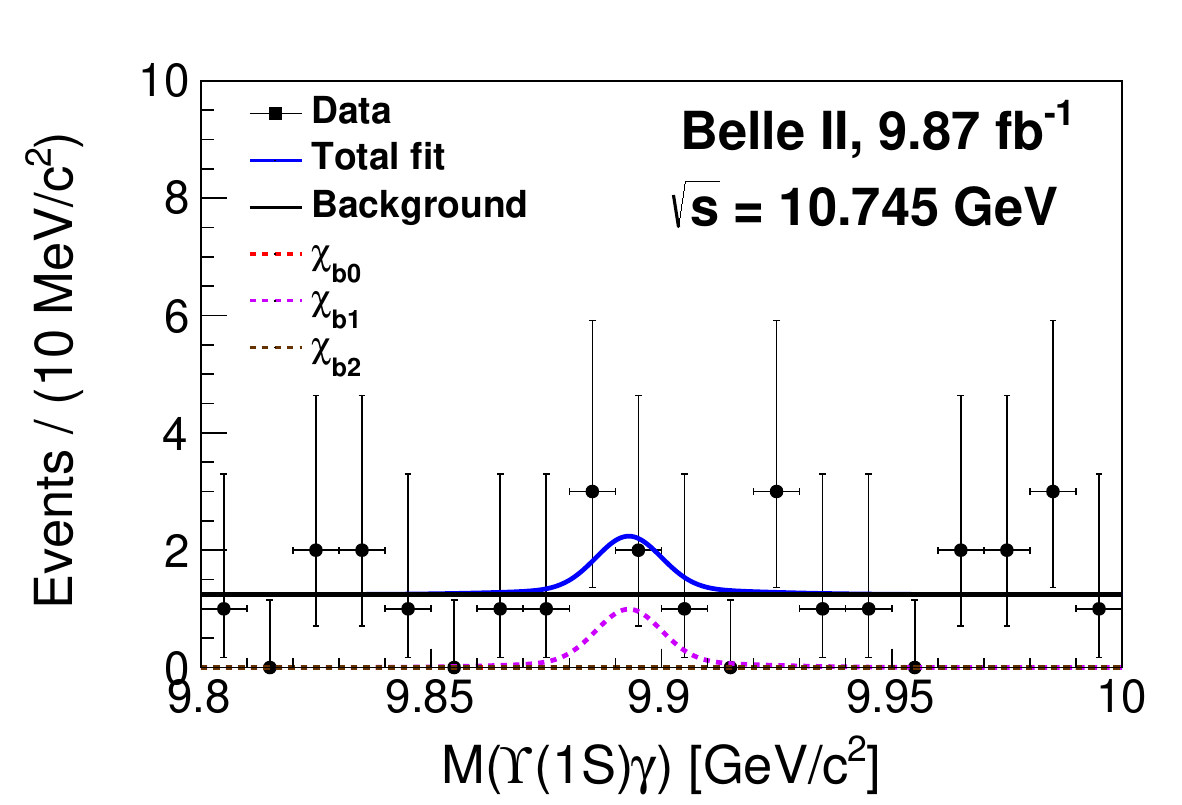}
\includegraphics[width=7cm]{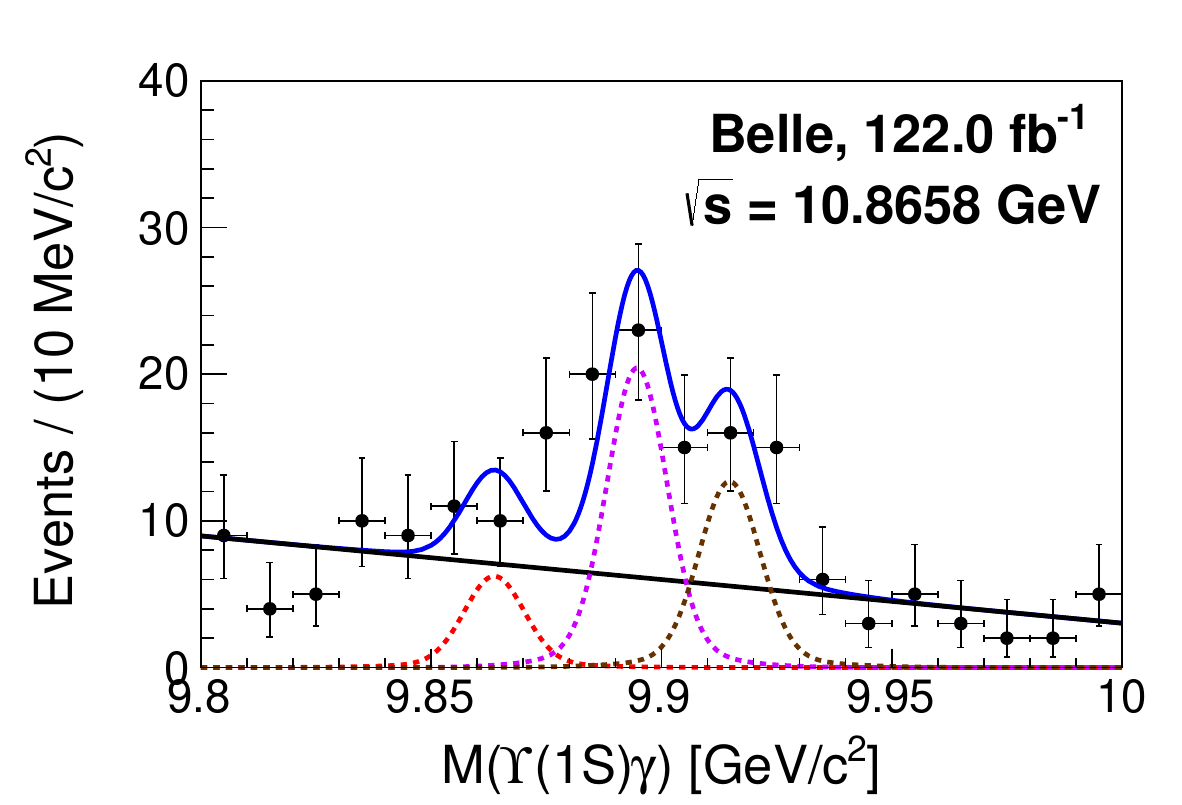}
\caption{Distributions of $M(\Upsilon(1S)\gamma)$ outside the $\omega$ signal region 
in Belle II data at $\sqrt{s}$ = 10.745 GeV and Belle data at $\sqrt{s}$ = 10.8658 GeV with fit results overlaid.
The solid blue and black curves show the total fit and total background; the dashed red, violet, and brown curves show the $\chi_{b0}$, $\chi_{b1}$, and $\chi_{b2}$ signal components, respectively.
}\label{non-fit}
\end{figure}

For $\sqrt{s}$ = 10.8658 GeV at Belle and $\sqrt{s}$ = 10.745 GeV at Belle II, 
the signal yields are extracted using an unbinned extended maximum-likelihood fit to the $M(\Upsilon(1S)\gamma)$ distributions, as shown in figure~\ref{non-fit}.
Each $\chi_{bJ}$ signal shape is described by a double Gaussian function, and signal shape parameters are fixed from signal MC simulations.
A first-order polynomial function is used to describe the background.

\begin{figure}[htbp]
\centering
\includegraphics[width=3.6cm]{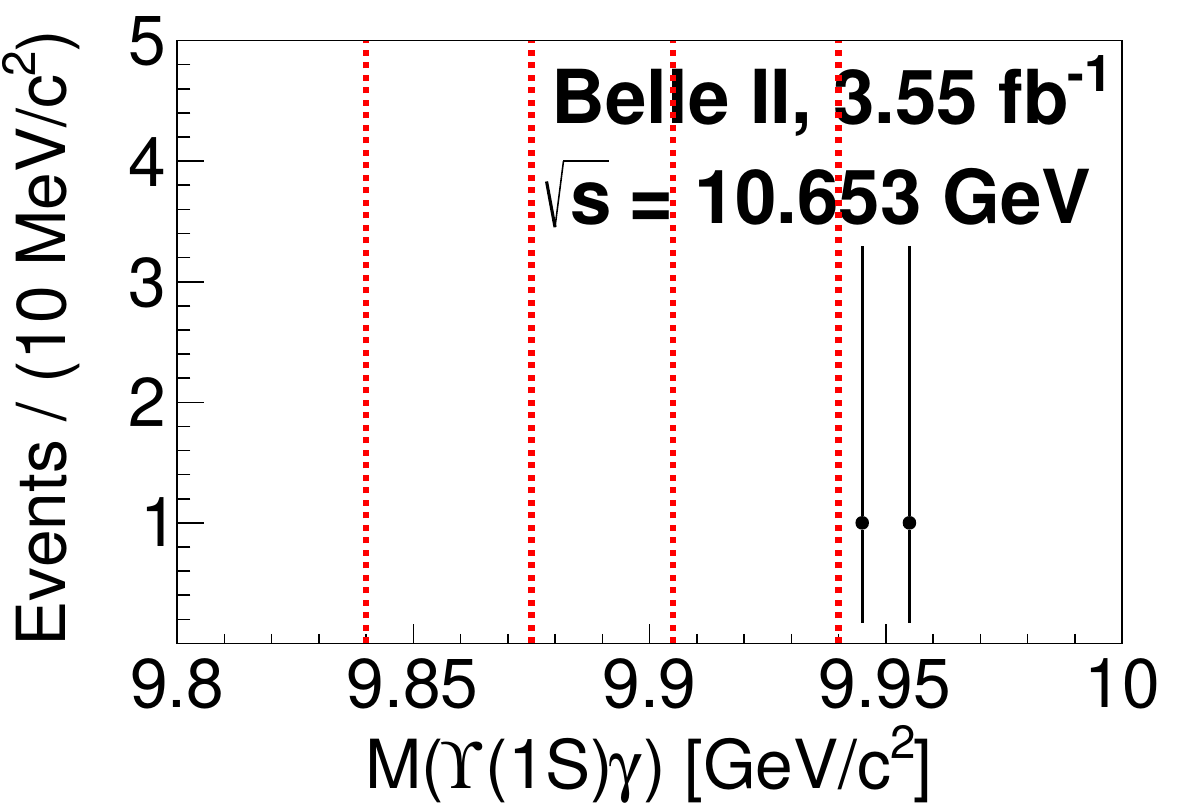}
\includegraphics[width=3.6cm]{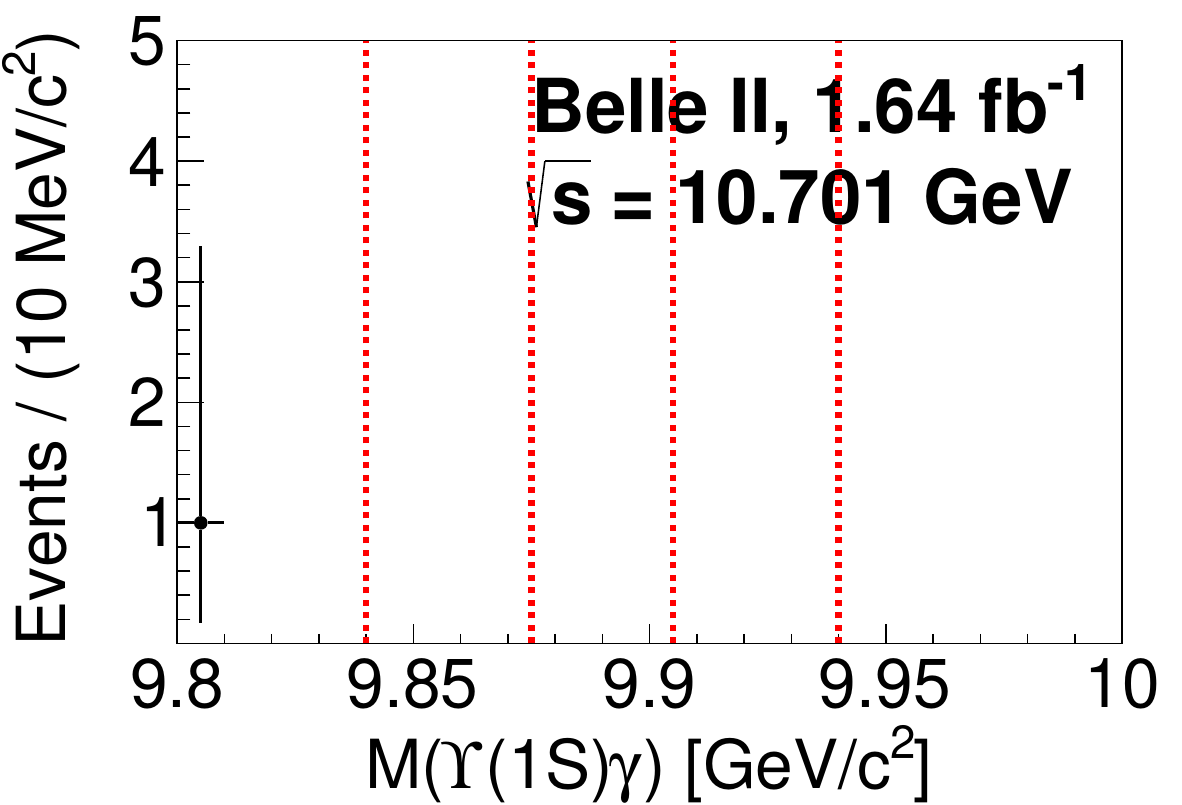}
\includegraphics[width=3.6cm]{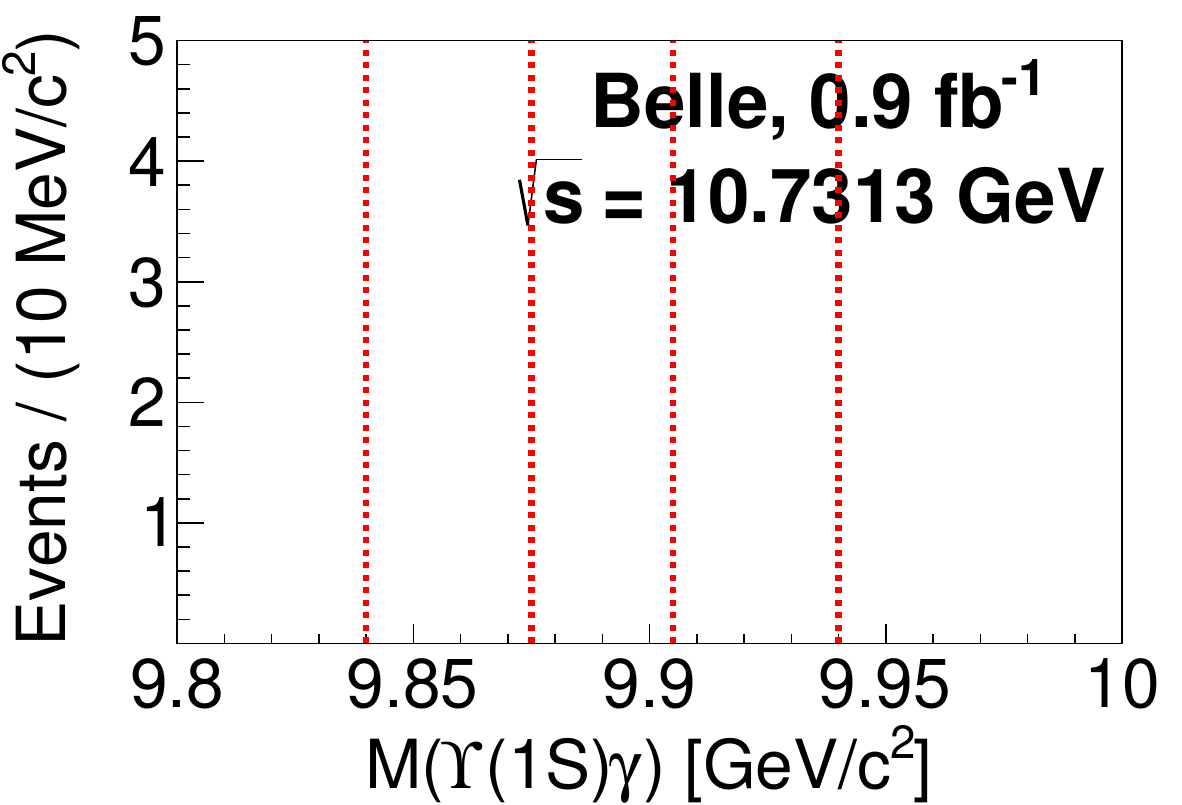}
\includegraphics[width=3.6cm]{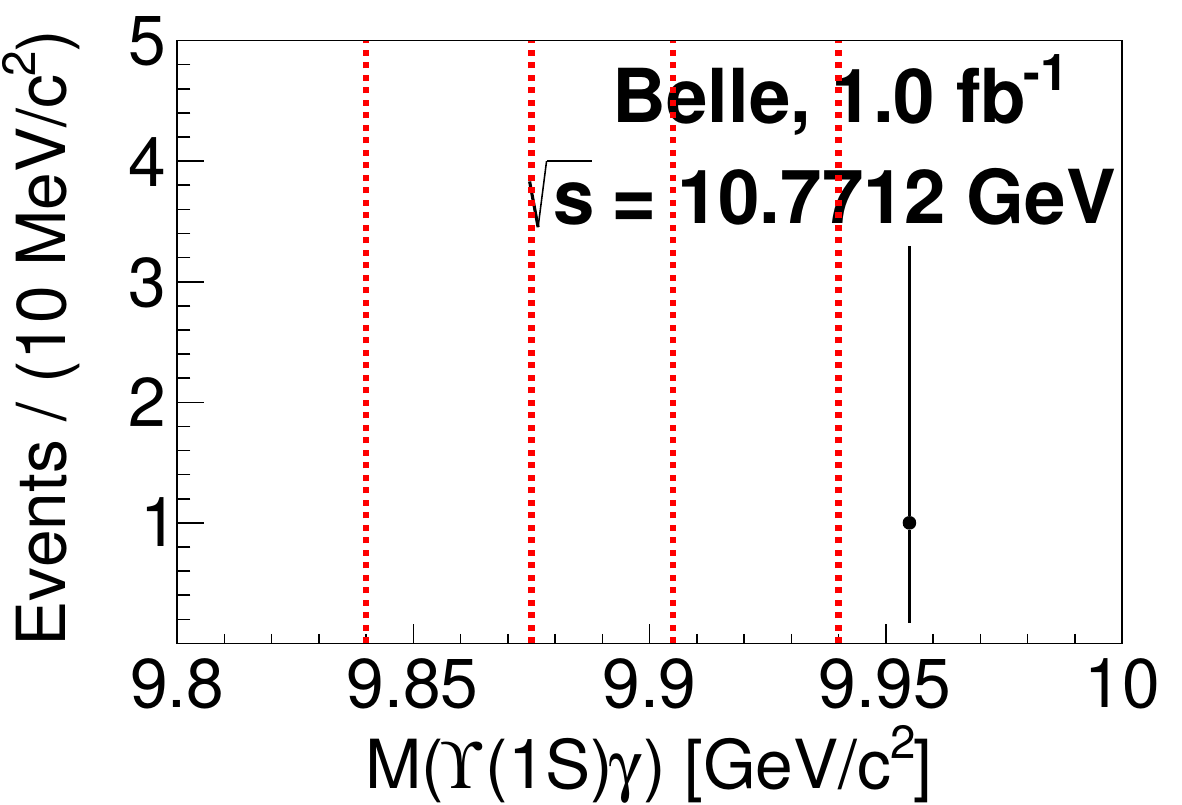}

\includegraphics[width=3.6cm]{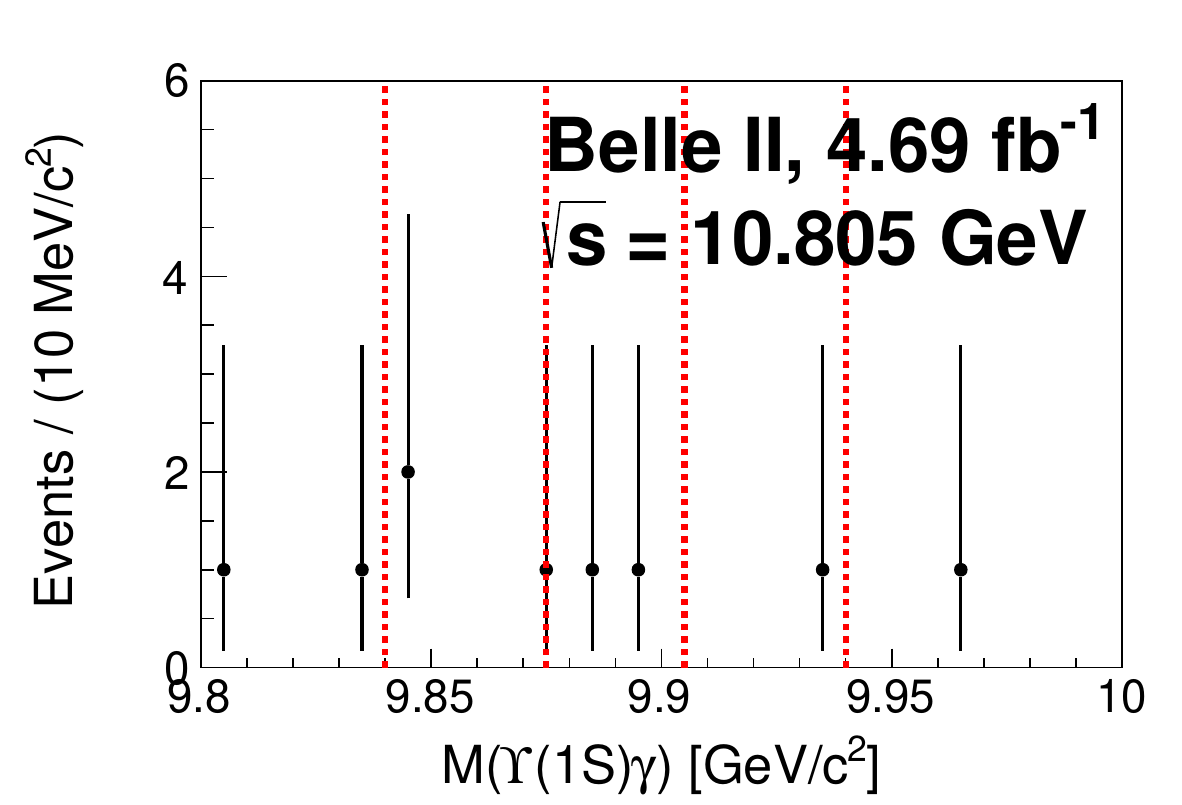}
\includegraphics[width=3.6cm]{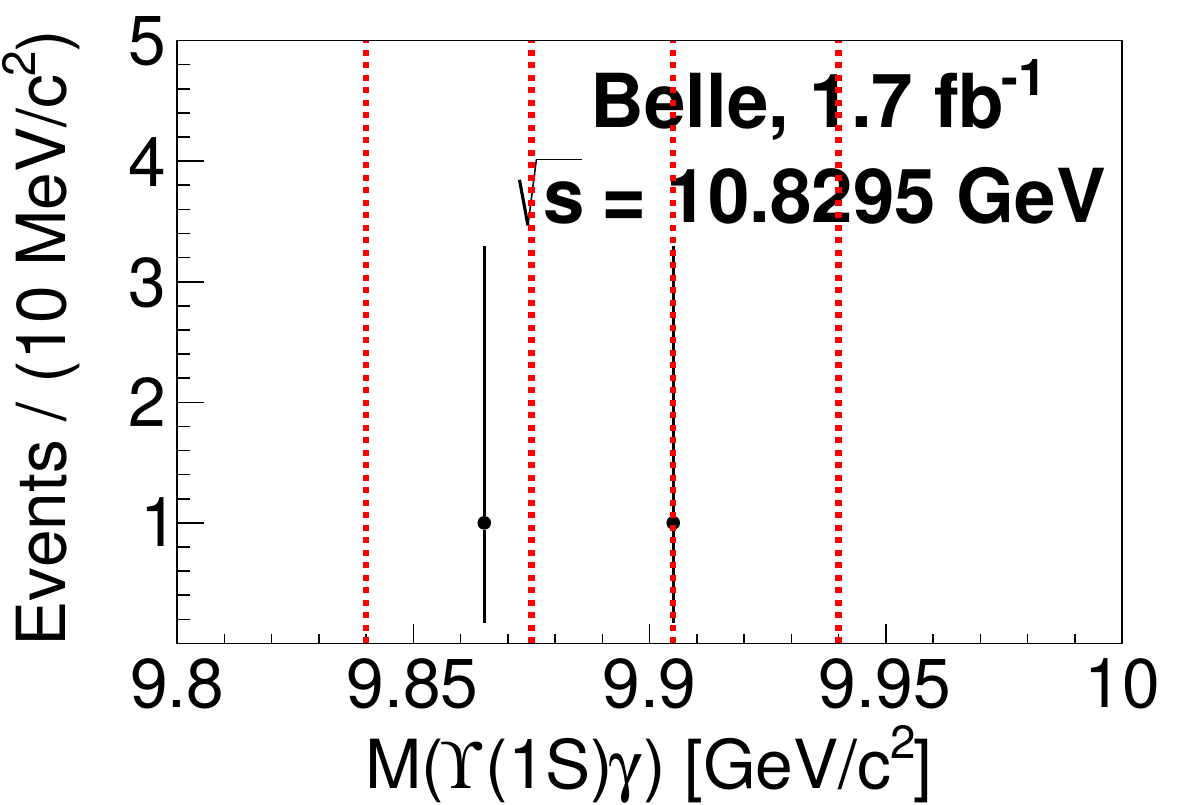}
\includegraphics[width=3.6cm]{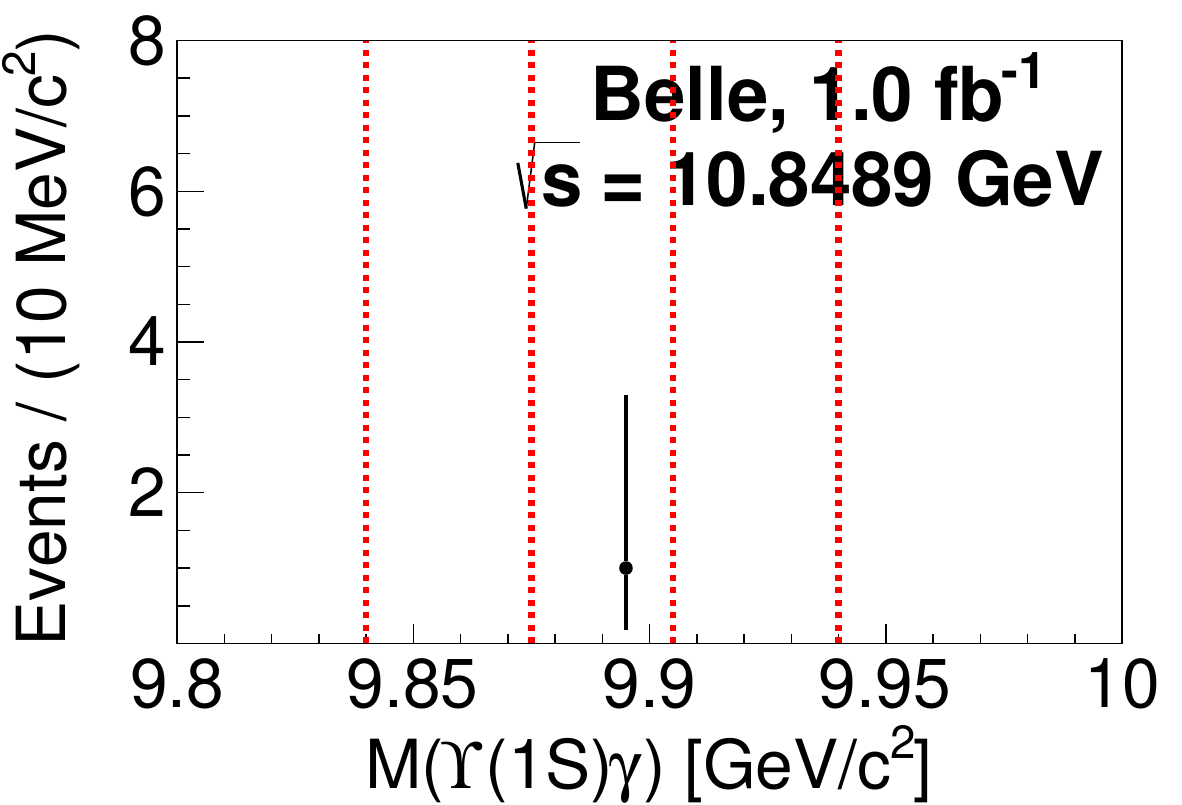}
\includegraphics[width=3.6cm]{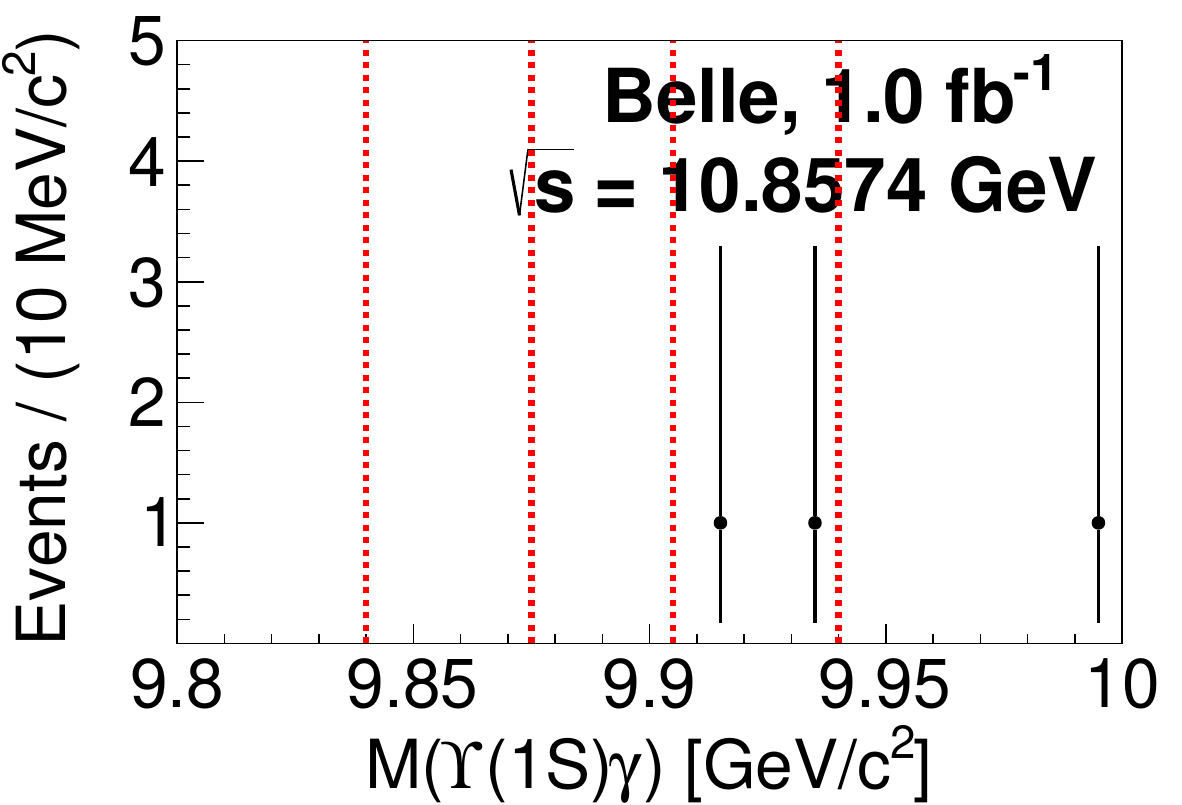}

\includegraphics[width=3.6cm]{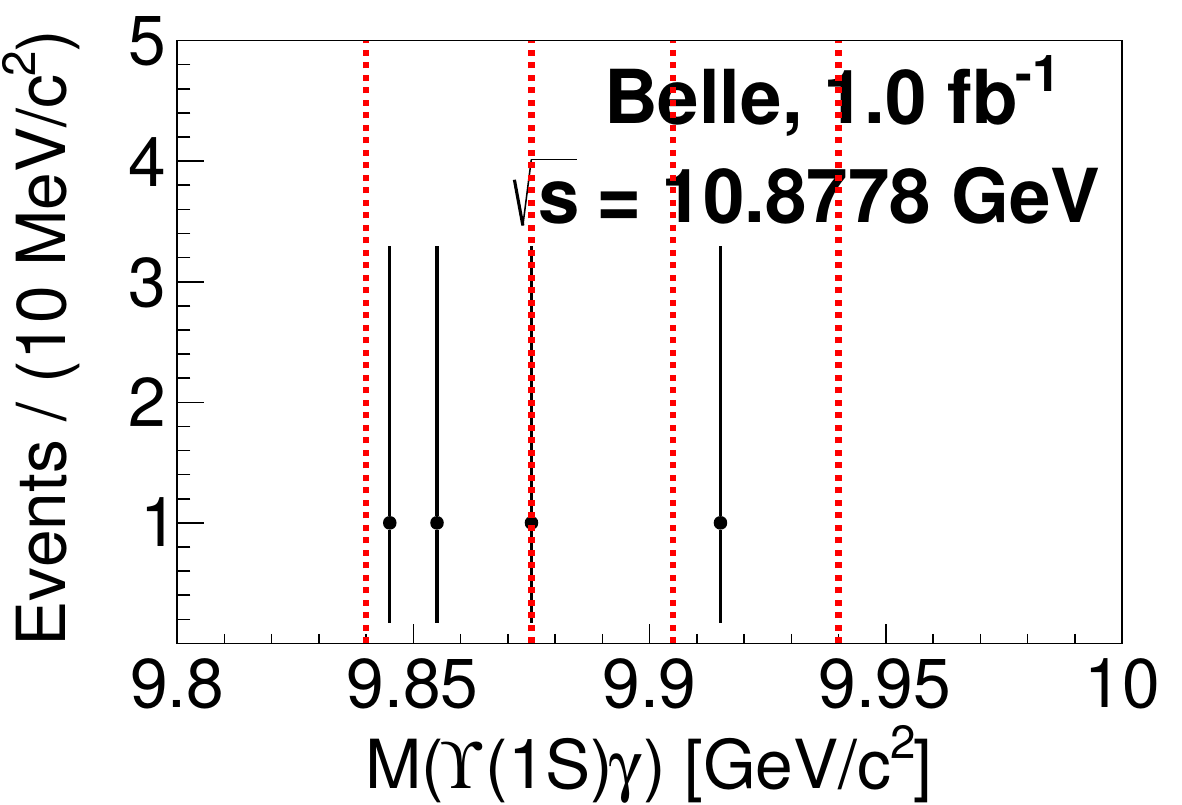}
\includegraphics[width=3.6cm]{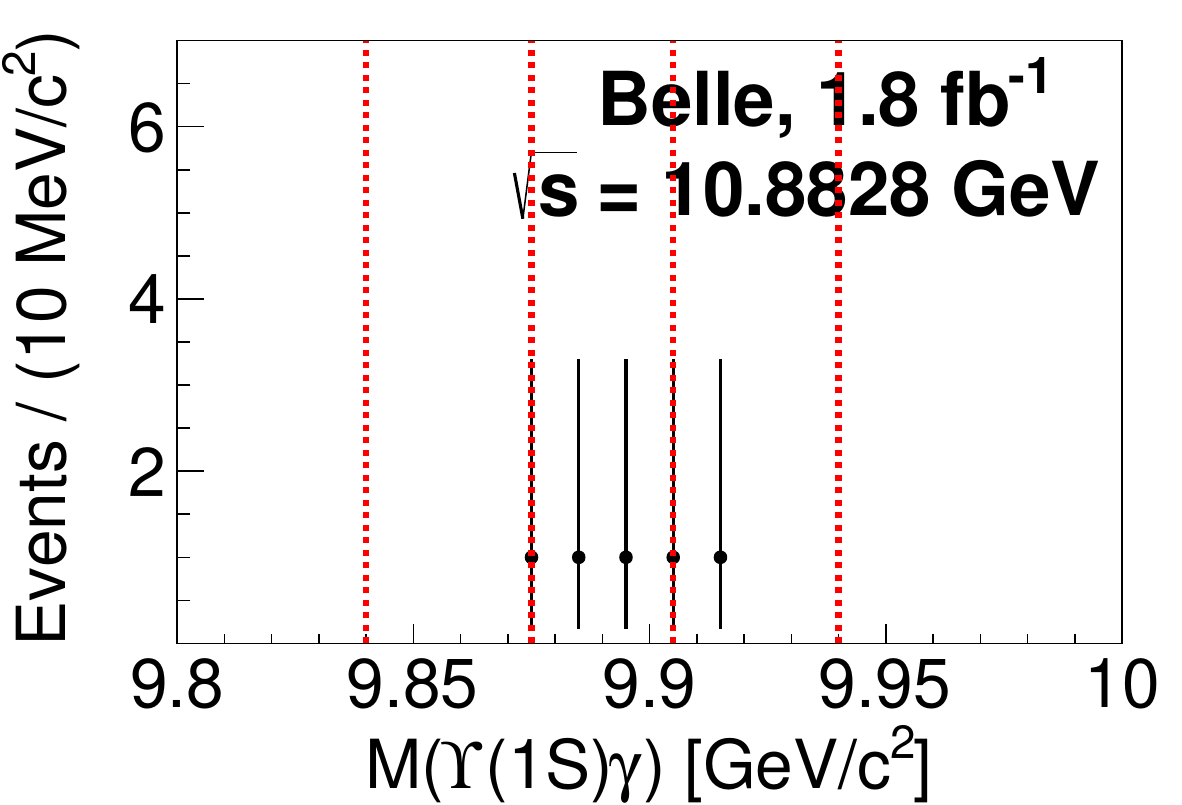}
\includegraphics[width=3.6cm]{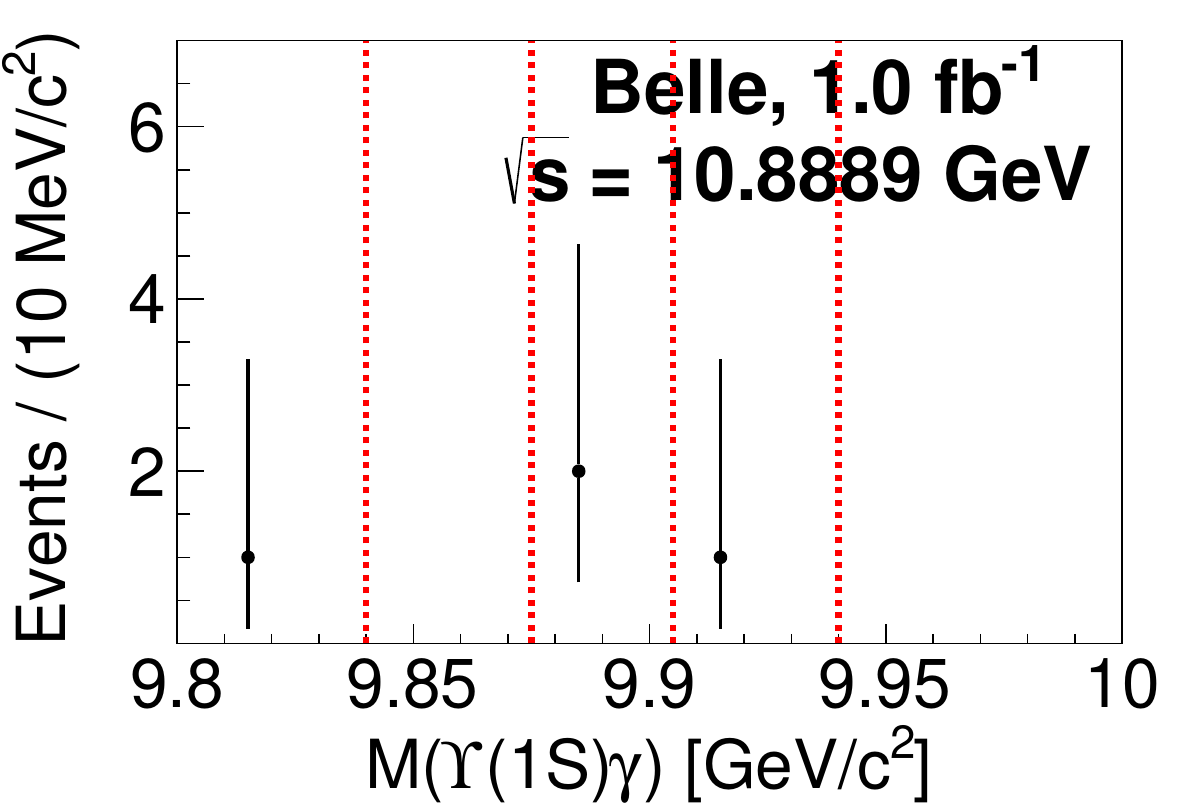}
\includegraphics[width=3.6cm]{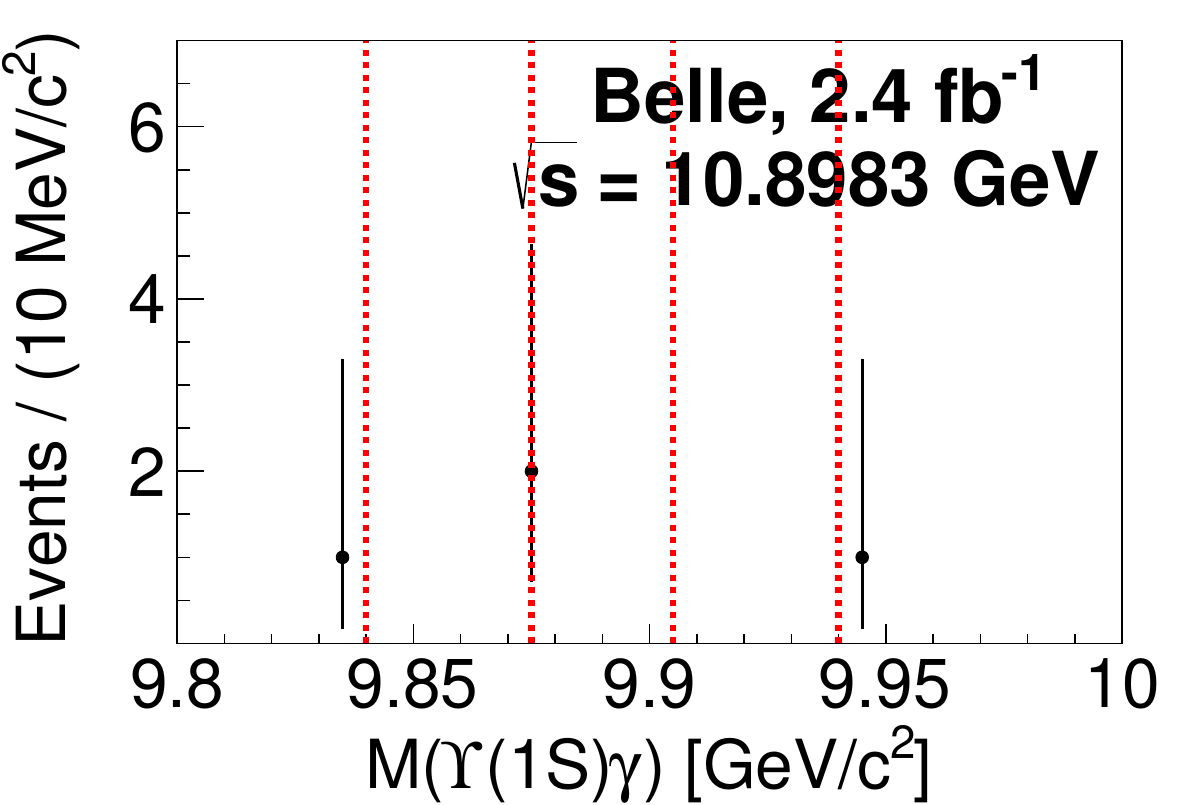}

\includegraphics[width=3.6cm]{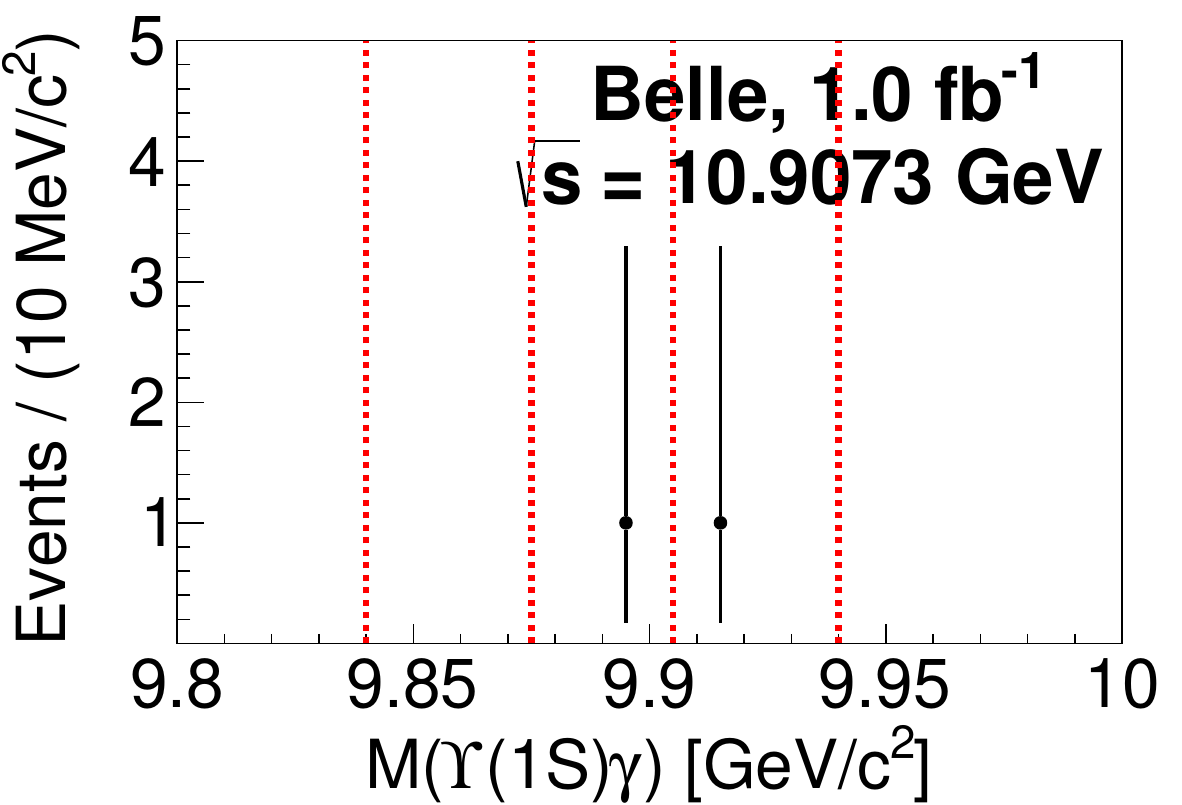}
\includegraphics[width=3.6cm]{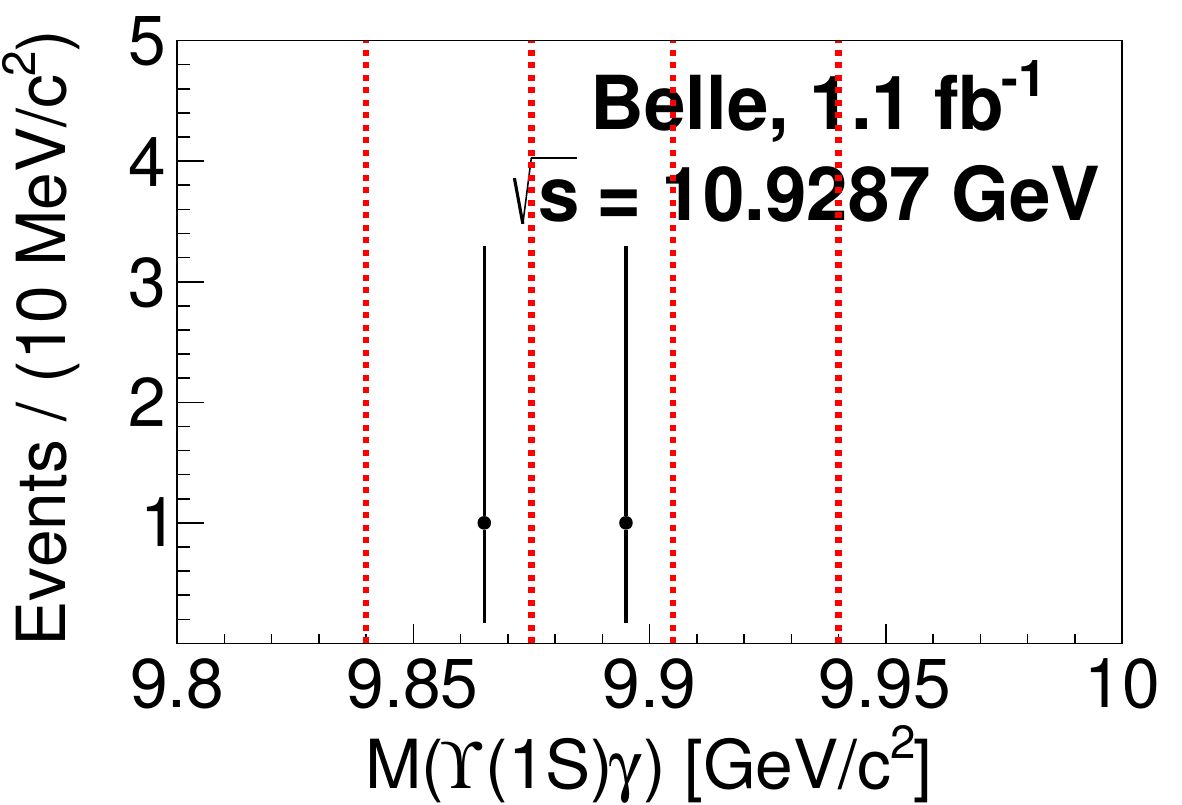}
\includegraphics[width=3.6cm]{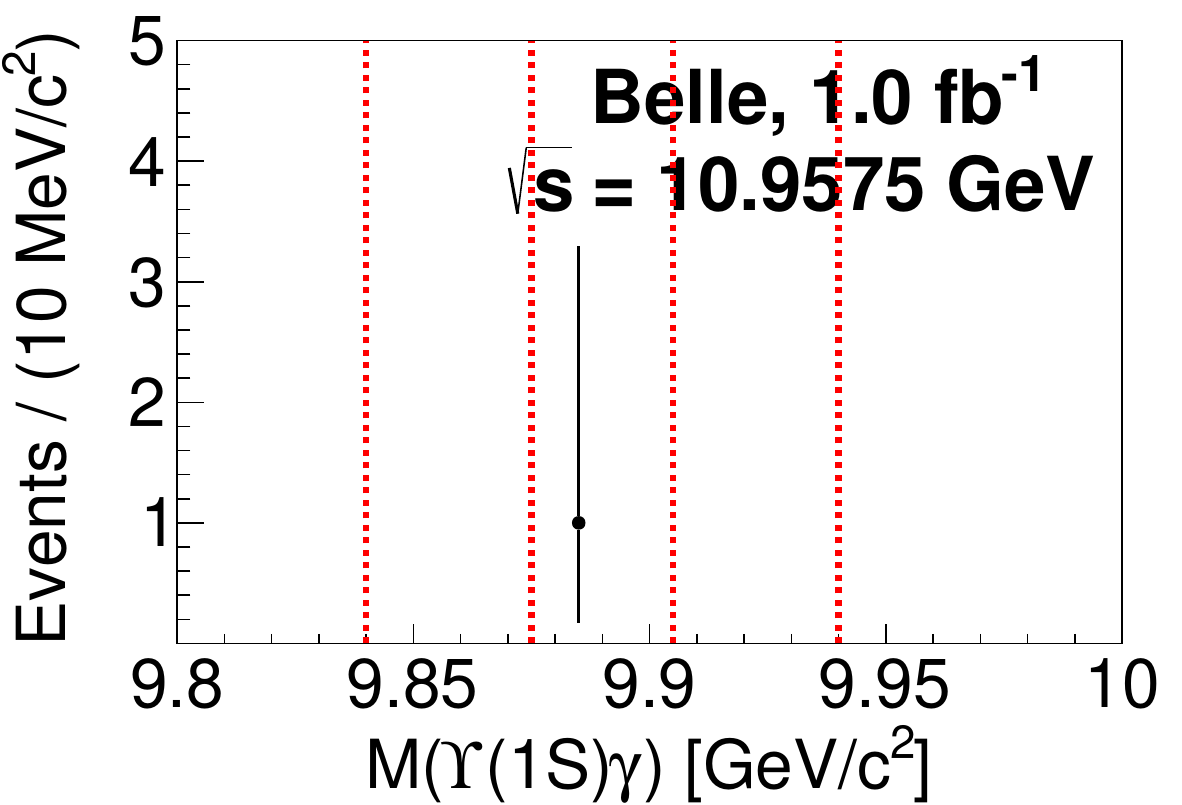}
\includegraphics[width=3.6cm]{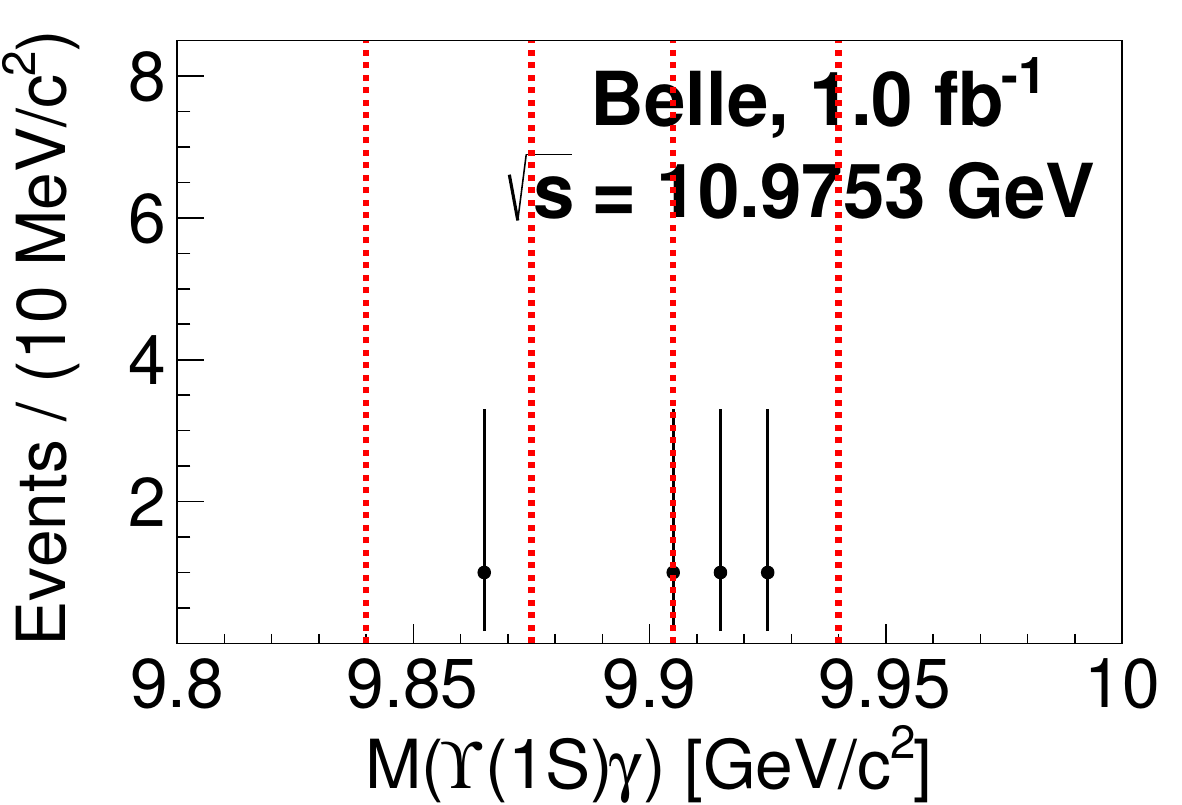}

\includegraphics[width=3.6cm]{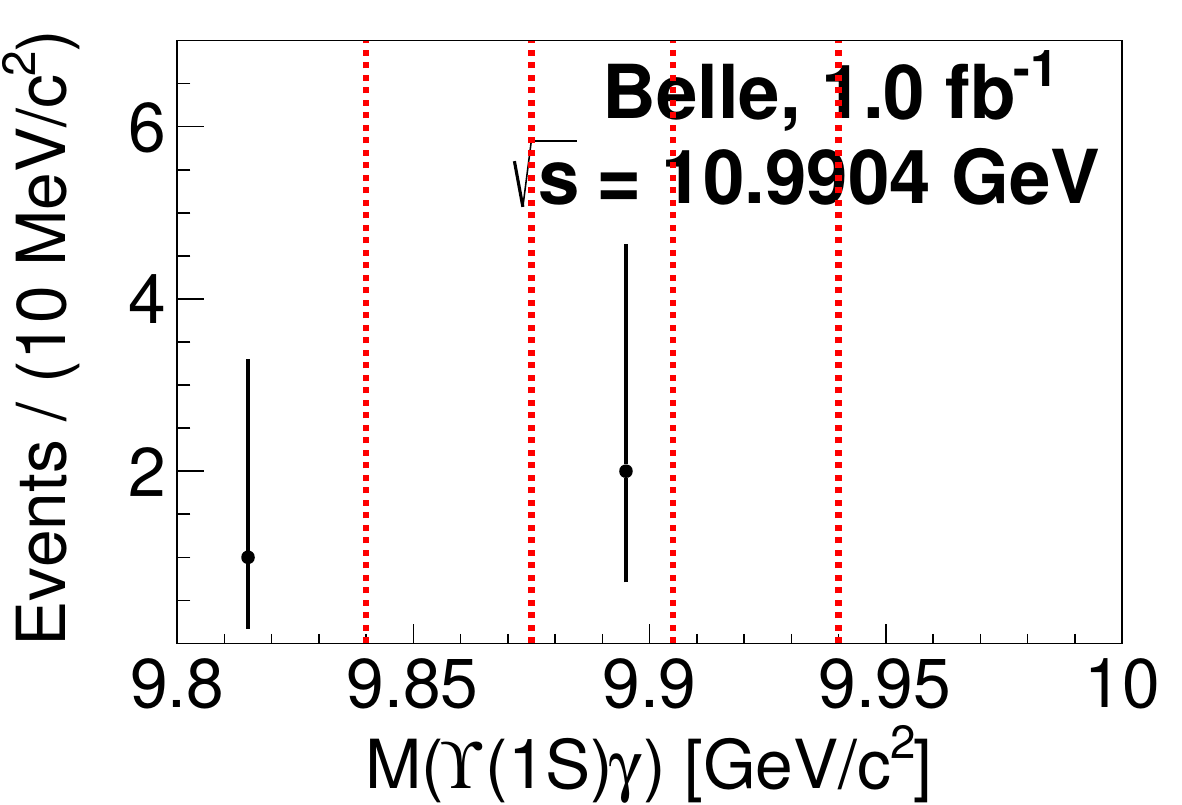}
\includegraphics[width=3.6cm]{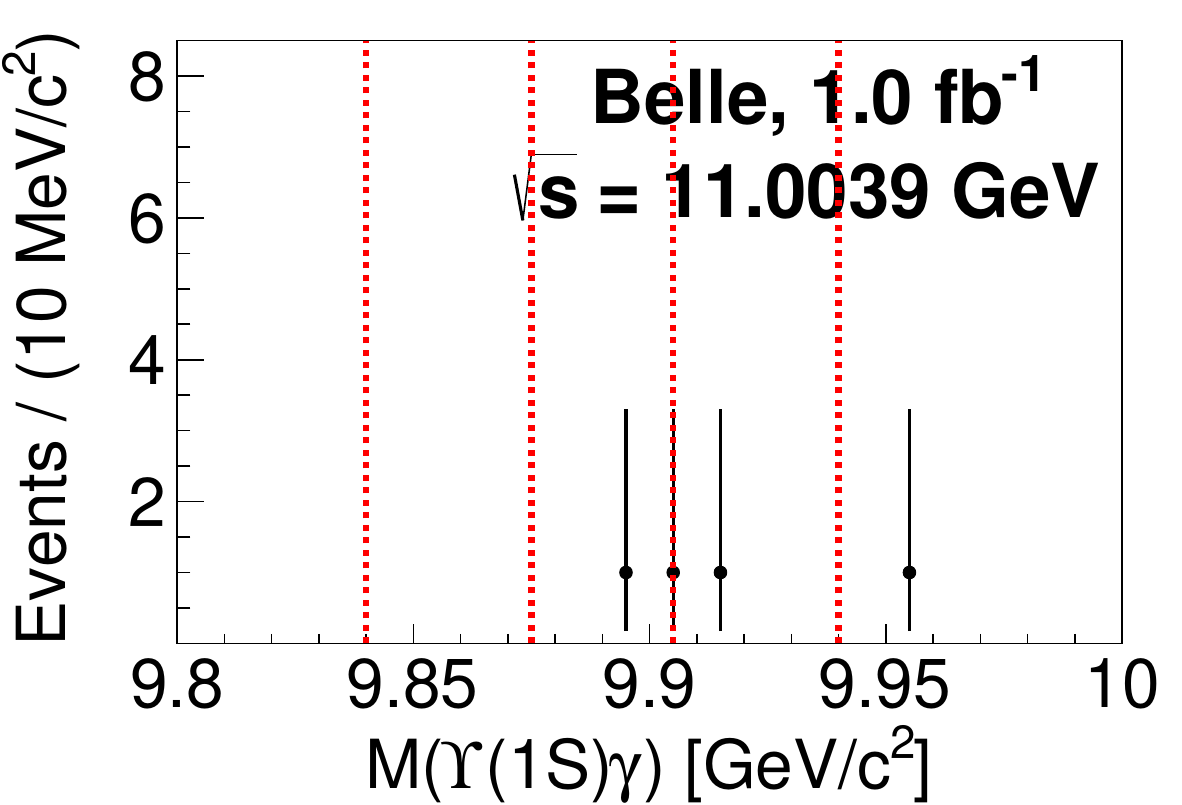}
\includegraphics[width=3.6cm]{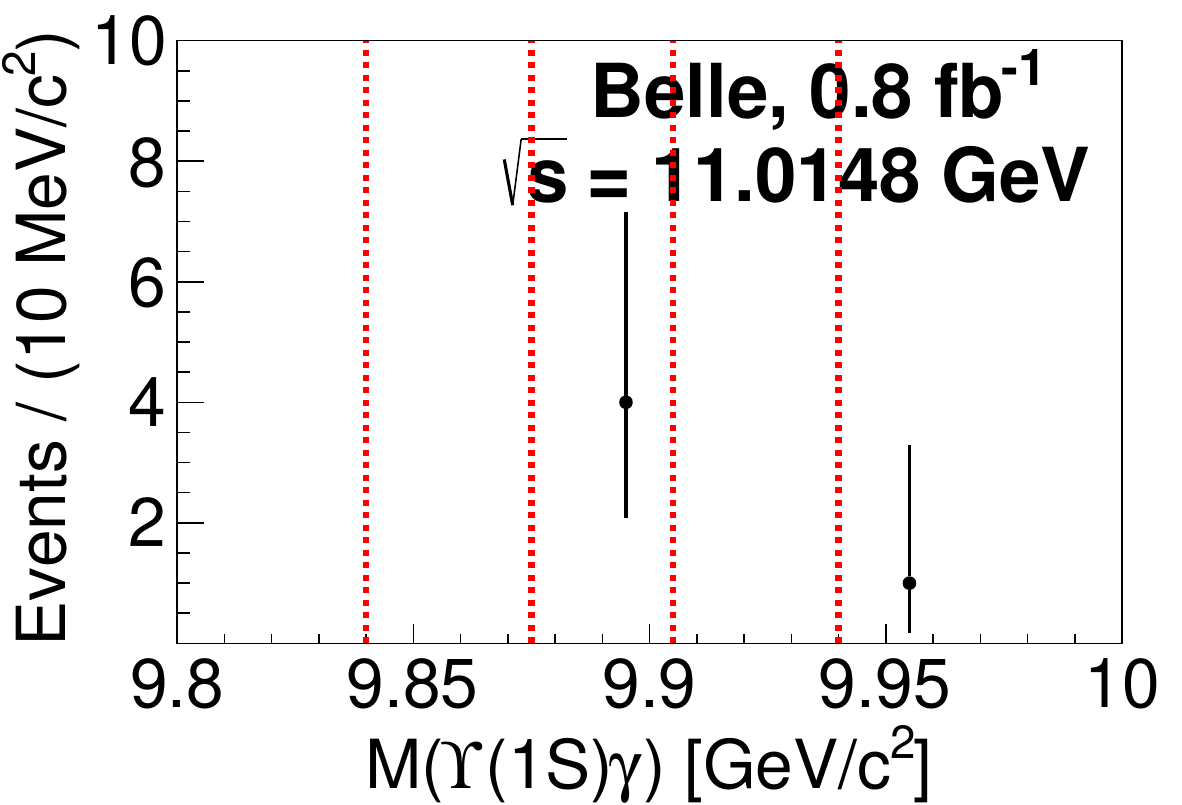}
\includegraphics[width=3.6cm]{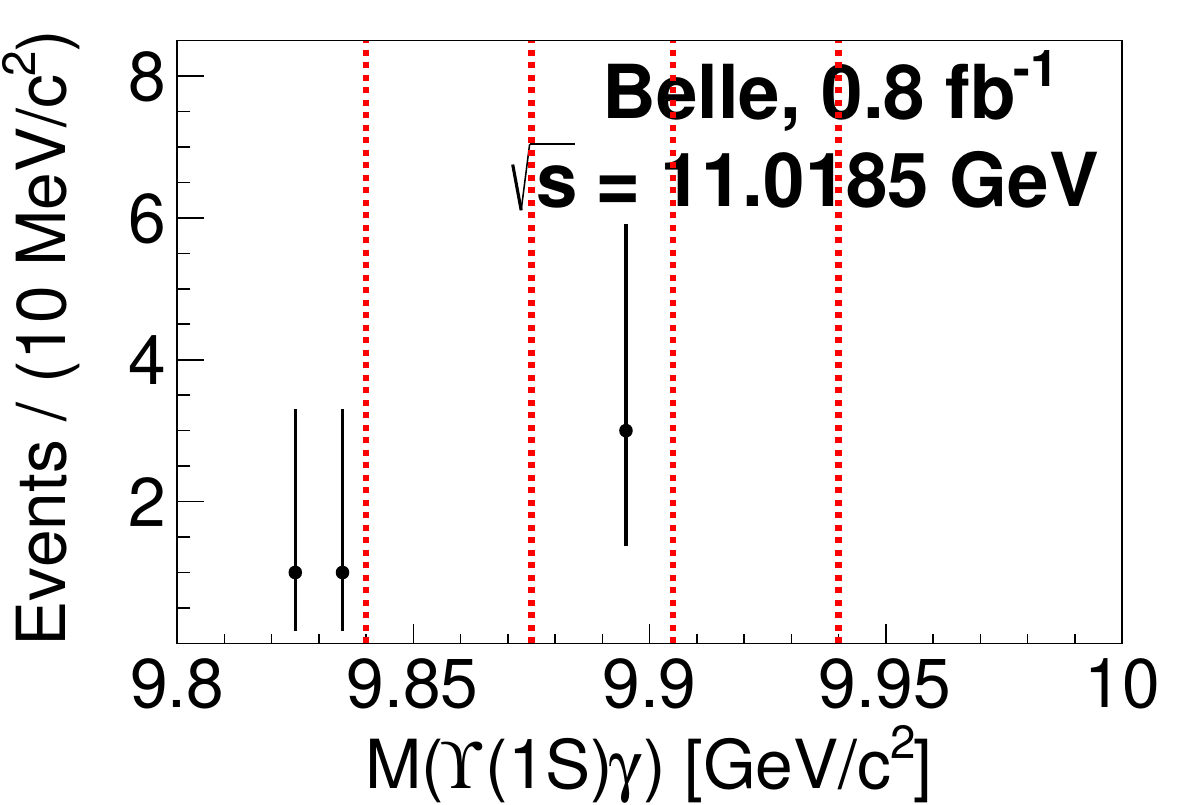}

\includegraphics[width=3.6cm]{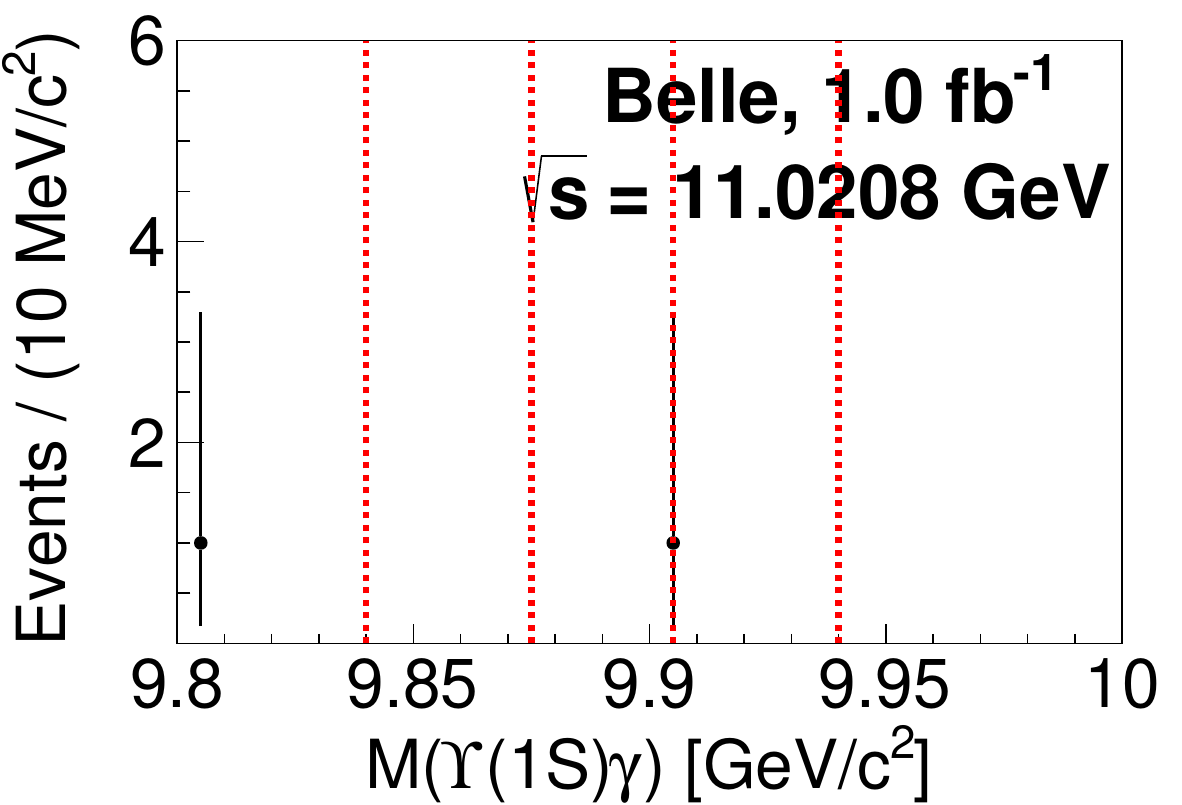}
\caption{$M(\Upsilon(1S)\gamma)$ distributions after requiring events outside the $\omega$ signal region in data at the lower-population energy points in the Belle and Belle II data samples. The vertical dashed lines (left to right) show the $\chi_{b0}$, $\chi_{b1}$, and $\chi_{b2}$ signal regions.
}\label{non-scan}
\end{figure}

\begin{figure}[htbp]
\centering
\includegraphics[width=7cm]{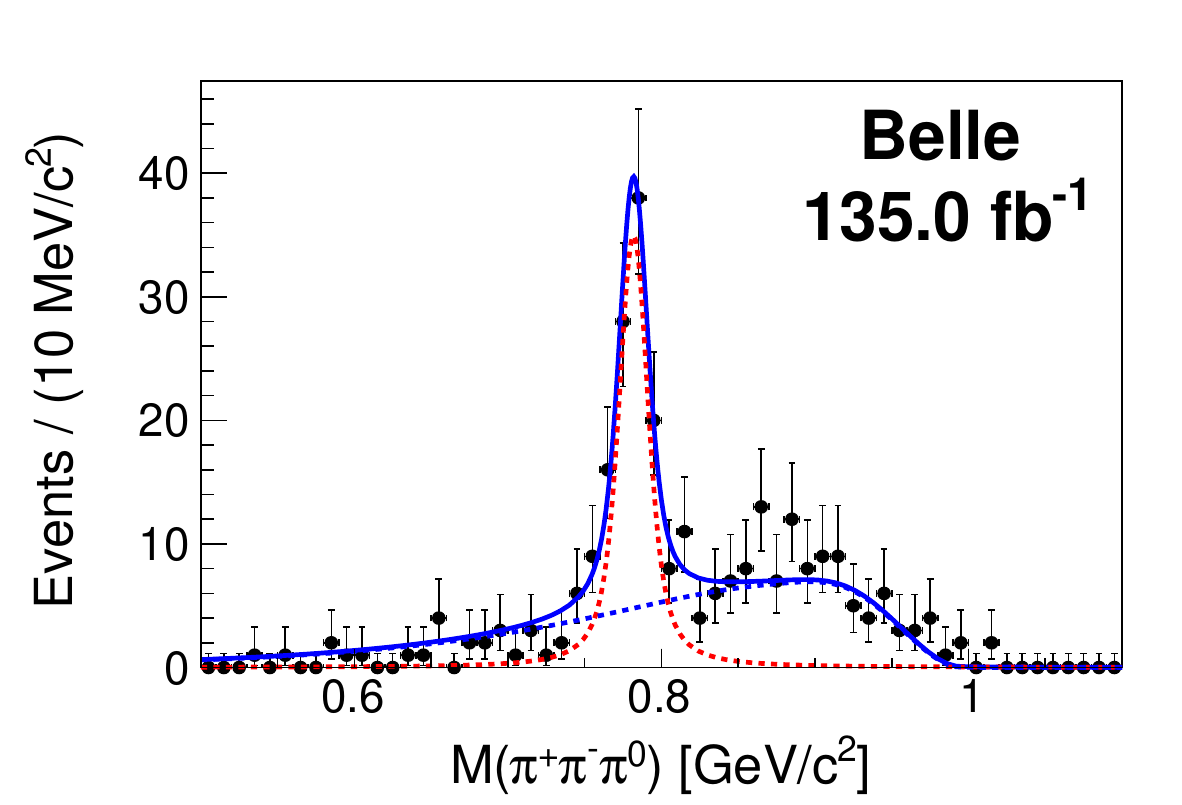}
\includegraphics[width=7cm]{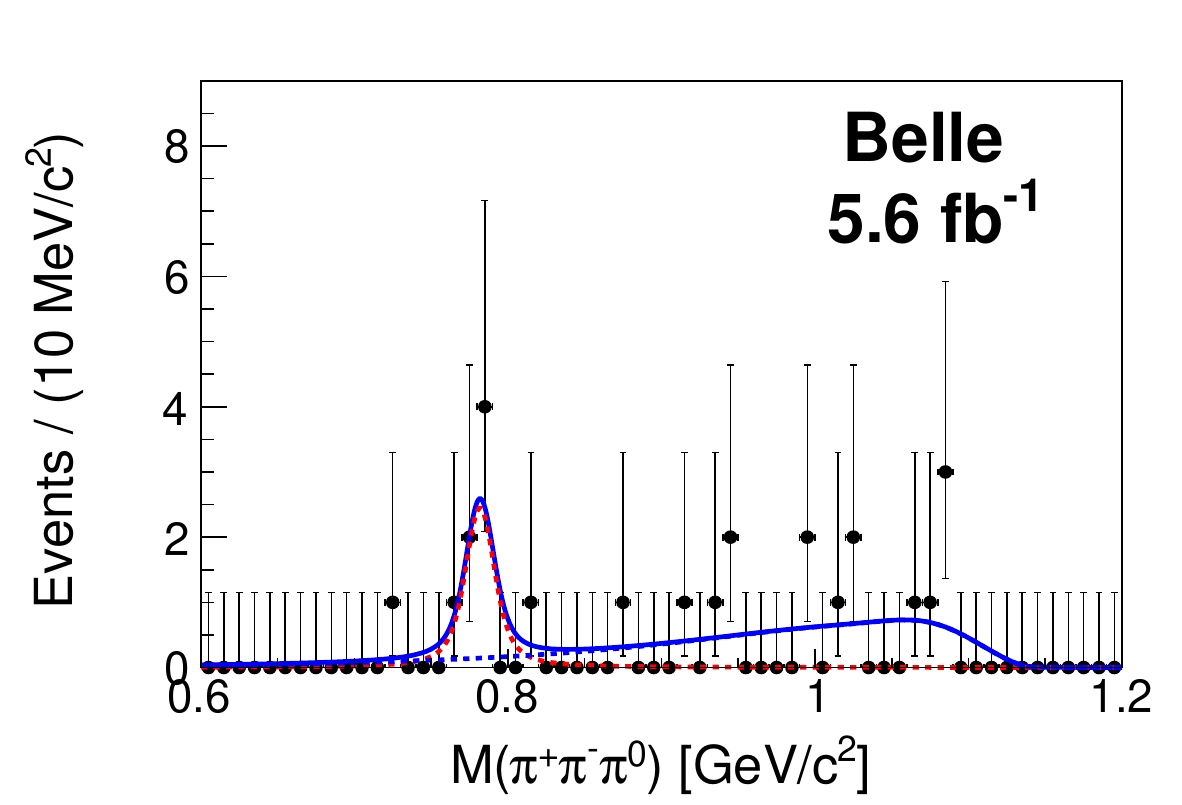}
\caption{A simultaneous fit to the $M(\pi^+\pi^-\pi^0)$ distributions in the well-populated $\Upsilon(5S)$ sample at $\sqrt{s}$ from 10829.5 MeV to 10957.5 MeV and the $\Upsilon(6S)$ sample obtained by combining the six highest-energy scan samples. 
The blue solid curves show the total fits.
The blue dashed curves show the $(\pi^+\pi^-\pi^0)_{\rm non-\omega}$ components.
The red dashed curves show the $\omega$ signal.
}\label{figadd2}
\end{figure}

For the other energy points, the $M(\Upsilon(1S)\gamma)$ distributions with $M(\pi^+\pi^-\pi^0)$ outside the $\omega$ signal region are shown in figure~\ref{non-scan}. 
For these sparse distributions, we use the same signal counting procedure as in the $\chi_{bJ}\,\omega$ case.
The signal yield and Born cross section, and upper limits on the Born cross section are determined 
using the method described for $e^+e^-\to\chi_{bJ}\,\omega$ above, treating components in the figure~\ref{figadd} fit other than ``$e^+e^-\to\chi_{bJ}\pi^+\pi^-\pi^0$'' as background. These values are listed in tables~\ref{tabsumnon2},~\ref{tabsumnon3}, and~\ref{tabsumnon1}. To determine
the efficiency of the $\omega$-veto requirement, we perform a simultaneous fit to the $M(\pi^+\pi^-\pi^0)$ distributions in the well-populated $\Upsilon(5S)$ sample and the $\Upsilon(6S)$ sample obtained by combining the six highest-energy scan samples, as shown in figure~\ref{figadd2}. The $(\pi^+\pi^-\pi^0)_{\rm non-\omega}$ components are described by products of a second-order polynomial that is common to the $\Upsilon(5S)$ and $\Upsilon(6S)$ samples and phase-space factors. This model describes the data well; therefore, we use it to determine the shapes of the $(\pi^+\pi^-\pi^0)_{\rm non-\omega}$ components at other energies. The inefficiency due to the $\omega$-veto requirement decreases with $\sqrt{s}$; it is 14\% at the $\Upsilon(5S)$ and 4\% at the $\Upsilon(6S)$.

As shown in table~\ref{tabsumnon3}, the $\chi_{b2}\,(\pi^+ \pi^- \pi^0)_{\rm non-\omega}$ fit
for the $\sqrt{s}$ = 10.745 GeV point returns a best-fit yield $N^{\rm sig}$ =
$0.0^{+1.1}_{-0.0}$. We find that the distribution of $-2\ln\mathcal{L}$ for
this fit is linear in $N^{\rm sig}$: accordingly, we fit the distribution
using Eq.~(\ref{eq:likelihood1}) and treat this as a poorly-populated point in the
energy-dependence fit of section~\ref{sec.8}, using a likelihood profile rather
than a $\chi^2$ term.

\begin{table*}[htbp]
\renewcommand\arraystretch{1.2}
\setlength{\tabcolsep}{2pt}
\small
\centering
\caption{Results for $e^+e^-\to\chi_{b1}\,(\pi^+\pi^-\pi^0)_{\rm non-\omega}$ as a function of c.m.\ energy at Belle and Belle II. 
Column descriptions are the same as in table~\ref{tabsumchib1}.}\label{tabsumnon2}
\vspace{0.2cm}
\begin{tabular}{ccccccccccc}
\hline\hline
&$\sqrt{s}$ (MeV) & $\cal L$ (fb$^{-1}$) & $\varepsilon$ & $1+\delta_{\rm ISR}$ & $N^{\rm sig}$ & $\Sigma(\sigma)$ & $\sigma_{\rm Born}$ (pb) & $\sigma^{\rm add}_{\rm syst}$ (pb) & $\sigma^{\rm UL}_{\rm Born}$ (pb) \\\hline
$\ast$	&	10653.0	&	3.550	&	0.117	&	0.644	&	$	0.0	^{+	0.5	}_{-	0.2	}$	& -&	$	0.0	^{+	0.1	}_{-	0.0	}	$	&	0.1			&	0.3	\\
$\ast$	&	10701.0	&	1.640	&	0.091	&	0.656	&	$	0.0	^{+	0.5	}_{-	0.2	}$	& -&	$	0.0	^{+	0.3	}_{-	0.1	}	$	&	0.1			&	0.9	\\
	&	10731.3	&	0.946	&	0.064	&	0.685	&	$	0.0	^{+	0.5	}_{-	0.2	}$	& -&	$	0.0	^{+	0.6	}_{-	0.3	}	$	&	0.1			&	2.7	\\
$\ast$	&	10745.0	&	9.870	&	0.126	&	0.690	&	\underline{$	2.1	^{+	3.1	}_{-	2.8	}$}	& 1.4&	$	0.1	^{+	0.2	}_{-	0.2	}	$	&	0.1	&	0.5	\\
	&	10771.2	&	0.955	&	0.084	&	0.692	&	$	0.0	^{+	0.5	}_{-	0.2	}$	& -&	$	0.0	^{+	0.5	}_{-	0.2	}	$	&	0.1			&	2.0	\\
$\ast$	&	10805.0	&	4.690	&	0.149	&	0.682	&	$	1.8	^{+	2.1	}_{-	1.5	}$	& -&	$	0.2	^{+	0.2	}_{-	0.2	}	$	&	0.1			&	0.7	\\
	&	10829.5	&	1.697	&	0.092	&	0.666	&	$	0.7	^{+	1.4	}_{-	0.7	}$	& -&	$	0.4	^{+	0.7	}_{-	0.4	}	$	&	0.1			&	2.0	\\
	&	10848.9	&	0.989	&	0.093	&	0.648	&	$	0.8	^{+	1.5	}_{-	0.7	}$	& -&	$	0.7	^{+	1.4	}_{-	0.6	}	$	&	0.1			&	3.7	\\
	&	10857.4	&	0.988	&	0.093	&	0.639	&	$	0.0	^{+	0.5	}_{-	0.2	}$	& -&	$	0.0	^{+	0.5	}_{-	0.2	}	$	&	0.1			&	1.9	\\
	&	10865.8	&	122.0	&	0.099	&	0.631	&	\underline{$	35.6	^{+	8.6\phantom{2}	}_{-	7.9	}$}	& 5.6&	$	0.3	^{+	0.1	}_{-	0.1	}	$	&	0.1			&	--	\\
	&	10877.8	&	0.978	&	0.095	&	0.631	&	$	0.0	^{+	0.5	}_{-	0.2	}$	& -&	$	0.0	^{+	0.5	}_{-	0.2	}	$	&	0.1			&	1.9	\\
	&	10882.8	&	1.848	&	0.095	&	0.642	&	$	3.6	^{+	2.4	}_{-	1.7	}$	& -&	$	1.7	^{+	1.2	}_{-	0.8	}	$	&	0.1			&	4.0	\\
	&	10888.9	&	0.990	&	0.095	&	0.670	&	$	1.8	^{+	1.8	}_{-	1.1	}$	& -&	$	1.5	^{+	1.6	}_{-	0.9	}	$	&	0.1			&	4.7	\\
	&	10898.3	&	2.408	&	0.095	&	0.743	&	$	0.5	^{+	1.4	}_{-	0.5	}$	& -&	$	0.2	^{+	0.4	}_{-	0.2	}	$	&	0.1			&	1.2	\\
	&	10907.3	&	0.980	&	0.097	&	0.826	&	$	0.8	^{+	1.5	}_{-	0.7	}$	& -&	$	0.6	^{+	1.0	}_{-	0.5	}	$	&	0.1			&	2.8	\\
	&	10928.7	&	1.149	&	0.098	&	0.964	&	$	0.8	^{+	1.5	}_{-	0.7	}$	& -&	$	0.4	^{+	0.7	}_{-	0.3	}	$	&	0.1			&	2.0	\\
	&	10957.5	&	0.969	&	0.101	&	0.852	&	$	0.8	^{+	1.5	}_{-	0.7	}$	& -&	$	0.5	^{+	1.0	}_{-	0.5	}	$	&	0.1			&	2.7	\\
	&	10975.3	&	0.999	&	0.103	&	0.707	&	$	0.0	^{+	0.5	}_{-	0.2	}$	& -&	$	0.0	^{+	0.4	}_{-	0.1	}	$	&	0.1			&	1.6	\\
	&	10990.4	&	0.985	&	0.105	&	0.627	&	$	1.8	^{+	1.8	}_{-	1.1	}$	& -&	$	1.5	^{+	1.5	}_{-	0.9	}	$	&	0.1			&	4.6	\\
	&	11003.9	&	0.976	&	0.106	&	0.673	&	$	0.8	^{+	1.4	}_{-	0.6	}$	& -&	$	0.6	^{+	1.1	}_{-	0.5	}	$	&	0.1			&	3.0	\\
	&	11014.8	&	0.771	&	0.106	&	0.843	&	$	3.8	^{+	2.4	}_{-	1.7	}$	& -&	$	3.0	^{+	1.9	}_{-	1.3	}	$	&	0.1			&	6.6	\\
	&	11018.5	&	0.859	&	0.106	&	0.912	&	$	2.8	^{+	1.9	}_{-	1.4	}$	& -&	$	1.8	^{+	1.2	}_{-	0.9	}	$	&	0.1			&	4.6	\\
	&	11020.8	&	0.982	&	0.107	&	0.956	&	$	0.8	^{+	1.5	}_{-	0.7	}$	& -&	$	0.4	^{+	0.8	}_{-	0.4	}	$	&	0.1			&	2.2	\\
\hline\hline
\end{tabular}
\end{table*}

\begin{table*}[htbp]
\renewcommand\arraystretch{1.2}
\setlength{\tabcolsep}{2pt}
\small
\centering
\caption{Results for $e^+e^-\to\chi_{b2}\,(\pi^+\pi^-\pi^0)_{\rm non-\omega}$ at each energy point at Belle and Belle II. 
Column descriptions are the same as in table~\ref{tabsumchib1}.
See the section~\ref{sec.7} text for the special treatment of the $\sqrt{s}$ = 10.745 GeV point.
}\label{tabsumnon3}
\vspace{0.2cm}
\begin{tabular}{ccccccccccc}
\hline\hline
&$\sqrt{s}$ (MeV) & $\cal L$ (fb$^{-1}$) & $\varepsilon$ & $1+\delta_{\rm ISR}$ & $N^{\rm sig}$ & $\Sigma(\sigma)$ & $\sigma_{\rm Born}$ (pb) & $\sigma^{\rm add}_{\rm syst}$ (pb) & $\sigma^{\rm UL}_{\rm Born}$ (pb) \\\hline
$\ast$	&	10653.0	&	3.550	&	0.153	&	0.597	&	$	0.0	^{+	0.5	}_{-	0.2	}$	&-&	$	0.0	^{+	0.2	}_{-	0.1	}	$	&	0.1			&	0.6	\\
$\ast$	&	10701.0	&	1.640	&	0.083	&	0.612	&	$	0.0	^{+	0.5	}_{-	0.2	}$	&-&	$	0.0	^{+	0.6	}_{-	0.3	}	$	&	0.1			&	2.2	\\
	&	10731.3	&	0.946	&	0.050	&	0.669	&	$	0.0	^{+	0.5	}_{-	0.2	}$	&-&	$	0.0	^{+	1.7	}_{-	0.7	}	$	&	0.1			&	7.0	\\
$\ast$	&	10745.0	&	9.870	&	0.112	&	0.679	&	\underline{$	0.0	^{+	1.1	}_{-	0.0	}$}	&-&	$	0.0	^{+	0.2	}_{-	0.0	}	$	&	0.1			&	0.6	\\
	&	10771.2	&	0.955	&	0.077	&	0.686	&	$	0.0	^{+	0.5	}_{-	0.2	}$	&-&	$	0.0	^{+	1.0	}_{-	0.4	}	$	&	0.1			&	4.4	\\
$\ast$	&	10805.0	&	4.690	&	0.143	&	0.679	&	$	0.0	^{+	1.3	}_{-	0.0	}$	&-&	$	0.0	^{+	0.3	}_{-	0.0	}	$	&	0.1			&	0.7	\\
	&	10829.5	&	1.697	&	0.088	&	0.665	&	$	0.0	^{+	0.5	}_{-	0.2	}$	&-&	$	0.0	^{+	0.5	}_{-	0.2	}	$	&	0.1			&	2.2	\\
	&	10848.9	&	0.989	&	0.088	&	0.648	&	$	0.0	^{+	0.5	}_{-	0.2	}$	&-&	$	0.0	^{+	0.9	}_{-	0.4	}	$	&	0.1			&	3.9	\\
	&	10857.4	&	0.988	&	0.089	&	0.639	&	$	1.8	^{+	1.8	}_{-	1.1	}$	&-&	$	3.4	^{+	3.4	}_{-	2.1	}	$	&	0.1			&	10.3\phantom{1}	\\
	&	10865.8	&	122.0	&	0.099	&	0.631	&	\underline{$	22.6	^{+	7.5\phantom{2}	}_{-	6.8	}$}	&3.9&	$	0.3	^{+	0.1	}_{-	0.1	}	$	&	0.1			&	--	\\
	&	10877.8	&	0.978	&	0.090	&	0.631	&	$	0.8	^{+	1.5	}_{-	0.7	}$	&-&	$	1.5	^{+	2.8	}_{-	1.3	}	$	&	0.1			&	7.7	\\
	&	10882.8	&	1.848	&	0.090	&	0.641	&	$	0.6	^{+	1.4	}_{-	0.6	}$	&-&	$	0.5	^{+	1.4	}_{-	0.6	}	$	&	0.1			&	4.0	\\
	&	10888.9	&	0.990	&	0.091	&	0.670	&	$	0.8	^{+	1.5	}_{-	0.7	}$	&-&	$	1.4	^{+	2.6	}_{-	1.2	}	$	&	0.1			&	7.1	\\
	&	10898.3	&	2.408	&	0.091	&	0.742	&	$	0.0	^{+	0.5	}_{-	0.2	}$	&-&	$	0.0	^{+	0.3	}_{-	0.1	}	$	&	0.1			&	1.2	\\
	&	10907.3	&	0.980	&	0.092	&	0.825	&	$	0.8	^{+	1.5	}_{-	0.7	}$	&-&	$	1.1	^{+	2.1	}_{-	1.0	}	$	&	0.1			&	5.8	\\
	&	10928.7	&	1.149	&	0.094	&	0.964	&	$	0.0	^{+	0.5	}_{-	0.2	}$	&-&	$	0.0	^{+	0.5	}_{-	0.2	}	$	&	0.1			&	2.1	\\
	&	10957.5	&	0.969	&	0.096	&	0.852	&	$	0.0	^{+	0.5	}_{-	0.2	}$	&-&	$	0.0	^{+	0.7	}_{-	0.3	}	$	&	0.1			&	2.8	\\
	&	10975.3	&	0.999	&	0.099	&	0.707	&	$	2.8	^{+	1.9	}_{-	1.4	}$	&-&	$	4.2	^{+	2.9	}_{-	2.1	}	$	&	0.1			&	11.7\phantom{1}	\\
	&	10990.4	&	0.985	&	0.100	&	0.627	&	$	0.0	^{+	0.5	}_{-	0.2	}$	&-&	$	0.0	^{+	0.9	}_{-	0.3	}	$	&	0.1			&	3.6	\\
	&	11003.9	&	0.976	&	0.101	&	0.673	&	$	1.8	^{+	1.8	}_{-	1.1	}$	&-&	$	2.9	^{+	2.9	}_{-	1.8	}	$	&	0.1			&	8.8	\\
	&	11014.8	&	0.771	&	0.101	&	0.843	&	$	0.0	^{+	0.5	}_{-	0.2	}$	&-&	$	0.0	^{+	0.8	}_{-	0.3	}	$	&	0.1			&	3.4	\\
	&	11018.5	&	0.859	&	0.102	&	0.912	&	$	0.0	^{+	0.5	}_{-	0.2	}$	&-&	$	0.0	^{+	0.7	}_{-	0.3	}	$	&	0.1			&	2.8	\\
	&	11020.8	&	0.982	&	0.103	&	0.956	&	$	0.0	^{+	0.5	}_{-	0.2	}$	&-&	$	0.0	^{+	0.5	}_{-	0.2	}	$	&	0.1			&	2.3	\\
\hline\hline
\end{tabular}
\end{table*}

\section{Systematic uncertainty}~\label{sec.7}

The systematic uncertainties in the measurements of Born cross sections for $e^+e^-\to\chi_{bJ}\,\omega$ and $e^+e^-\to\chi_{bJ}\,(\pi^+\pi^-\pi^0)_{\rm non-\omega}$ 
include contributions from the photon energy calibration, fit model, reconstruction efficiency, radiative correction factor,
angular distributions, beam-energy calibration, trigger simulation, integrated luminosity, and branching fractions of intermediate states. The additive systematic uncertainties are those from the photon energy calibration and fit model.
The other sources of systematic uncertainties are multiplicative.

We study the additive systematic uncertainties from the photon energy calibration and fitting procedure as follows.
A photon energy calibration has been performed with three processes, $D^{*0} \to D^0\gamma$, $\pi^0 \to \gamma\gamma$, and $\eta \to \gamma\gamma$~\cite{232002}.
The resulting uncertainties of the peak positions are 1.0, 1.1, and 1.2 MeV/$c^2$ for $\chi_{b0}$, $\chi_{b1}$, and $\chi_{b2}$ decays, respectively; the uncertainty in the width is 0.09 MeV/$c^2$ for all states.
For the results obtained with the fitting method, we change the $M(\Upsilon(1S)\gamma)$ peak positions and resolutions by $\pm1\sigma$.
We increase the order of the polynomial describing the background by one, change the fit interval, and exclude the $\chi_{b0}$ component from the fit. 

In each of the fits shown in figures~\ref{fig4} and~\ref{non-fit}, we perform all possible combinations of the above variations and conservatively consider the largest deviation of fit results as the additive systematic uncertainty for the corresponding energy point. The same variations are performed in the 
fit shown in figure~\ref{figadd} that sets the background estimates $N^{\rm bg}$ for the poorly-populated energy points. At each such point, the largest resulting deviation in $N^{\rm bg}$ is taken as the additive systematic uncertainty. These uncertainties are listed in the $\sigma_{\text{syst}}^{\text{add}}$ columns of tables~\ref{tabsumchib1}--\ref{tabsumnon3},~\ref{tabsumchib0}, and~\ref{tabsumnon1}, and are assumed to be uncorrelated for different energy points.
The effects of the photon calibration and the changes to the background shape dominate these uncertainties.

The contributions to the multiplicative systematic uncertainty are given in Table~\ref{sys1}. Detection efficiency uncertainties for Belle include momentum-dependent tracking uncertainties (1\% per pion and 0.35\% per lepton, as derived from $D^{*+}\to D^0(\to K_S^0\pi^+\pi^-)\pi^+$), pion identiﬁcation (0.9\% per pion, as derived from $D^{*+}\to D^0(\to K^-\pi^+)\pi^{+}$), lepton identification (1.6\% per electron and 1.2\% per muon, as derived from $\gamma\gamma \to \ell^+\ell^-$ ($\ell$ = e, $\mu$)), photon reconstruction (2.0\% per photon, as derived from $e^+e^- \to \gamma e^+e^-$), and $\pi^0$ reconstruction (2.3\% per $\pi^0$, as derived from $\tau^- \to \pi^-\pi^0\nu_\tau$). 
Detection efficiency uncertainties for Belle II include momentum-dependent tracking uncertainties (1.3\% per pion and 0.3\% per lepton, as derived from $\bar B^0\to D^{*+}(\to D^0\pi^+)\pi^-$ and $e^+e^-\to \tau^+\tau^-$), pion identification (1.1\% per pion, as derived from $D^{*+} \to D^0(\to K^-\pi^+)\pi^+$), lepton identification (0.4\% per electron and 0.7\% per muon, as derived from $J/\psi$ decays, Bhabha, dimuon, and two-photon processes), photon reconstruction (3.5\% per photon, as derived from $e^+e^-\to \gamma\mu^+\mu^-$), and $\pi^0$ reconstruction (4.8\% per $\pi^0$, as derived from $\eta\to\pi^0\pi^0\pi^0$).
The total uncertainty in the reconstruction efficiency is 5.5\% for Belle and 7.2\% for Belle II; the higher value in Belle II is due to the $\gamma$ and $\pi^0$ contributions.

The distribution in the $\pi^+\pi^-\pi^0$ polar angle measured in the $e^+e^-$ rest frame, $\theta_{3\pi}$, is uniform in the nominal simulation. When the distribution is changed to $1\pm {\rm cos}^2\theta_{3\pi}$, the maximal deviations in the efficiency are 2.7\% and 2.4\% for $e^+e^-\to \chi_{bJ}\,\omega$ and $e^+e^-\to\chi_{bJ}\,(\pi^+\pi^-\pi^0)_{\rm non-\omega}$, respectively; these deviations are considered as a systematic uncertainty.
We generated MC samples of $e^+e^-\to Z_b^+\pi^-$ at $\sqrt{s}$ = 10.866 GeV, with $Z_b^+ \to \chi_{bJ} \rho^+$ and $\rho^+ \to
\pi^+ \pi^0$. The efficiency for this sample differs from that of $e^+ e^-
\to \chi_{bJ}\,(\pi^+ \pi^- \pi^0)_{\rm non-\omega}$ at 10.866 GeV by only
0.9\%. We neglect the resulting uncertainty.
A 1.4\% systematic uncertainty is assigned due to the trigger simulation. 
The uncertainties in the center of mass energies are about 1 MeV.
We change the collision energies in the 6C kinematic fit by $\pm1\sigma$ and find the deviations in the signal yield are less than 1.0\%, which is negligible.
In calculating the radiative correction factor, the measured energy dependence of the Born cross sections is used.~We change all of the parameters in eq.~(\ref{eq:BW}) below by $\pm1\sigma$ according to the fitted results from the distributions of $\sigma(e^+e^- \to \chi_{b1,b2}\,\omega)$ and $\sigma(e^+e^-\to\chi_{bJ}\,(\pi^+\pi^-\pi^0)_{\rm non-\omega})$ as a function of $e^+e^-$ c.m.\ energy, and take the maximum difference in the radiative correction factor across all energy points, 5.1\%, as the resulting uncertainty. This value is dominated by the uncertainties on the widths of the $\Upsilon(10753)$ and $\Upsilon(11020)$; these uncertainties are uncorrelated.
Belle measures luminosity at 1.4\% precision using wide angle Bhabha events.
Belle II measures luminosity at 0.6\% precision using Bhabha and digamma events~\cite{013001}.
The branching fractions of all intermediate decays are taken from ref.~\cite{PDG}.

For the Born cross section measurements reported in sections~\ref{sec.5} and~\ref{sec.6}, we add all
the multiplicative systematic uncertainties in quadrature to obtain the
final multiplicative systematic uncertainty, listed in table~\ref{sys1}. 
For the energy dependence fits in section 8, multiplicative systematic uncertainties are treated as follows. 
Uncertainties due to the radiative corrections at different energies
are assumed to be uncorrelated; uncertainties due to the branching
fractions are correlated for all energies; the remaining uncertainties
are assumed to be correlated separately for Belle and Belle II
points. To obtain the total uncorrelated uncertainty, we combine the
statistical uncertainty, additive systematic uncertainty, and
uncertainty due to the radiative corrections. For the results obtained
with the fitting method, we add the contributions in quadrature (to
include the multiplicative uncertainty, we use eq.~(3) from
ref.~\cite{072013}). For the results obtained with the
counting method, the contributions are combined using the POLE program;
the profile likelihoods shown in the Supplemental Material contain contributions
of all the uncorrelated uncertainties. The correlated uncertainties for 
Belle and Belle~II points are taken into account in the energy-dependence fit
function as described in the next section.

\begin{table}
\centering
\caption{The multiplicative systematic uncertainties (\%) in the measurements of Born cross sections for $e^+e^-\to\chi_{bJ}\,(\pi^+\pi^-\pi^0)_{\rm non-\omega}$ and $e^+e^-\to\chi_{bJ}\,\omega$ at Belle and Belle II. 
When using individual $\sigma_{\text{Born}}$
values from tables~\ref{tabsumchib1}--\ref{tabsumnon3} and~\ref{tabsumchib0}--\ref{tabsumnon1}, the total multiplicative uncertainty
should be taken into account; it is already included in the upper limits
$\sigma_{\text{Born}}^{\text{UL}}$. In the energy dependence fits,
correlations between uncertainties are taken into account: see the text.}
\vspace{0.2cm}
\label{sys1}
\footnotesize
\begin{tabular}{lcccccc}
\hline\hline
\multirow{2}{*}{Source}  & \multicolumn{2}{c}{$e^+e^-\to\chi_{b0,b1,b2}\,(\pi^+\pi^-\pi^0)_{\rm non-\omega}$} &  \multicolumn{2}{c}{$e^+e^-\to\chi_{b0,b1,b2}\omega$} \\
& Belle & Belle~II& Belle & Belle~II \\\hline
Efficiency 	               & 5.5 & 7.2 & 5.5 & 7.2 \\
Angular distributions & 2.4 & 1.0 & 2.7 & 1.0 \\
Trigger & 1.4 & 1.0 & 1.4 & 1.0 \\
Radiative correction factor & 5.1 & 5.1 & 5.1 & 5.1 \\
Luminosity & 1.4 & 1.0 & 1.4 & 1.0 \\
Branching fractions &  14.6, 7.3, 7.2 & 14.6, 7.3, 7.2 & 14.7, 7.4, 7.3 & 14.7, 7.4, 7.3 \\
\hline
Total &16.7, 10.9, 10.9 &17.1, 11.6, 11.5 &16.8, 11.0, 10.9 &17.1, 11.6, 11.5 \\
\hline\hline 
\end{tabular}
\end{table}

\section{Energy dependence of Born cross sections}~\label{sec.8}

\begin{figure}[htbp]
\centering
\includegraphics[width=15cm]{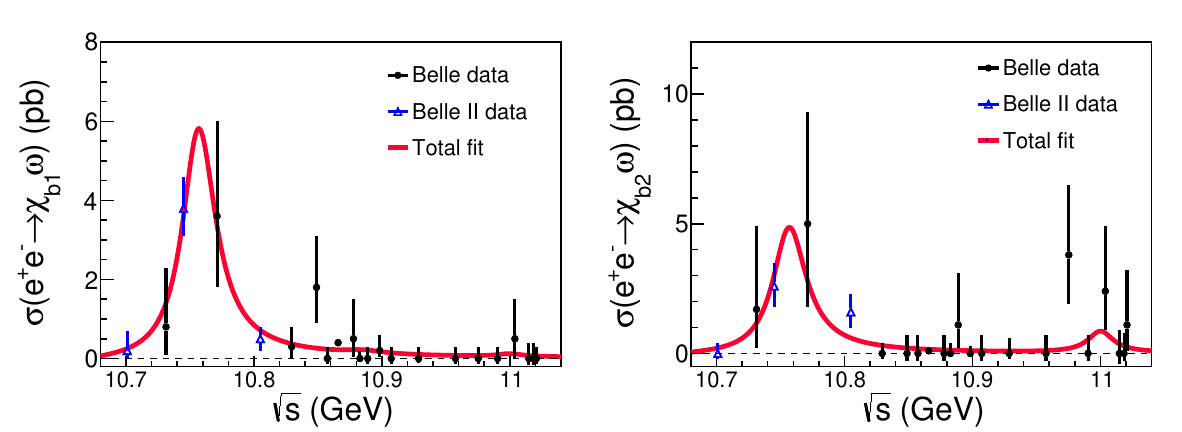}
\put(-380, 125){\large \bf (a)}
\put(-165, 125){\large \bf (b)}

\includegraphics[width=15cm]{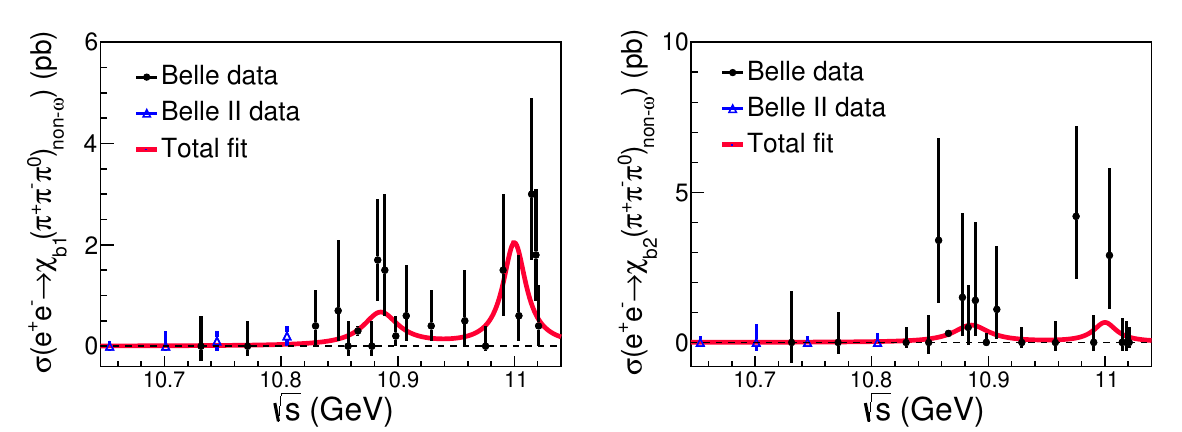}
\put(-250, 125){\large \bf (c)}
\put(-35, 125){\large \bf (d)}
\caption{Energy dependences of the Born cross sections for (a) $e^+e^-\to \chi_{b1}\,\omega$, (b) $e^+e^-\to \chi_{b2}\,\omega$, (c) $e^+e^-\to \chi_{b1}\,(\pi^+\pi^-\pi^0)_{\rm non-\omega}$, and (d) $e^+e^-\to \chi_{b2}\,(\pi^+\pi^-\pi^0)_{\rm non-\omega}$. Filled circles show the measurements at Belle, and triangles show the measurements at Belle II. 
Error bars represent the total uncorrelated uncertainties.
Curves show the fit results.}\label{dependency1}
\end{figure}

The Born cross sections for $e^+e^-\to\chi_{bJ}\,\omega$ and $e^+e^-\to\chi_{bJ}\,\,(\pi^+\pi^-\pi^0)_{\rm non-\omega}$ as a function of 
c.m.\ energy are shown in figure~\ref{dependency1}. 
For visualization, the low-population points are adjusted for 
the effects of migration using the inverse of the matrix shown in equation~(\ref{eq.6.1}).
In the $\chi_{bJ}\,\omega$ channel, a pronounced $\Upsilon(10753)$ signal is visible, but there are no clear $\Upsilon(10860)$ or $\Upsilon(11020)$ signals. In contrast, in the $\chi_{bJ}\,(\pi^+\pi^-\pi^0)_{\rm non-\omega}$ channel, 
$\Upsilon(10860)$ and $\Upsilon(11020)$ peaks are seen (with low significance), but there is no $\Upsilon(10753)$ signal.

We fit the cross sections as a function of energy by minimizing the sum of the $\chi^2$ values for the well-populated points and the $\Delta(-2\ln\mathcal{L})$ values (obtained using equations~\ref{eq:likelihood1} or~\ref{eq:likelihood2}) for the poorly-populated ones.
The fit function is a coherent sum of the $\Upsilon(10753)$, $\Upsilon(10860)$, and $\Upsilon(11020)$ Breit-Wigner amplitudes:

\vspace{-0.2cm}
\begin{equation} \label{eq:BW}
{\Bigl |}\sum _{i=1}^3\frac{\sqrt{12\pi\Gamma^{(0)}_{ee,i}\BR_{ik}\Gamma_i}}{s-M_i^2-iM_i\Gamma_i}\sqrt{ \frac{\Phi(\sqrt{s})}{\Phi(M_i)}}e^{i\phi_{ik}}{\Bigr |}^2,
\end{equation}where the
index $i$ runs over the $\Upsilon(10753)$, $\Upsilon(10860)$, and $\Upsilon(11020)$ states; the index $k$ runs over the channels $\chi_{b1}\,\omega$, $\chi_{b2}\,\omega$, $\chi_{b1}\,(\pi^+\pi^-\pi^0)_{\rm non-\omega}$, and $\chi_{b2}\,(\pi^+\pi^-\pi^0)_{\rm non-\omega}$; $M_i$, $\Gamma_i$, and $\Gamma^{(0)}_{ee,i}$ are the mass, total width, and Born level electron width of the $\Upsilon$ states; $\Phi$ is the phase-space factor; and $\phi_{ik}$ are complex phases.
The physical electron width $\Gamma_{ee,i}$ is obtained as $\Gamma_{ee,i}=\Gamma^{(0)}_{ee,i}\,|1-\Pi|^2$.
We use two-body phase space factors for both $\chi_{bJ}\,\omega$ and
$\chi_{bJ}\,(\pi^+\pi^-\pi^0)_{\rm non-\omega}$. For $\chi_{bJ}\,(\pi^+\pi^-\pi^0)_{\rm non-\omega}$, at each energy, the invariant
mass of the $\pi^+ \pi^-\pi^0$ system is set to $(\sqrt{s} - 10.866 + 0.9)$ GeV/$c^2$. The motivation is that in the well-populated sample at $\sqrt{s}$ = 10.866 GeV,  we see that $M(\pi^+ \pi^-\pi^0)$ peaks
at 0.9 GeV/$c^2$, as shown in figure~\ref{figadd2} (left).
At each energy, we considered several lower $M(\pi^+\pi^-\pi^0)$ values down to the $\pi^+\pi^-\pi^0$ mass threshold. There is no change in the results. The phases of the $\Upsilon(10753)$ and $\Upsilon(10860)$ amplitudes are set to zero for the $\chi_{bJ}\,\omega$ and $\chi_{bJ}\,(\pi^+\pi^-\pi^0)_{\rm non-\omega}$ channels, respectively.
The fit functions for the poorly-populated samples are corrected for migration between the $\chi_{b1}$ and $\chi_{b2}$ signal regions using eq.~(\ref{eq.6.1}); the $\chi_{b1}$ and $\chi_{b2}$ channels are then fitted simultaneously.
The mass and width of $\Upsilon(10753)$ are free parameters for the $\chi_{b1,b2}\,\omega$ channels and are fixed to the results of ref.~\cite{2401.12021} for the $\chi_{b1,b2}\,(\pi^+\pi^-\pi^0)_{\rm non-\omega}$ channels. The masses and widths of the $\Upsilon(10860)$ and $\Upsilon(11020)$ are fixed to the world-average values~\cite{PDG}.
In $e^+e^-\to\chi_{b1,b2}\,(\pi^+\pi^-\pi^0)_{\rm non-\omega}$, the contribution from the $\Upsilon(10753)$ decay is consistent with zero.
The fitted results are shown in figure~\ref{dependency1} and summarized in table~\ref{dependency1tab}.
For the $\chi_{b1,b2}\,(\pi^+\pi^-\pi^0)_{\rm non-\omega}$ channels, we find two solutions that correspond to constructive and destructive interference between the $\Upsilon(10860)$ and $\Upsilon(11020)$ amplitudes.

Systematic uncertainties in the cross section energy dependence fit results include the beam-energy calibration, the fit model, and the correlated systematic uncertainties of cross-section measurements.
\begin{itemize}
\item We change the c.m.\ energy by $\pm$1 MeV for each energy point independently, and take the differences in the mass, width, and $\Gamma_{ee}\BR_f$ as the uncertainties due to the beam-energy calibration.
\item 

We change the masses and widths of the $\Upsilon(10860)$ and $\Upsilon(11020)$ by $\pm1\sigma$~\cite{PDG}. We include an additional 
non-resonant component which has the shape of phase space; we find its contribution is consistent with zero and the change in the fit results is negligible. For the $\chi_{b1,b2}\,\omega$ channels, we exclude the $\Upsilon(10860)$ and $\Upsilon(11020)$ components from the fit. 

\item
We re-scale the fit function Eq.~(\ref{eq:BW}) for the Belle points to take
the Belle-specific correlated multiplicative uncertainties into account,
and repeat the fits; we do the same for the Belle II points; and
finally, we re-scale the fit function for all points to take the
uncertainties due to branching fractions into account.

\end{itemize}
We estimate the systematic uncertainty of a given source as the maximum deviation of the fit result. 
The uncertainties due to the masses and widths of the $\Upsilon(10860)$ and $\Upsilon(11020)$ are dominant.
The total systematic uncertainty is obtained by adding the contributions of the various sources in quadrature.

The measured mass and width of $\Upsilon(10753)\to \chi_{b1,b2}\,\omega$ are consistent with those in refs.~\cite{220,2401.12021}.
The products $\Gamma_{ee}\BR(\Upsilon(10753)\to\chi_{b1}\,\omega)$ and $\Gamma_{ee}\BR(\Upsilon(10753)\to\chi_{b2}\,\omega)$ are consistent with the results reported in ref.~\cite{091902} and supersede them. 
The statistical significances of the decays $\Upsilon(10753)\to\chi_{b1}\,\omega$ and $\Upsilon(10753)\to\chi_{b2}\,\omega$ are 6.0$\sigma$ and 4.1$\sigma$, respectively.
We measure $\BR(\Upsilon(10753)\to\chi_{b1}\,\omega)/\BR(\Upsilon(10753)\to\chi_{b2}\,\omega)$ = $1.13\pm0.38\pm0.34$.

The statistical significances of the decays $\Upsilon(10860)\to\chi_{bJ}\,(\pi^+\pi^-\pi^0)_{\rm non-\omega}$ and $\Upsilon(11020)$ $\to\chi_{bJ}\,(\pi^+\pi^-\pi^0)_{\rm non-\omega}$, where the $J=1$ and $J=2$ channels are combined, are 4.5$\sigma$ and 2.5$\sigma$, respectively. 
For the combined $J=1$ and $2$ channels, we find $\Gamma_{ee}\BR(\Upsilon(10860)\to\chi_{bJ}\,(\pi^+\pi^-\pi^0)_{\rm non-\omega})$ = $(0.46\pm0.10\pm0.07)$ eV / $(0.33\pm0.07\pm0.07)$ eV and $\Gamma_{ee}\BR(\Upsilon(11020)\to\chi_{bJ}\,(\pi^+\pi^-\pi^0)_{\rm non-\omega})$ = $(0.67\pm0.24\pm0.16)$ eV / $(0.51\pm0.21\pm0.14)$ eV from destructive / constructive solutions.

\begin{table*}[htbp]
\tabcolsep=0.22cm
\centering
\caption{The fitted mass and width of $\Upsilon(10753)$, and $\Gamma_{ee}\BR_f$ for $\Upsilon(10753)$, $\Upsilon(10860)$, and $\Upsilon(11020)$ decaying into $\chi_{b1,b2}\,\omega$ and $\chi_{b1,b2}\,(\pi^+\pi^-\pi^0)_{\rm non-\omega}$. The upper limits on $\Gamma_{ee}\BR(\Upsilon(10753,10860)\to\chi_{b1,b2}\,\omega)$ and $\Gamma_{ee}\BR(\Upsilon(10753)\to\chi_{b1,b2}\,(\pi^+\pi^-\pi^0)_{\rm non-\omega})$ at 90\% C.L. are shown in parentheses.
For $\Gamma_{ee}\BR(\Upsilon(10860,11020)\to\chi_{b1,b2}\,(\pi^+\pi^-\pi^0)_{\rm non-\omega})$, the destructive / constructive solutions are shown.
}\label{dependency1tab}
\vspace{0.2cm}
\begin{tabular}{cc}
\hline\hline
$M(\Upsilon(10753))$ & $(10756.1\pm3.4\pm2.7)$ MeV/$c^2$  \\
$\Gamma(\Upsilon(10753))$ & ($32.2\pm11.3\pm 14.9$) MeV  \\
$\Gamma_{ee}\BR(\Upsilon(10753)\to\chi_{b1}\,\omega)$ & ($1.57\pm0.27\pm 0.22$) eV \\
$\Gamma_{ee}\BR(\Upsilon(10753)\to\chi_{b2}\,\omega)$ & ($1.39\pm0.41\pm 0.33$) eV \\
$\Gamma_{ee}\BR(\Upsilon(10860)\to\chi_{b1}\,\omega)$ & ($0.02\pm0.04\pm 0.04$) eV ($<0.10$ eV)\\
$\Gamma_{ee}\BR(\Upsilon(10860)\to\chi_{b2}\,\omega)$ & ($0.00\pm0.04\pm 0.02$) eV ($<0.07$ eV)\\
$\Gamma_{ee}\BR(\Upsilon(11020)\to\chi_{b1}\,\omega)$ & ($0.01\pm0.02\pm 0.03$) eV ($<0.07$ eV)\\
$\Gamma_{ee}\BR(\Upsilon(11020)\to\chi_{b2}\,\omega)$ & ($0.18\pm0.17\pm 0.05$) eV ($<0.46$ eV)\\
\hline
$\Gamma_{ee}\BR(\Upsilon(10753)\to\chi_{b1}\,(\pi^+\pi^-\pi^0)_{\rm non-\omega})$ & ($0.00\pm0.05\pm 0.02$) eV ($<0.08$ eV) \\
$\Gamma_{ee}\BR(\Upsilon(10753)\to\chi_{b2}\,(\pi^+\pi^-\pi^0)_{\rm non-\omega})$ & ($0.00\pm0.03\pm 0.02$) eV ($<0.07$ eV) \\
$\Gamma_{ee}\BR(\Upsilon(10860)\to\chi_{b1}\,(\pi^+\pi^-\pi^0)_{\rm non-\omega})$ & ($0.28\pm0.09\pm 0.06$) eV / ($0.16\pm0.05\pm 0.06$) eV \\
$\Gamma_{ee}\BR(\Upsilon(10860)\to\chi_{b2}\,(\pi^+\pi^-\pi^0)_{\rm non-\omega})$ & ($0.18\pm0.05\pm 0.04$) eV / ($0.17\pm0.05\pm 0.04$) eV \\
$\Gamma_{ee}\BR(\Upsilon(11020)\to\chi_{b1}\,(\pi^+\pi^-\pi^0)_{\rm non-\omega})$ & ($0.52\pm0.20\pm 0.12$) eV / ($0.37\pm0.16\pm 0.10$) eV \\
$\Gamma_{ee}\BR(\Upsilon(11020)\to\chi_{b2}\,(\pi^+\pi^-\pi^0)_{\rm non-\omega})$ & ($0.15\pm0.13\pm 0.10$)  eV / ($0.14\pm0.13\pm 0.10$) eV \\
\hline\hline
\end{tabular}
\end{table*}

\section{Summary}~\label{sec.9}

In summary, we study the processes $e^+e^-\to\chi_{bJ}\,\omega$ and $e^+e^-\to\chi_{bJ}\,(\pi^+\pi^-\pi^0)_{\rm non-\omega}$ ($J$ = 0, 1, 2) using Belle datasets with c.m.\ energies from 10.73 to 11.02 GeV and Belle~II datasets at $\sqrt{s}$ = 10.653, 10.701, 10.745, and 10.805 GeV. 
The results for the Born cross sections are given in tables~\ref{tabsumchib1},~\ref{tabsumchib2},~\ref{tabsumnon2},~\ref{tabsumnon3},~\ref{tabsumchib0}, and~\ref{tabsumnon1}. The energy dependences of the $e^+e^-\to\chi_{bJ}\,\omega$ cross sections show a prominent $\Upsilon(10753)$ signal, but no clear signals of the $\Upsilon(10860)$ or $\Upsilon(11020)$. The energy dependences of the $e^+e^-\to\chi_{bJ}\,(\pi^+\pi^-\pi^0)_{\rm non-\omega}$ cross sections show low-significance peaks for the $\Upsilon(10860)$ and $\Upsilon(11020)$, but no signal for the $\Upsilon(10753)$. We measure the mass and width of the $\Upsilon(10753)$ to be $M=(10756.1\pm3.4\pm2.7)$ MeV/$c^2$ and $\Gamma=(32.2\pm11.3\pm14.9)$ MeV. The results for electron widths times branching fractions are shown in table~\ref{dependency1tab}. The ratio $\BR(\Upsilon(10753)\to\chi_{b1}\,\omega)/\BR(\Upsilon(10753)\to\chi_{b2}\,\omega)$ = $1.13\pm0.38\pm0.34$ is consistent with the prediction for an $S-D$ mixed state of 0.2~\cite{034036} at the $1.8\,\sigma$ level.

Using the results of ref.~\cite{220}, we estimate $\sum_{n=1}^3\BR(\Upsilon(10753)\to\Upsilon(nS)\pi^+\pi^-)/\sum_{J=1}^2$ $\BR(\Upsilon(10753)\to\chi_{bJ}\,\omega)<0.9$ and $\sum_{n=1}^3\BR(\Upsilon(10860)\to\Upsilon(nS)\pi^+\pi^-)/\sum_{J=1}^2\BR(\Upsilon(10860)\to\chi_{bJ}\,\omega)>28$. Such a distinct decay pattern for two states with the same $J^{PC}$ quantum numbers, separated by 100\,MeV, indicates that they have different internal structures.

The $\Upsilon(10860)$ state, and possibly the $\Upsilon(11020)$ state, decay with a noticeable probability to $\chi_{bJ}\,(\pi^+\pi^-\pi^0)_{\rm non-\omega}$, while the corresponding decay of $\Upsilon(10753)$ is not observed. This decay pattern is expected if the $\chi_{bJ}\,(\pi^+\pi^-\pi^0)_{\rm non-\omega}$ final states are produced via intermediate $Z_b(10610)$ and $Z_b(10650)$ states~\cite{122001}, $\Upsilon(10860, 11020) \to Z_b\,\pi \to \chi_{bJ}\,\rho\,\pi$.
The decay $Z_b \to \chi_{bJ}\,\rho$ was predicted in ref.~\cite{0140362014}.
With larger data samples collected around $\Upsilon(10860)$ and $\Upsilon(11020)$ resonances at Belle II in the future, the $\chi_{bJ}\,\rho$ mass spectrum will be further investigated to search for $Z_b$ states.

\acknowledgments

This work, based on data collected using the Belle II detector, which was built and commissioned prior to March 2019,
and data collected using the Belle detector, which was operated until June 2010,
was supported by
Higher Education and Science Committee of the Republic of Armenia Grant No.~23LCG-1C011;
Australian Research Council and Research Grants
No.~DP200101792, 
No.~DP210101900, 
No.~DP210102831, 
No.~DE220100462, 
No.~LE210100098, 
and
No.~LE230100085; 
Austrian Federal Ministry of Education, Science and Research,
Austrian Science Fund (FWF) Grants
DOI:~10.55776/P34529,
DOI:~10.55776/J4731,
DOI:~10.55776/J4625,
DOI:~10.55776/M3153,
and
DOI:~10.55776/PAT1836324,
and
Horizon 2020 ERC Starting Grant No.~947006 ``InterLeptons'';
Natural Sciences and Engineering Research Council of Canada, Digital Research Alliance of Canada, and Canada Foundation for Innovation;
Fundamental Research Funds of China for the Central Universities No.~2242025RCB0014 and No.~RF1028623046;
National Key R\&D Program of China under Contract No.~2024YFA1610503, No.~2024YFA1610504, and No.~2022YFA1601903,
National Natural Science Foundation of China and Research Grants
No.~12475076,
No.~11575017,
No.~11761141009,
No.~11705209,
No.~11975076,
No.~12135005,
No.~12150004,
No.~12161141008,
No.~12405099,
No.~12475093,
and
No.~12175041,
and Shandong Provincial Natural Science Foundation Project~ZR2022JQ02;
the Czech Science Foundation Grant No. 22-18469S,  Regional funds of EU/MEYS: OPJAK
FORTE CZ.02.01.01/00/22\_008/0004632 
and
Charles University Grant Agency project No. 246122;
European Research Council, Seventh Framework PIEF-GA-2013-622527,
Horizon 2020 ERC-Advanced Grants No.~267104 and No.~884719,
Horizon 2020 ERC-Consolidator Grant No.~819127,
Horizon 2020 Marie Sklodowska-Curie Grant Agreement No.~700525 ``NIOBE''
and
No.~101026516,
and
Horizon 2020 Marie Sklodowska-Curie RISE project JENNIFER2 Grant Agreement No.~822070 (European grants);
L'Institut National de Physique Nucl\'{e}aire et de Physique des Particules (IN2P3) du CNRS
and
L'Agence Nationale de la Recherche (ANR) under Grant No.~ANR-21-CE31-0009 (France);
BMFTR, DFG, HGF, MPG, and AvH Foundation (Germany);
Department of Atomic Energy under Project Identification No.~RTI 4002,
Department of Science and Technology,
and
UPES SEED funding programs
No.~UPES/R\&D-SEED-INFRA/17052023/01 and
No.~UPES/R\&D-SOE/20062022/06 (India);
Israel Science Foundation Grant No.~2476/17,
U.S.-Israel Binational Science Foundation Grant No.~2016113, and
Israel Ministry of Science Grant No.~3-16543;
Istituto Nazionale di Fisica Nucleare and the Research Grants BELLE2,
and
the ICSC – Centro Nazionale di Ricerca in High Performance Computing, Big Data and Quantum Computing, funded by European Union – NextGenerationEU;
Japan Society for the Promotion of Science, Grant-in-Aid for Scientific Research Grants
No.~16H03968,
No.~16H03993,
No.~16H06492,
No.~16K05323,
No.~17H01133,
No.~17H05405,
No.~18K03621,
No.~18H03710,
No.~18H05226,
No.~19H00682, 
No.~20H05850,
No.~20H05858,
No.~22H00144,
No.~22K14056,
No.~22K21347,
No.~23H05433,
No.~26220706,
and
No.~26400255,
and
the Ministry of Education, Culture, Sports, Science, and Technology (MEXT) of Japan;  
National Research Foundation (NRF) of Korea Grants
No.~2021R1-F1A-1064008, 
No.~2022R1-A2C-1003993,
No.~2022R1-A2C-1092335,
No.~RS-2016-NR017151,
No.~RS-2018-NR031074,
No.~RS-2021-NR060129,
No.~RS-2023-00208693,
No.~RS-2024-00354342
and
No.~RS-2025-02219521,
Radiation Science Research Institute,
Foreign Large-Size Research Facility Application Supporting project,
the Global Science Experimental Data Hub Center, the Korea Institute of Science and
Technology Information (K25L2M2C3 ) 
and
KREONET/GLORIAD;
Universiti Malaya RU grant, Akademi Sains Malaysia, and Ministry of Education Malaysia;
Frontiers of Science Program Contracts
No.~FOINS-296,
No.~CB-221329,
No.~CB-236394,
No.~CB-254409,
and
No.~CB-180023, and SEP-CINVESTAV Research Grant No.~237 (Mexico);
the Polish Ministry of Science and Higher Education and the National Science Center;
the Ministry of Science and Higher Education of the Russian Federation
and
the HSE University Basic Research Program, Moscow;
University of Tabuk Research Grants
No.~S-0256-1438 and No.~S-0280-1439 (Saudi Arabia), and
Researchers Supporting Project number (RSPD2025R873), King Saud University, Riyadh,
Saudi Arabia;
Slovenian Research Agency and Research Grants
No.~J1-50010
and
No.~P1-0135;
Ikerbasque, Basque Foundation for Science,
State Agency for Research of the Spanish Ministry of Science and Innovation through Grant No. PID2022-136510NB-C33, Spain,
Agencia Estatal de Investigacion, Spain
Grant No.~RYC2020-029875-I
and
Generalitat Valenciana, Spain
Grant No.~CIDEGENT/2018/020;
the Swiss National Science Foundation;
The Knut and Alice Wallenberg Foundation (Sweden), Contracts No.~2021.0174 and No.~2021.0299;
National Science and Technology Council,
and
Ministry of Education (Taiwan);
Thailand Center of Excellence in Physics;
TUBITAK ULAKBIM (Turkey);
National Research Foundation of Ukraine, Project No.~2020.02/0257,
and
Ministry of Education and Science of Ukraine;
the U.S. National Science Foundation and Research Grants
No.~PHY-1913789 
and
No.~PHY-2111604, 
and the U.S. Department of Energy and Research Awards
No.~DE-AC06-76RLO1830, 
No.~DE-SC0007983, 
No.~DE-SC0009824, 
No.~DE-SC0009973, 
No.~DE-SC0010007, 
No.~DE-SC0010073, 
No.~DE-SC0010118, 
No.~DE-SC0010504, 
No.~DE-SC0011784, 
No.~DE-SC0012704, 
No.~DE-SC0019230, 
No.~DE-SC0021274, 
No.~DE-SC0021616, 
No.~DE-SC0022350, 
No.~DE-SC0023470; 
and
the Vietnam Academy of Science and Technology (VAST) under Grants
No.~NVCC.05.02/25-25
and
No.~DL0000.05/26-27.

These acknowledgements are not to be interpreted as an endorsement of any statement made
by any of our institutes, funding agencies, governments, or their representatives.

We thank the SuperKEKB team for delivering high-luminosity collisions;
the KEK cryogenics group for the efficient operation of the detector solenoid magnet and IBBelle on site;
the KEK Computer Research Center for on-site computing support; the NII for SINET6 network support;
and the raw-data centers hosted by BNL, DESY, GridKa, IN2P3, INFN, PNNL/EMSL, 
and the University of Victoria.

\renewcommand{\baselinestretch}{1.2}

\section{Appendix A}

Values of the inputs, resulting Born cross sections, and their upper limits for $e^+e^-\to \chi_{b0}\,\omega$ and $e^+e^-\to \chi_{b0}\,(\pi^+\pi^-\pi^0)_{\rm non-\omega}$ are listed in tables~\ref{tabsumchib0} and~\ref{tabsumnon1}, respectively.

\begin{table*}[htbp]
\renewcommand\arraystretch{1.2}
\setlength{\tabcolsep}{2pt}
\small
\centering
\caption{Results for $e^+e^-\to \chi_{b0}\omega$ at each energy point at Belle and Belle II. Column descriptions are the same in table~\ref{tabsumchib1}.}\label{tabsumchib0}
\vspace{0.2cm}
\begin{tabular}{cccccccccccc}
\hline\hline
&$\sqrt{s}$ (MeV) & $\cal L$ (fb$^{-1}$) & $\varepsilon$ & $1+\delta_{\rm ISR}$ & $N^{\rm sig}$ & $\Sigma(\sigma)$ & $\sigma_{\rm Born}$ (pb) & $\sigma^{\rm add}_{\rm syst}$ (pb) &  $\sigma^{\rm UL}_{\rm Born}$ (pb) \\\hline
$\ast$&10701.0	&	1.640	&	0.150	&	0.650	&	$	0.0	^{+	0.5	}_{-	0.2	}$	&- &	$	0.0	^{+	3.3	}_{-	1.3	}			$	&0.8&	11.4\phantom{1} \\
&10731.3	&	0.946	&	0.101	&	0.637	&	$	0.0	^{+	0.5	}_{-	0.2	}$	&-&	$	0.0	^{+	9.0	}_{-	3.6	}			$	&0.5&	41.6\phantom{1}	\\
$\ast$&10745.0    &	9.870	&	0.178	&	0.630	&	\underline{$	0.0	^{+	1.1	}_{-	0.0	}$}	&-&	$	0.0	^{+	1.1	}_{-	0.0	}	$	&0.3&	4.1	\\
&10771.2	&	0.955	&	0.106	&	0.786	&	\underline{$	1.0	^{+	1.4	}_{-	0.7	}$}	&1.5&	$	14.0	^{+	19.4	}_{-	10.0	}	$	&3.8&	55.7\phantom{1}	\\
$\ast$&10805.0    &	4.690  	&	0.177	&	0.940  	&	\underline{$	0.0	^{+	1.0	}_{-	0.0	}$}	&-&	$	0.0	^{+	1.4	}_{-	0.0	}	$	&0.3&	5.1	\\
&10829.5	&	1.697	&	0.104	&	0.943	&	$	0.0	^{+	0.5	}_{-	0.2	}$	&-&	$	0.0	^{+	3.3	}_{-	1.3	}			$	&0.3&	13.9\phantom{1}	\\
&10848.9	&	0.989	&	0.104	&	0.927	&	$	0.0	^{+	0.5	}_{-	0.2	}$	&-&	$	0.0	^{+	5.8	}_{-	2.3	}			$	&0.3&	24.3\phantom{1}	\\
&10857.4	&	0.988	&	0.104	&	0.919	&	$	0.0	^{+	0.5	}_{-	0.2	}$	&-&	$	0.0	^{+	5.8	}_{-	2.3	}			$	&0.3&	24.5\phantom{1}	\\
&10865.8	&	122.0	&	0.108	&	0.911	&	\underline{$	0.0	^{+	3.4	}_{-	0.0	}$}	&-&	$	0.0	^{+	0.3	}_{-	0.0	}	$	&0.1&	0.6	\\
&10877.8	&	0.978	&	0.106	&	0.901	&	$	0.0	^{+	0.5	}_{-	0.2	}$	&-&	$	0.0	^{+	5.8	}_{-	2.3	}			$	&0.3&	24.5\phantom{1}	\\
&10882.8	&	1.848	&	0.106	&	0.897	&	$	0.0	^{+	0.5	}_{-	0.2	}$	&-&	$	0.0	^{+	3.1	}_{-	1.2	}			$	&0.3&	13.0\phantom{1}	\\
&10888.9	&	0.990	&	0.106	&	0.893	&	$	0.0	^{+	0.5	}_{-	0.2	}$	&-&	$	0.0	^{+	5.8	}_{-	2.3	}			$	&0.3&	24.5\phantom{1}	\\
&10898.3	&	2.408	&	0.107	&	0.886	&	$	0.0	^{+	0.5	}_{-	0.2	}$	&-&	$	0.0	^{+	2.4	}_{-	1.0	}			$	&0.3&	9.1	\\
&10907.3	&	0.980	&	0.107	&	0.881	&	$	0.0	^{+	0.5	}_{-	0.2	}$	&-&	$	0.0	^{+	5.9	}_{-	2.4	}			$	&0.3&	24.8\phantom{1}	\\
&10928.7	&	1.149	&	0.108	&	0.871	&	$	0.0	^{+	0.5	}_{-	0.2	}$	&-&	$	0.0	^{+	5.1	}_{-	2.0	}			$	&0.3&	23.3\phantom{1}	\\
&10957.5	&	0.969	&	0.108	&	0.860	&	$	0.0	^{+	0.5	}_{-	0.2	}$	&-&	$	0.0	^{+	6.1	}_{-	2.4	}			$	&0.3&	27.9\phantom{1}	\\
&10975.3	&	0.999	&	0.109	&	0.856	&	$	0.0	^{+	0.5	}_{-	0.2	}$	&-&	$	0.0	^{+	5.9	}_{-	2.3	}			$	&0.3&	24.6\phantom{1}	\\
&10990.4	&	0.985	&	0.109	&	0.852	&	$	0.0	^{+	0.5	}_{-	0.2	}$	&-&	$	0.0	^{+	6.0	}_{-	2.4	}			$	&0.3&	25.1\phantom{1}	\\
&11003.9	&	0.976	&	0.110	&	0.850	&	$	0.0	^{+	0.5	}_{-	0.2	}$	&-&	$	0.0	^{+	6.0	}_{-	2.4	}			$	&0.3&	25.1\phantom{1}	\\
&11014.8	&	0.771	&	0.110	&	0.849	&	$	0.0	^{+	0.5	}_{-	0.2	}$	&-&	$	0.0	^{+	7.6	}_{-	3.0	}			$	&0.3&	34.9\phantom{1}	\\
&11018.5	&	0.859	&	0.110	&	0.848	&	$	0.0	^{+	0.5	}_{-	0.2	}$	&-&	$	0.0	^{+	6.8	}_{-	2.7	}			$	&0.3&	31.4\phantom{1}	\\
&11020.8	&	0.982	&	0.111	&	0.848	&	$	0.0	^{+	0.5	}_{-	0.2	}$	&-&	$	0.0	^{+	5.9	}_{-	2.4	}			$	&0.3&	24.8\phantom{1}	\\
\hline\hline
\end{tabular}
\end{table*}

\begin{table*}[htbp]
\renewcommand\arraystretch{1.2}
\setlength{\tabcolsep}{2pt}
\small
\centering
\caption{Results for $e^+e^-\to\chi_{b0}(\pi^+\pi^-\pi^0)_{\rm non-\omega}$ at each energy point at Belle and Belle II. 
Column descriptions are the same as in table~\ref{tabsumchib1}.}\label{tabsumnon1}
\vspace{0.2cm}
\begin{tabular}{ccccccccccc}
\hline\hline
&$\sqrt{s}$ (MeV) & $\cal L$ (fb$^{-1}$) & $\varepsilon$ & $1+\delta_{\rm ISR}$ & $N^{\rm sig}$ & $\Sigma(\sigma)$ & $\sigma_{\rm Born}$ (pb) & $\sigma^{\rm add}_{\rm syst}$ (pb) & $\sigma^{\rm UL}_{\rm Born}$ (pb) \\\hline
$\ast$	&	10653.0	&	3.550	&	0.086	&	0.682	&	$	0.0	^{+	0.5	}_{-	0.2	}$	&-&	$	0.0	^{+	2.4	}_{-	0.9	}	$	&	0.6			&	8.1	\\
$\ast$	&	10701.0	&	1.640	&	0.102	&	0.692	&	$	0.0	^{+	0.5	}_{-	0.2	}$	&-&	$	0.0	^{+	4.2	}_{-	1.7	}	$	&	0.6			&	14.4\phantom{1}	\\
	&	10731.3	&	0.946	&	0.077	&	0.701	&	$	0.0	^{+	0.5	}_{-	0.2	}$	&-&	$	0.0	^{+	9.6	}_{-	3.8	}	$	&	0.4			&	40.2\phantom{1}	\\
$\ast$	&	10745.0	&	9.870	&	0.143	&	0.701	&	\underline{$	0.0	^{+	0.9	}_{-	0.0	}$}	&-&	$	0.0	^{+	0.9	}_{-	0.0	}	$	&	0.3			&	3.6	\\
	&	10771.2	&	0.955	&	0.091	&	0.698	&	$	0.0	^{+	0.5	}_{-	0.2	}$	&-&	$	0.0	^{+	8.1	}_{-	3.2	}	$	&	0.4			&	33.9\phantom{1}	\\
$\ast$	&	10805.0	&	4.690	&	0.157	&	0.684	&	$	0.5	^{+	2.2	}_{-	0.5	}$	&-&	$	1.0	^{+	4.3	}_{-	1.0	}	$	&	0.6			&	8.5	\\
	&	10829.5	&	1.697	&	0.095	&	0.668	&	$	0.6	^{+	1.4	}_{-	0.6	}$	&-&	$	\phantom{1}5.5^{+12.7}_{-5.5}	$	&	0.3			&	37.3\phantom{1}	\\
	&	10848.9	&	0.989	&	0.095	&	0.649	&	$	0.0	^{+	0.5	}_{-	0.2	}$	&-&	$	0.0	^{+	8.0	}_{-	3.2	}	$	&	0.3			&	33.7\phantom{1}	\\
	&	10857.4	&	0.988	&	0.095	&	0.640	&	$	0.0	^{+	0.5	}_{-	0.2	}$	&-&	$	0.0	^{+	8.2	}_{-	3.3	}	$	&	0.3			&	34.3\phantom{1}	\\
	&	10865.8	&	122.0	&	0.098	&	0.631	&	\underline{$	10.8	^{+	7.0\phantom{2}	}_{-	6.3	}$}	&1.8&	$	1.4	^{+	0.9	}_{-	0.8	}	$	&	0.3			&	2.7	\\
	&	10877.8	&	0.978	&	0.097	&	0.631	&	$	2.8	^{+	1.9	}_{-	1.4	}$	&-&	$	45.6	^{+	30.9	}_{-	22.8	}	$	&	0.3			&	115.5\phantom{5}~~	\\
	&	10882.8	&	1.848	&	0.097	&	0.642	&	$	0.0	^{+	0.5	}_{-	0.2	}$	&-&	$	0.0	^{+	4.2	}_{-	1.7	}	$	&	0.3			&	16.1\phantom{1}	\\
	&	10888.9	&	0.990	&	0.097	&	0.670	&	$	0.0	^{+	0.5	}_{-	0.2	}$	&-&	$	0.0	^{+	7.6	}_{-	3.0	}	$	&	0.3			&	31.8\phantom{1}	\\
	&	10898.3	&	2.408	&	0.098	&	0.743	&	$	0.4	^{+	1.5	}_{-	0.4	}$	&-&	$	2.2	^{+	8.3	}_{-	2.2	}	$	&	0.3			&	20.6\phantom{1}	\\
	&	10907.3	&	0.980	&	0.098	&	0.826	&	$	0.0	^{+	0.5	}_{-	0.2	}$	&-&	$	0.0	^{+	6.1	}_{-	2.5	}	$	&	0.3			&	25.8\phantom{1}	\\
	&	10928.7	&	1.149	&	0.101	&	0.964	&	$	0.7	^{+	1.4	}_{-	0.7	}$	&-&	$	\phantom{1}6.1	^{+	12.2	}_{-	6.1	}	$	&	0.3			&	34.1\phantom{1}	\\
	&	10957.5	&	0.969	&	0.104	&	0.853	&	$	0.0	^{+	0.5	}_{-	0.2	}$	&-&	$	0.0	^{+	5.7	}_{-	2.3	}	$	&	0.3			&	24.0\phantom{1}	\\
	&	10975.3	&	0.999	&	0.107	&	0.708	&	$	0.8	^{+	1.5	}_{-	0.7	}$	&-&	$	10.4	^{+	19.5	}_{-	9.1	}	$	&	0.3			&	53.2\phantom{1}	\\
	&	10990.4	&	0.985	&	0.108	&	0.627	&	$	0.0	^{+	0.5	}_{-	0.2	}$	&-&	$	0.0	^{+	7.4	}_{-	2.9	}	$	&	0.3			&	30.9\phantom{1}	\\
	&	11003.9	&	0.976	&	0.109	&	0.674	&	$	0.0	^{+	0.5	}_{-	0.2	}$	&-&	$	0.0	^{+	6.8	}_{-	2.7	}	$	&	0.3			&	28.6\phantom{1}	\\
	&	11014.8	&	0.771	&	0.110	&	0.843	&	$	0.0	^{+	0.5	}_{-	0.2	}$	&-&	$	0.0	^{+	6.9	}_{-	2.7	}	$	&	0.3			&	28.8\phantom{1}	\\
	&	11018.5	&	0.859	&	0.110	&	0.912	&	$	0.0	^{+	0.5	}_{-	0.2	}$	&-&	$	0.0	^{+	5.7	}_{-	2.3	}	$	&	0.3			&	23.8\phantom{1}	\\
	&	11020.8	&	0.982	&	0.111	&	0.956	&	$	0.0	^{+	0.5	}_{-	0.2	}$	&-&	$	0.0	^{+	4.7	}_{-	1.9	}	$	&	0.3			&	19.7\phantom{1}	\\
\hline\hline
\end{tabular}
\end{table*} 

\section{Appendix B}\label{AppendixB}

The fitted values of $a$, $p_1$, $p_2$, $p_3$, $q_1$, $q_2$, $q_3$, and $q_4$ (see eqs.~(\ref{eq:likelihood1}) and (\ref{eq:likelihood2})) for the processes $e^+e^- \to \chi_{bJ}\, \omega$ and
$e^+e^-\to\chi_{bJ}\,(\pi^+\pi^-\pi^0)_{\rm non-\omega}$ at each energy point are provided
in the supplemental material~\cite{SM}.

For the convenience of the reader, we also provide profile likelihoods
for well-populated energy points. Uncorrelated systematic uncertainties
are included in these profile likelihoods, allowing them to be used
directly in fits to the energy dependence on the same footing as
sparsely populated points.

This procedure yields results identical to those obtained by treating
well-populated points with asymmetric statistical uncertainties, adding
uncorrelated systematic uncertainties in quadrature, and including the
corresponding $\chi^2$ contributions in the energy-dependence fit.

\end{document}